\documentclass[manuscript]{aastex62}

\usepackage{bbm}
\usepackage{amsmath}
\usepackage{algorithm}
\usepackage{algorithmic}
\usepackage[caption=false]{subfig}
\usepackage{soul}

\newcommand{\angstrom}{\mbox{\normalfont\AA}}

\DeclareMathOperator*{\argmin}{argmin}

\received{March 3, 2019}
\revised{April 11, 2019}
\accepted{April 20, 2019}
\submitjournal{AJ}

\shorttitle{Modeling the Echelle Spectra Continuum}
\shortauthors{Xu et al.}

\begin{document}

\title{Modeling the Echelle Spectra Continuum with Alpha Shapes and Local Regression Fitting}

\correspondingauthor{Xin Xu}
\email{xin.xu@yale.edu}

\author[0000-0001-5574-745X]{Xin Xu}
\affil{Department of Statistics and Data Science, Yale University\\ 24 Hillhouse Ave, New Haven, CT 06511, USA}

\author[0000-0002-9656-2272]{Jessi Cisewski-Kehe}
\affil{Department of Statistics and Data Science, Yale University\\ 24 Hillhouse Ave, New Haven, CT 06511, USA}

\author[0000-0002-5070-8395]{Allen B. Davis}
\affil{Department of Astronomy, Yale University\\ 52 Hillhouse Ave, New Haven, CT 06511, USA}

\author[0000-0003-2221-0861]{Debra A. Fischer}
\affil{Department of Astronomy, Yale University\\ 52 Hillhouse Ave, New Haven, CT 06511, USA}

\author[0000-0002-9873-1471]{John M. Brewer}
\affil{Department of Astronomy, Yale University\\ 52 Hillhouse Ave, New Haven, CT 06511, USA}

\begin{abstract}

Continuum normalization of echelle spectra is an important data analysis step that is difficult to automate. Polynomial fitting requires a reasonably high order model to follow the steep slope of the blaze function. However, in the presence of deep spectral lines, a high order polynomial fit can result in ripples in the normalized continuum that increase errors in spectral analysis. Here, we present two algorithms for flattening the spectrum continuum. The Alpha-shape Fitting to Spectrum algorithm (AFS) is completely data-driven, using an alpha shape to obtain an initial estimate of the blaze function. The Alpha-shape and Lab Source Fitting to Spectrum algorithm (ALSFS) incorporates a continuum constraint from a lab source reference spectrum for the blaze function estimation. These algorithms are tested on a simulated spectrum, where we demonstrate improved normalization compared to polynomial regression for continuum fitting. 
We show an additional application, using the algorithms for mitigation of spatially correlated quantum efficiency variations and fringing in the CCD detector of the EXtreme PREcision Spectrometer (EXPRES).

\end{abstract}


\section{Introduction} \label{sec:intro}

Spectroscopy is a powerful observational technique for understanding fundamental astrophysics. A high dispersion stellar spectrum contains detailed information about individual atomic transitions that enables the derivation of parameters such as effective temperature, surface gravity, elemental abundances, and the measurement of Doppler shifts in stellar spectra reveals the presence of stellar and planetary companions \citep[and references within]{fischer2016state}.  
Most spectroscopic analysis techniques require a flat, continuum normalized spectrum \citep{blanco2014gaia}. For example, the precision of equivalent width measurements for abundance analysis or cross-correlation for exoplanet detection is very sensitive to even small errors in continuum normalization \citep[and references within]{torres2012improved}.

An echelle spectrograph disperses light so that high spectral orders can be recorded. The higher orders have greater dispersion and therefore provide higher resolution spectra over a broad range of wavelengths. However, most of the brightness of a spectrum is concentrated in the zeroth and lower spectral orders; higher orders are intrinsically fainter. The optical grating in an echelle spectrograph has angled facets designed to shift the intensity envelope of dispersed light to high spectral orders. When this phase shift is introduced, the grating is said to be blazed and with cross dispersion, several dozen spectral orders can be stacked onto the detector. Each order has an intensity distribution characterized by the blaze function so that the continuum intensity is strongest in the center of the order and drops off steeply toward the edges of the order. The term ``blaze function'' is often used to describe the shape of the continuum across an echelle order and indeed, the blaze distribution is the dominant effect. 

One approach for flattening the spectrum is to divide by the theoretical blaze function \citep{barker1984ripple}, which depends on grating parameters (incident angle, the angle of the blazed facets, and the grating facet width and spacing) and can be calculated for each order of the spectrum. However, this method for normalization or flattening of the spectrum will leave residual variations in the continuum because of departures from the theoretical blaze function: manufacturing defects in the grating, chromatic aberrations in the optics of a spectrograph, or quantum efficiency variations in the electronic detector. In addition to the instrumental blaze, additional wavelength-dependent intensity variations will occur because of the black-body temperature of stars or calibration lamps. 

Another approach for normalizing the continuum is to divide by an extracted flat-field calibration source \citep[e.g.,][]{skoda2008investigation}, such as a quartz lamp. However, the quartz lamp will also have a black-body curve superimposed on the blaze function. 
One of the most common methods for normalizing the continuum of a spectrum is to fit a polynomial to high points along the order. This is a completely agnostic approach that does not require prior knowledge about the blaze function and while it works fairly well, polynomial fitting can fail near broad and deep lines, especially if they are close to the edges of the orders. Here, we describe a new approach for continuum fitting and compare our method with polynomial fitting in Section~\ref{simulation}.

\section{Description of methods}

In this work, the goal is to estimate the blaze function of a target spectrum so that it may be removed, leaving a flattened spectrum. Two algorithms are proposed to accommodate different scenarios: (i) the \emph{Alpha-shape Fitting to Spectrum algorithm (AFS)} (the baseline approach), (ii) the \emph{Alpha-shape and Lab Source Fitting to Spectrum algorithm (ALSFS)} (when a lab continuum source is available)\footnote{The implementation code of AFS and ALSFS, as well as example data can be
found and downloaded from: https://github.com/xinxuyale/AFS or https://zenodo.org/badge/latestdoi/173169370.}.

In the proposed algorithms, we first fit an \textit{alpha shape} \citep{edelsbrunner1983shape}, which is a polygon enclosing a dataset, to capture the general shape of the blaze function of a target spectrum. Then we use this preliminary estimation to select a set of pixels that are ultimately used to fit the final blaze function model. The pixels are selected so that they are generally on or near the continuum and not in the absorption lines. With the selected set of pixels, we use local polynomial regression \citep{cleveland1979robust} to estimate the blaze function. The two algorithms are introduced next, followed by a discussion on how to select the tuning parameters.

\subsection{AFS Algorithm}\label{algorithm1}

\begin{figure}
\centering 
\subfloat[Alpha hull]{%
  \includegraphics[width=0.45\columnwidth]{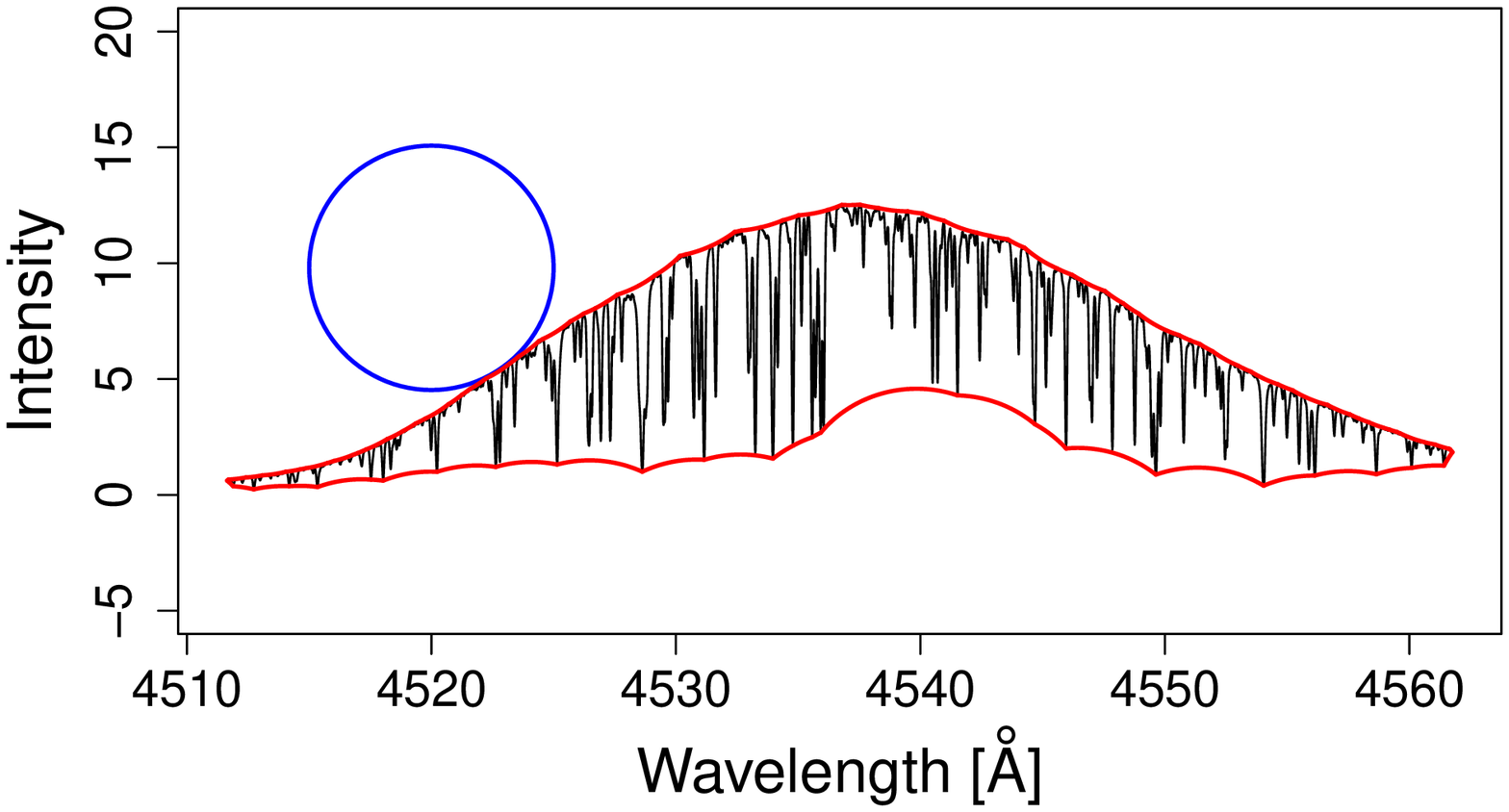}%
  \label{alpha_hull}%
}\qquad
\subfloat[Alpha shape]{%
  \includegraphics[width=0.45\columnwidth]{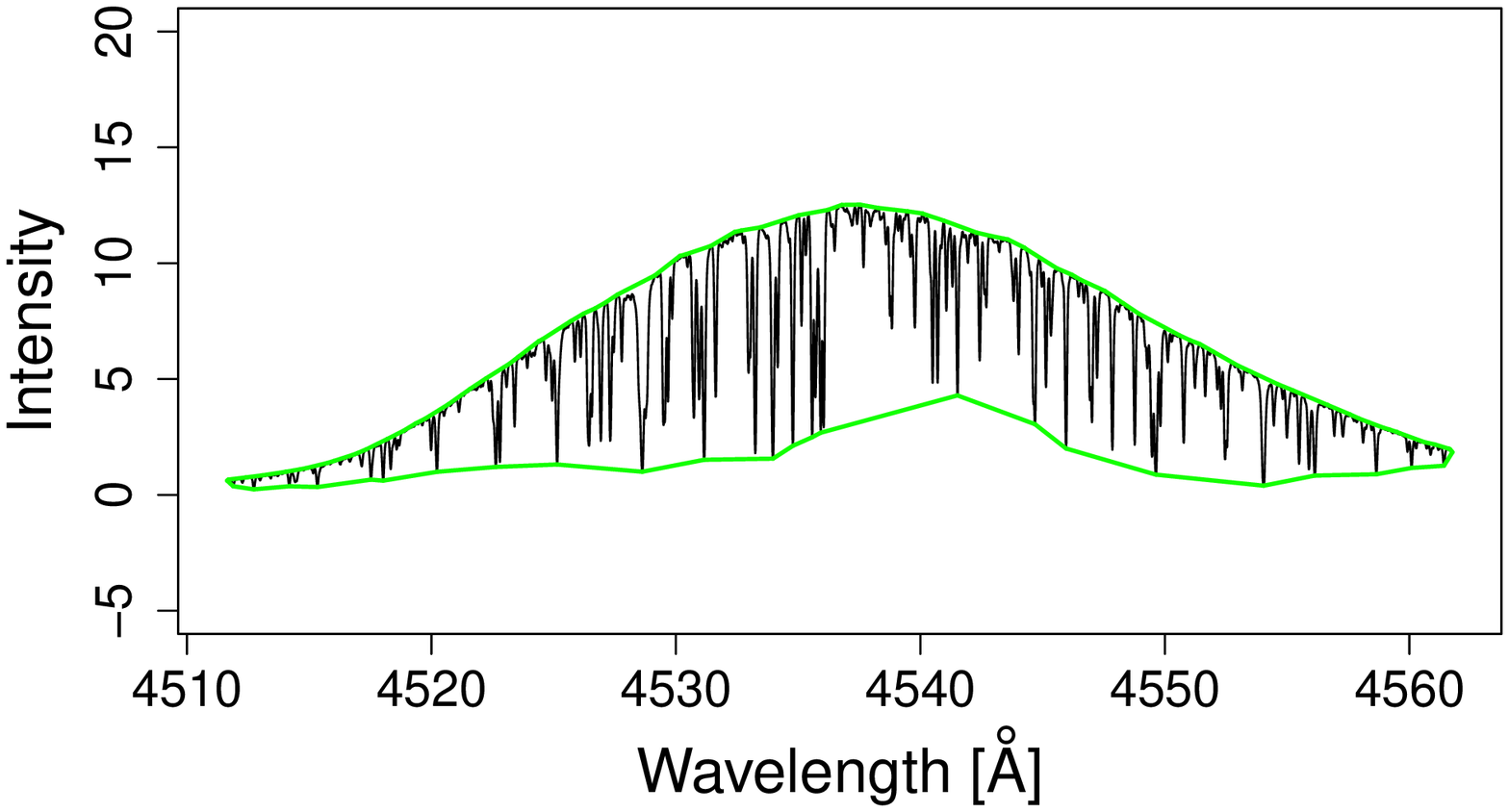}%
  \label{alpha_shape}%
}
\caption{(a) An alpha hull with $\alpha=5$ in red. The blue circle is an example of $\alpha$-ball illustrating the process of generating the alpha hull. (b) The resulting alpha shape in green, by straightening the red arcs from (a).}
\label{alpha_shape_illustration}
\end{figure}

The proposed AFS algorithm is a versatile algorithm that can be used to remove the blaze function whether or not a corresponding lab source spectrum is available. 
In this algorithm, an alpha shape is used to obtain a preliminary estimation of the blaze function's high-level shape. An alpha shape is a generalization of a convex hull, but is not required to be a convex set; it is a region bounded by a set of segments generated from a set of points.
To generate an alpha shape, imagine a piece of paper with a plot of spectrum printed on it. Consider a special paper cutter that can only cut out circles with radius $\alpha$, known as \textit{$\alpha$-balls}. An incomplete circle is allowed, but the cutter cannot cut anything from the spectrum. In Figure\autoref{alpha_hull}, the blue circle is an $\alpha$-ball that can be cut out from the paper, but we cannot move the $\alpha$-ball any lower vertically since it would cut the spectrum. 
Continue to cut as many $\alpha$-balls as possible, and the remaining paper is called an \textit{alpha hull}. Figure\autoref{alpha_hull} shows the resulting alpha hull with $\alpha=5$. By connecting the points where the alpha hull touches the spectrum with straight segments, it becomes an alpha shape, as displayed in Figure\autoref{alpha_shape}. An alpha shape can capture the general shape of a spectrum, and its upper boundary is used as a starting model for the blaze function estimation.

Another technique used in the AFS algorithm is local polynomial regression \citep{cleveland1979robust}, which is a non-parametric method for estimating functions given a set of points. At each point, $(\lambda_i, y_i)$, a p-degree polynomial model is fitted to a subset of neighboring points of $\lambda_i$. The p-degree polynomial regression is fitted by weighted least squares, giving more weight to points closer to $\lambda_i$ and less weight to points further away. 
Consider a dataset $(\lambda_1, y_1), (\lambda_2, y_2), \dots, (\lambda_n, y_n)$, and a p-degree polynomial function in the neighborhood of $\lambda_i$ is $f_{\beta^{(\lambda_i)}}(\lambda_j)=\beta_0^{(\lambda_i)}+\beta_1^{(\lambda_i)}(\lambda_j-\lambda_i)+\beta_2^{(\lambda_i)}(\lambda_j-\lambda_i)^2+\dots+\beta_p^{(\lambda_i)}(\lambda_j-\lambda_i)^p$, where $\lambda_j$ is in the neighborhood of $\lambda_i$. Then the estimation of $\beta^{(\lambda_i)}$,  $\hat{\beta}^{(\lambda_i)}=(\hat{\beta}_0^{(\lambda_i)}, \hat{\beta}_1^{(\lambda_i)}, \dots, \hat{\beta}_p^{(\lambda_i)})$, is obtained by minimizing
\begin{equation}
\sum_{\substack{\lambda_j \in N_{m_0}(\lambda_i)}} \omega_i(\lambda_j)( y_j-f_{\beta^{(\lambda_i)}}(\lambda_j))^2,
\end{equation}
where $N_{m_0}(\lambda_i)$ is the neighboring set of $\lambda_i$ containing the nearest $\lfloor m_0n\rfloor$ pixels to $\lambda_i$, $m_0$ is a smoothing parameter, and $\omega_i(\lambda_j)=K\Big(\frac{|\lambda_j-\lambda_i|}{\max\limits_{\substack{\lambda_{j^*}\in N_{m_0}(\lambda_i)}} |\lambda_{j^*}-\lambda_i|}\Big)$, with $K(x)=(1-x^3)^3$ as a common weighting function.
The local polynomial estimate at $\lambda_i$ is $\hat{\beta}_0^{(\lambda_i)}$. 
An advantage of local polynomial regression over ordinary polynomial regression is its ability to adapt to local characteristics of a dataset rather than fitting all the data points using one single model. This flexibility makes it useful for modeling complex data sets where regular polynomial regression fails.

\begin{figure}
\centering 
\subfloat[Alpha shape]{%
  \includegraphics[width=0.45\columnwidth]{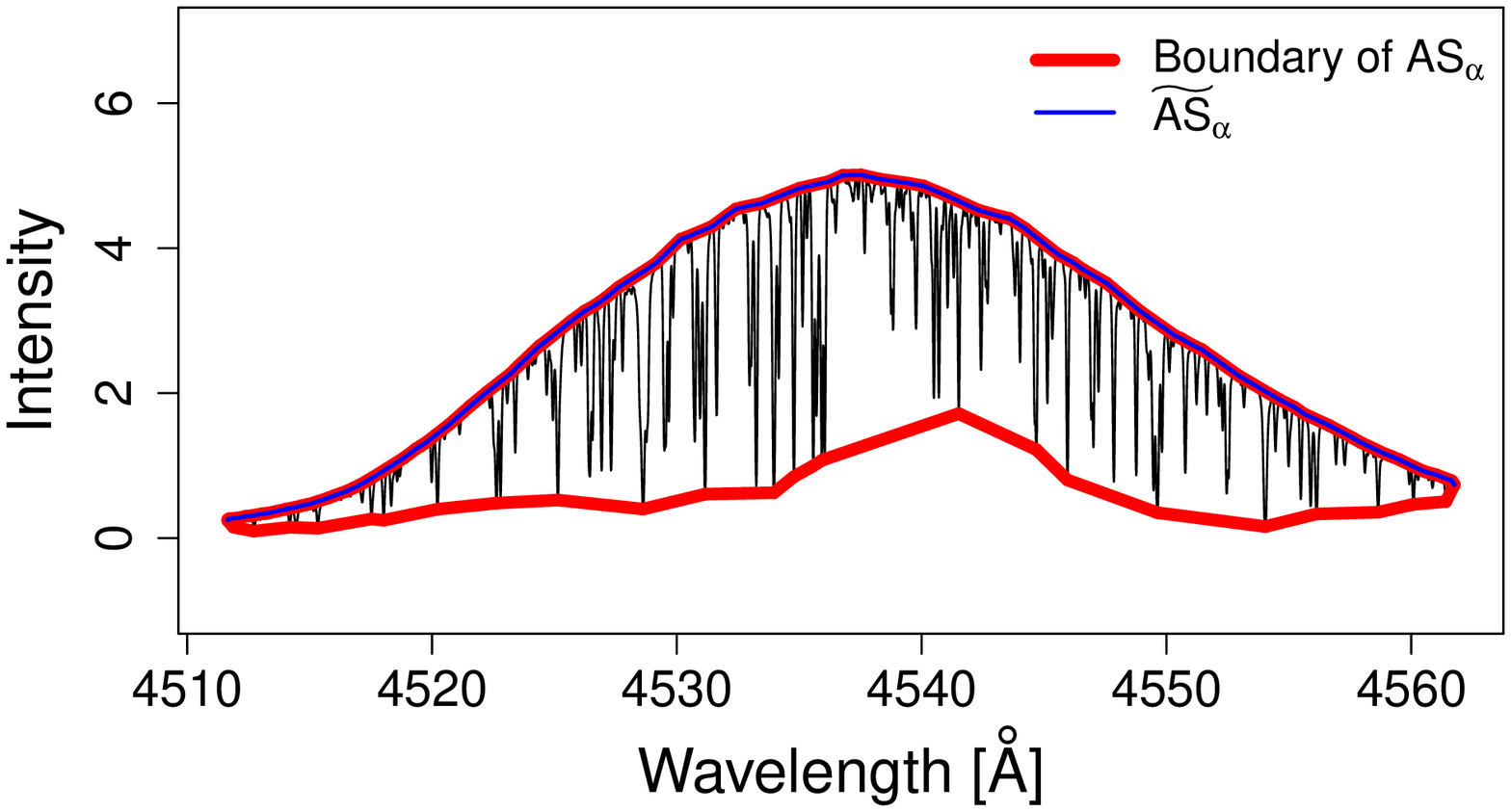}%
  \label{algo1_1}%
}\qquad
\subfloat[Primary estimation]{%
  \includegraphics[width=0.45\columnwidth]{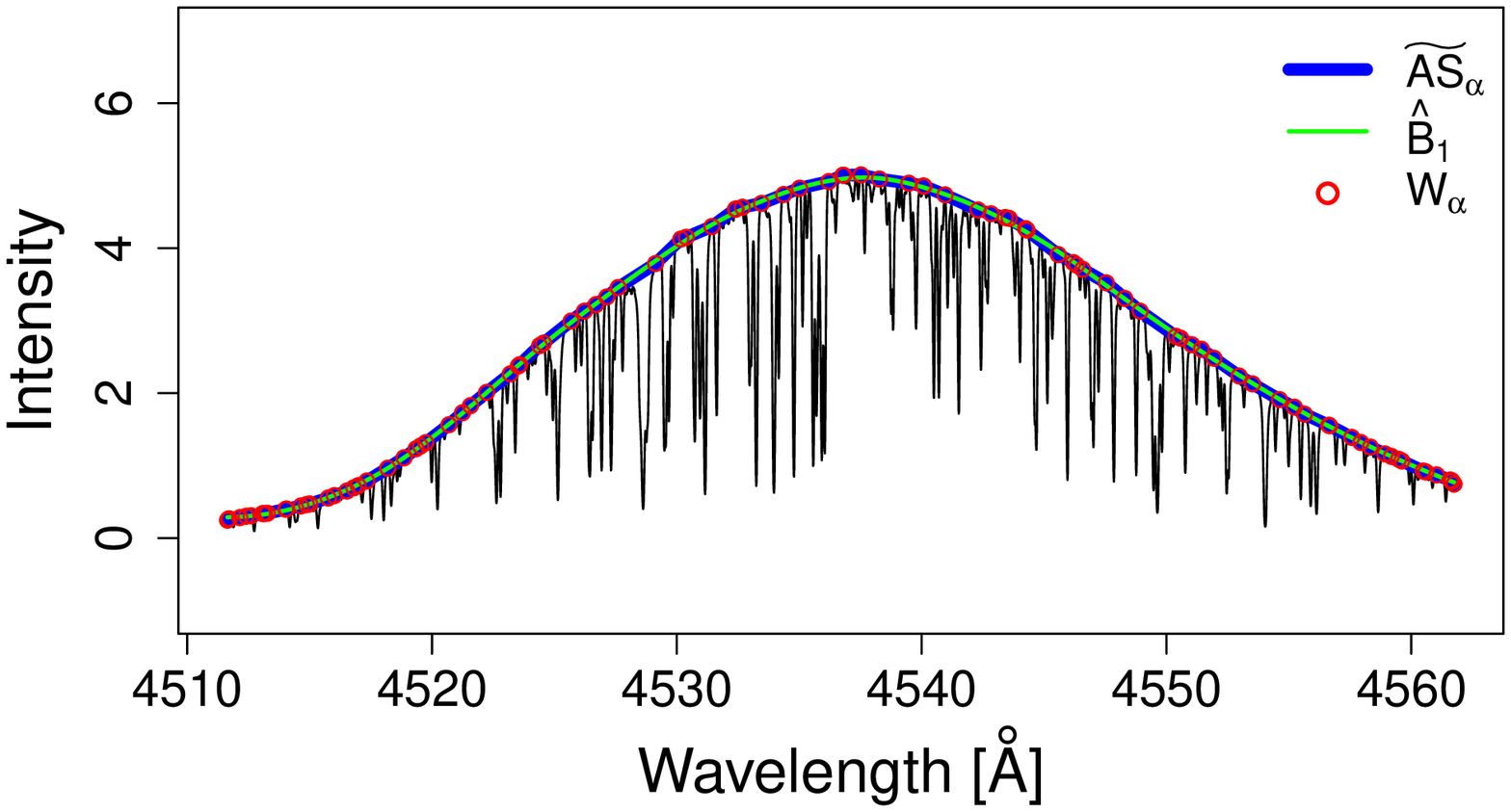}%
  \label{algo1_2}%
}\qquad
\subfloat[Points selected]{%
  \includegraphics[width=0.45\columnwidth]{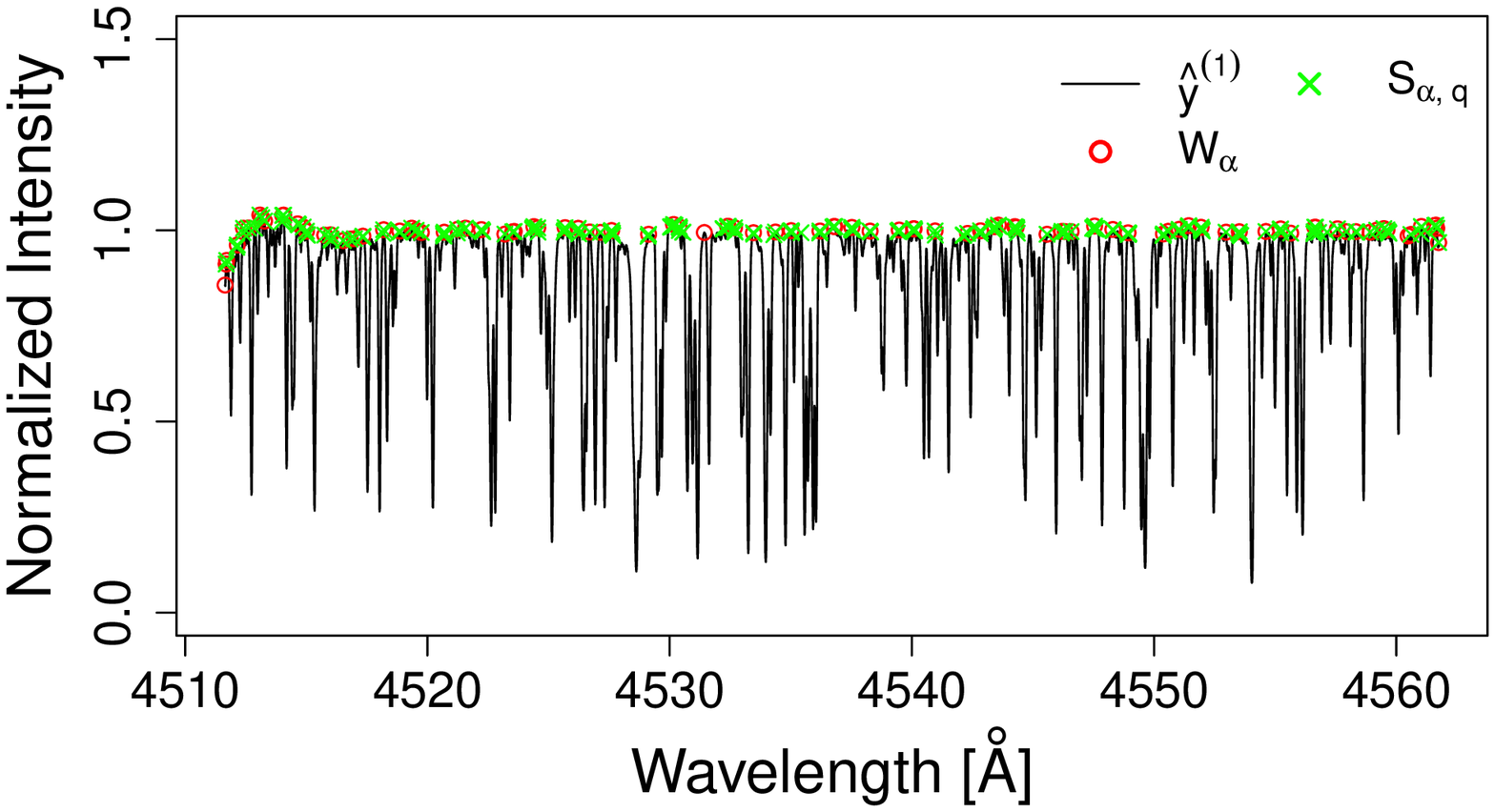}%
  \label{algo1_3}%
}\qquad
\subfloat[Local polynomial fitting]{%
  \includegraphics[width=0.45\columnwidth]{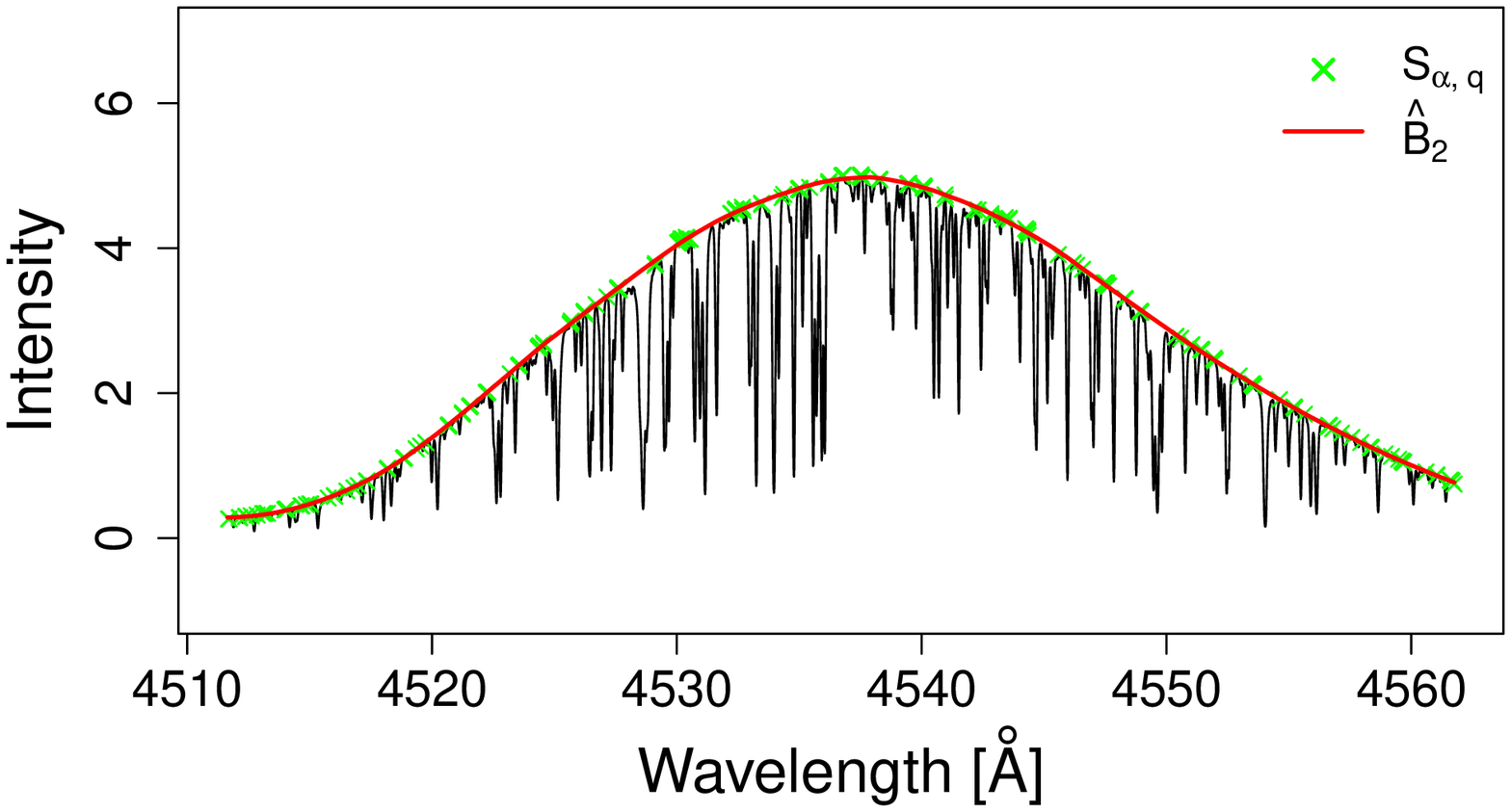}%
  \label{algo1_4}%
}\qquad
\subfloat[Final estimation]{%
  \includegraphics[width=0.45\columnwidth]{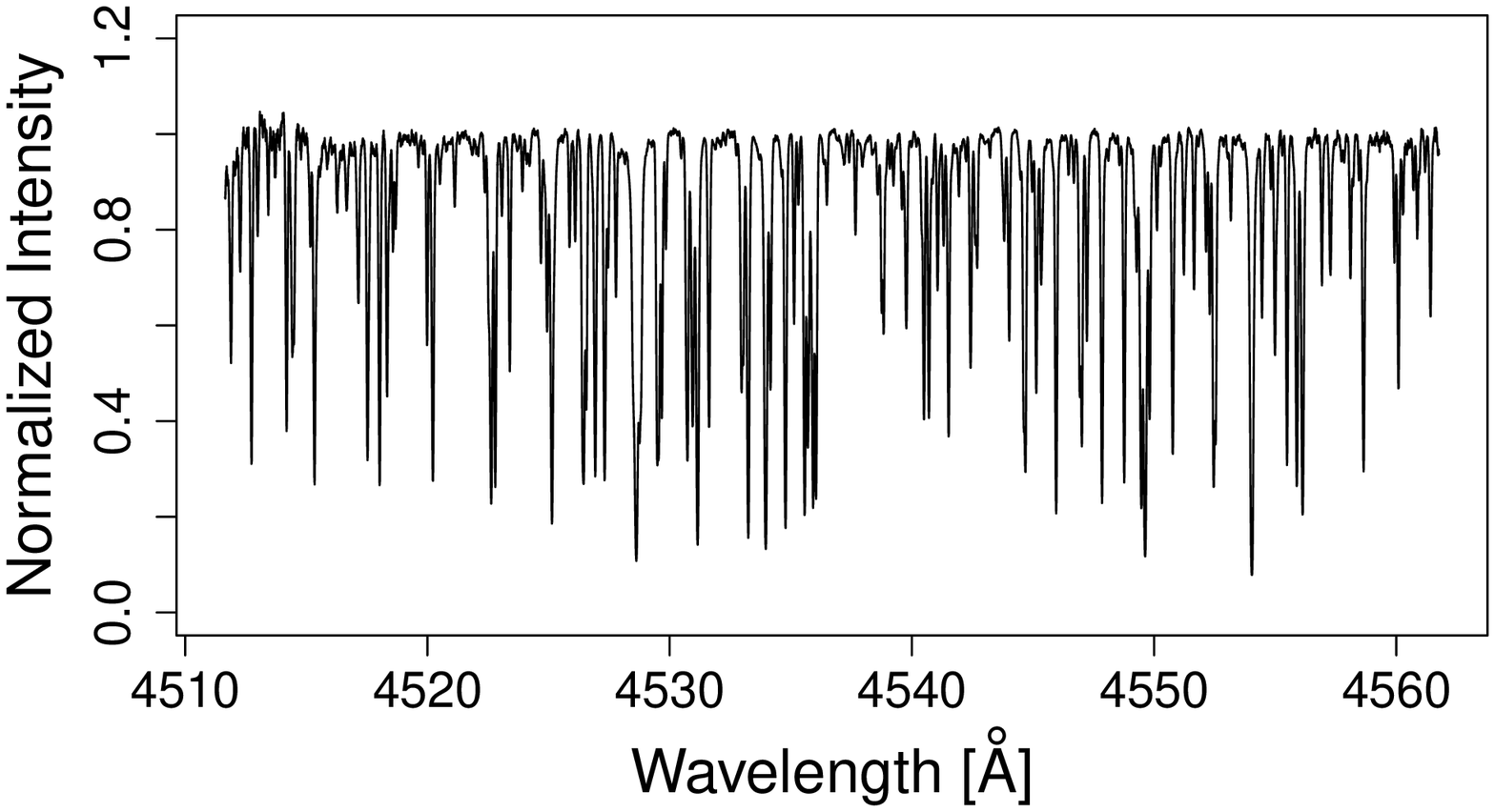}%
  \label{algo1_5}%
}
\caption{Illustration of the AFS \autoref{AFS_algo}.  (a) Steps 1 and 2: the alpha shape of the whole spectrum $AS_{\alpha}$ and its upper boundary $\widetilde{AS}_{\alpha}$. The red arcs represent the boundary of the alpha shape $AS_{\alpha}$, and the blue line connects the points in $\widetilde{AS}_{\alpha}$. (b) Step 3: a smoothed version of $\widetilde{AS}_{\alpha}$, denoted as $\hat{B}_1$ (in green). The red circles are $W_{\alpha}$: the intersection of $\widetilde{AS}_{\alpha}$ and the spectrum $\{(\lambda_i, y_i)\}_{i=1}^n$. (c) The last part of step 3, and step 4: divide $y$ by $\hat{B}_1$ to get a primary blaze-removed spectrum $\hat{y}^{(1)}$ (in black) and select points to the set $S_{\alpha, q}$ (green exes) using $\hat{y}^{(1)}$. (d) Step 5: a local polynomial fitting $\hat{B}_{2}$ (in red) using points in $S_{\alpha, q}$ (green exes), which is the final estimation of the blaze function. (e) Step 6: final blaze-removed spectrum.} 
\label{algo_figure}
\end{figure}

Let $\{(\lambda_i, y_i)\}_{i=1}^n$ be an observed spectrum, where $\lambda_i$ is the wavelength of pixel $i$, and $y_i$ is the intensity of pixel $i$. Our method is summarized in \autoref{AFS_algo} and illustrated in \autoref{algo_figure} with the details of the AFS algorithm presented next.

In step 1, the intensity vector, $y=(y_1, y_2, \dots, y_n)$, is rescaled by multiplying by a value $u=\frac{\max(\lambda) - \min(\lambda)}{10\times \max(y)}$. This $u$ corresponds to the $\alpha$ value for the alpha shape. Since the construction of an alpha shape depends on coverage areas of small circles, the relative range of $\lambda$ and $y$ affects results:  
if the range of $y$ is too large or too small compared to the range of $\lambda$, larger alpha values should be used for the alpha shape. For a common blaze function, $u=\frac{\max(\lambda) - \min(\lambda)}{10\times \max(y)}$ works well by scaling the range of $y$ and $\lambda$ to be $1:10$, in coordination with a recommended value for $\alpha$ later.
In step 2, the alpha shape, $AS_{\alpha}$, is constructed with radius $\alpha$, which is an infinite point set containing all the points within the boundary of the alpha shape (see Figure\autoref{algo1_1}). 
In $AS_{\alpha}$, for each $\lambda_i$, there are infinite $y^*_i$ such that $(\lambda_i, y^*_i) \in AS_{\alpha}$. The upper boundary of $AS_{\alpha}$ is defined as $\widetilde{AS}_{\alpha}$, which is a finite point set only including the largest $y^*_i$ such that $(\lambda_i, y^*_i) \in AS_{\alpha}$ for each $\lambda_i$, as displayed in Figure\autoref{algo1_1}.
In step 3, we fit a local polynomial regression using all the points in $\widetilde{AS}_{\alpha}$, denoted as $\hat{B}_1$, such that $\hat{B}_1$ is a smoothed version of $\widetilde{AS}_{\alpha}$ and is the initial model of the blaze function (see Figure\autoref{algo1_2}). 
$\hat{B}_1$ is not an accurate estimate of the blaze function, but rather an approximation of its shape. 
Next, $y$ is divided by $\hat{B}_1$ to get the first estimate of the flattened spectrum, denoted as $\hat{y}^{(1)}$, displayed in Figure\autoref{algo1_3}.

In step 4, we denote the intersection of $\widetilde{AS}_{\alpha}$ and the spectrum $\{(\lambda_i, y_i)\}_{i=1}^n$ as $W_{\alpha}$, shown in Figure\autoref{algo1_2} and\autoref{algo1_3}. Notice that $W_{\alpha}$ contains points mostly near the continuum. The blue line in Figure\autoref{algo1_2}, which connects the points in $\widetilde{AS}_{\alpha}$, can be thought of as a collection of segments, and points in $W_{\alpha}$ are vertices of those segments. 
Each point in $W_{\alpha}$ is a vertex of a window where splits are defined - 
the spectrum is cut into small windows by the points in $W_{\alpha}$ in order to get the local quantiles.
The purpose of the next steps is to select a subset of points that do not fall into an absorption feature. This is accomplished by selecting $\hat{y}_i^{(1)}$ in the upper quantiles of these windows. 
In particular, in the j-th window, we select the points whose $\hat{y}^{(1)}$ values are larger than the $q$ quantile of the $\hat{y}^{(1)}$ in the window, and the selected points make up the set $S_{j, \alpha, q}$. For $j = 1, 2, \dots, |W_{\alpha}|-1$, where $|W_{\alpha}|$ is the number of points in $W_{\alpha}$, the combined set of $S_{j, \alpha, q}$ for different j's is defined as $S_{\alpha, q}$, displayed in Figure\autoref{algo1_3}. $S_{\alpha, q}$ contains points that are locally in the upper $1-q$ quantile, which will be used to estimate the blaze function since these points generally do not fall into an absorption line. In step 5 we run a local polynomial regression on $S_{\alpha, q}$ and fit it to the whole spectrum. This regression is our final estimate of the blaze function, denoted as $\hat{B}_2$. The red line in Figure\autoref{algo1_4} shows the final estimate of the blaze function. Then, in step 6, $y$ is divided by $\hat{B}_2$ to get the blaze-removed spectrum, denoted as $\hat{y}^{(2)}$ and shown in Figure\autoref{algo1_5}.

\begin{algorithm}[H]
\caption{AFS Algorithm}\label{AFS_algo}
\begin{algorithmic}
\STATE Step 0: Let $\{(\lambda_i, y_i)\}_{i=1}^n$ be an observed spectrum.
\STATE Step 1: Let $u=\frac{max(\lambda)-min(\lambda)}{10\times max(y)}$. Multiply $y$ by $u$.
\STATE Step 2: Let $AS_{\alpha}=alpha~shape(\{(\lambda_i, y_i), i=1, \dots, n\})$ with radius value $\alpha$. Then $\widetilde{AS}_{\alpha}=\left\{(\lambda_i, \tilde{y}(\lambda_i)): \lambda_i \in \{\lambda_i, i=1, \dots, n\}, ~\tilde{y}(\lambda_i)=\max\limits_{\substack{\forall (\lambda_i, y^*_i) \in AS_{\alpha}}}y^*_i\right\}$. 
\STATE Step 3: Run a local polynomial regression on $\widetilde{AS}_{\alpha}$ with smoothing parameter $m_0$, denoting the fit model as $\hat{B}_1$. Calculate $\hat{y}^{(1)}=\frac{y}{\hat{B}_1}$. 
\STATE Step 4: Let $W_{\alpha}=\widetilde{AS}_{\alpha}\cap\{(\lambda_i, y_i)\}_{i=1}^n=\{(\lambda_i, y_i), i=w_1, w_2, \dots, w_{|W_{\alpha}|}\}$.
Let $S_{j, \alpha, q}=\Bigg\{w_j \leq i \leq w_{j+1}: \frac{\sum\limits_{k=w_j}^{w_{j+1}}\mathbbm{1}(\hat{y}^{(1)}_i \geq \hat{y}^{(1)}_k)}{w_{j+1}-w_j+1}\geq q\Bigg\}$. The $S_{\alpha, q}=\bigcup\limits_{\substack{j=1, \dots, |W_{\alpha}|-1}} S_{j, \alpha, q}$. 
\STATE Step 5: Run a local polynomial regression on set $\{(\lambda_i, y_i)\}_{i\in S_{\alpha, q}}$ with $m_0$ and fit to the whole spectrum, denoted as $\hat{B}_2$. 
\STATE Step 6: Calculate $\hat{y}^{(2)}=\frac{y}{\hat{B}_2}$. Output $\{(\lambda_i, \hat{y}_i^{(2)})\}_{i=1}^n$.
\end{algorithmic}
\end{algorithm}

\subsection{ALSFS Method}
The ALSFS algorithm is a method for removing the blaze function when a lab source spectrum, such as an LED or quartz lamp spectrum, is available as a reference. Generally, when a reliable reference spectrum is available, it can be used as the preliminary estimate of the blaze function shape. The use of a reference spectrum can be particularly advantageous in situations when the science spectrum contains wide absorption lines, which often make estimation of blaze function shape especially challenging.

The AFS algorithm in \autoref{algorithm1} can be adapted to take advantage of this additional information in the reference spectrum. 
To a first approximation, a lab source continuum spectrum should trace the instrumental blaze function for each order. However, the calibration source will typically have some effective blackbody temperature -- that is, its intrinsic intensity will peak at a specific wavelength. Over the limited wavelength range covered by a single spectral order, this effective blackbody function is approximately linear. Therefore, an observation of a lab source spectrum can be modeled as a linear transformation of the instrumental blaze function, and therefore the difference between the blaze function and the corresponding lab source spectrum is only a location-scale transformation. Thus, the blaze estimation problem can be translated into an optimization problem to find the best intercept, scale, and slope of the lab source. 

\subsubsection{ALSFS Algorithm}
The ALSFS algorithm is initialized in the same manner as the AFS algorithm in steps 1 and 2, but it differs slightly in steps 3 and 4 and greatly in step 5. Let the reference lab source spectrum be $\{(\lambda_i, l_i)\}_{i=1}^n$. The ALSFS algorithm is summarized in \autoref{ALSFS_algo}.

\begin{figure}
\centering 
\subfloat[Points selected]{%
  \includegraphics[width=0.45\columnwidth]{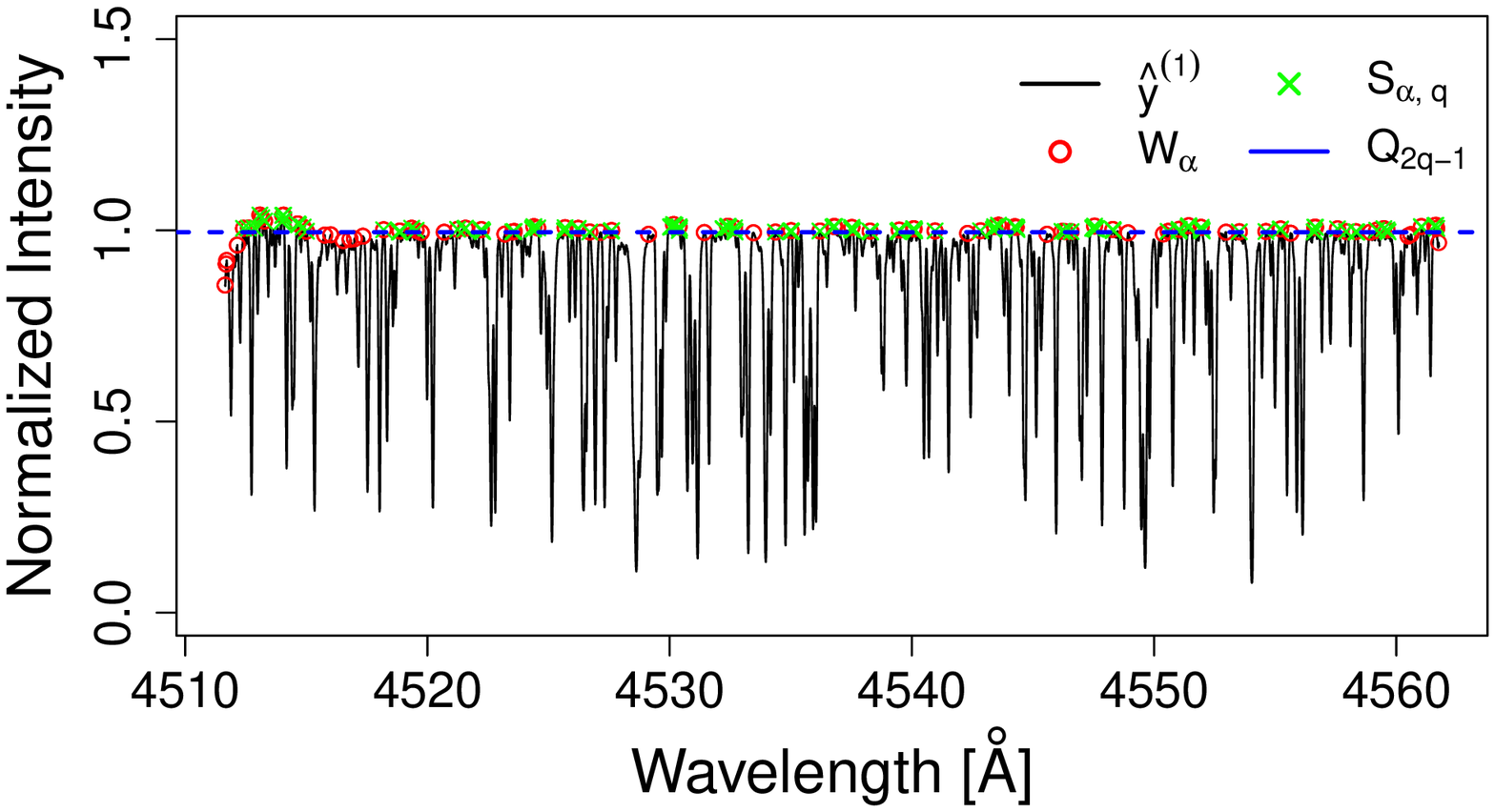}%
  \label{algo2_1}%
}\qquad
\subfloat[Local polynomial fitting]{%
  \includegraphics[width=0.45\columnwidth]{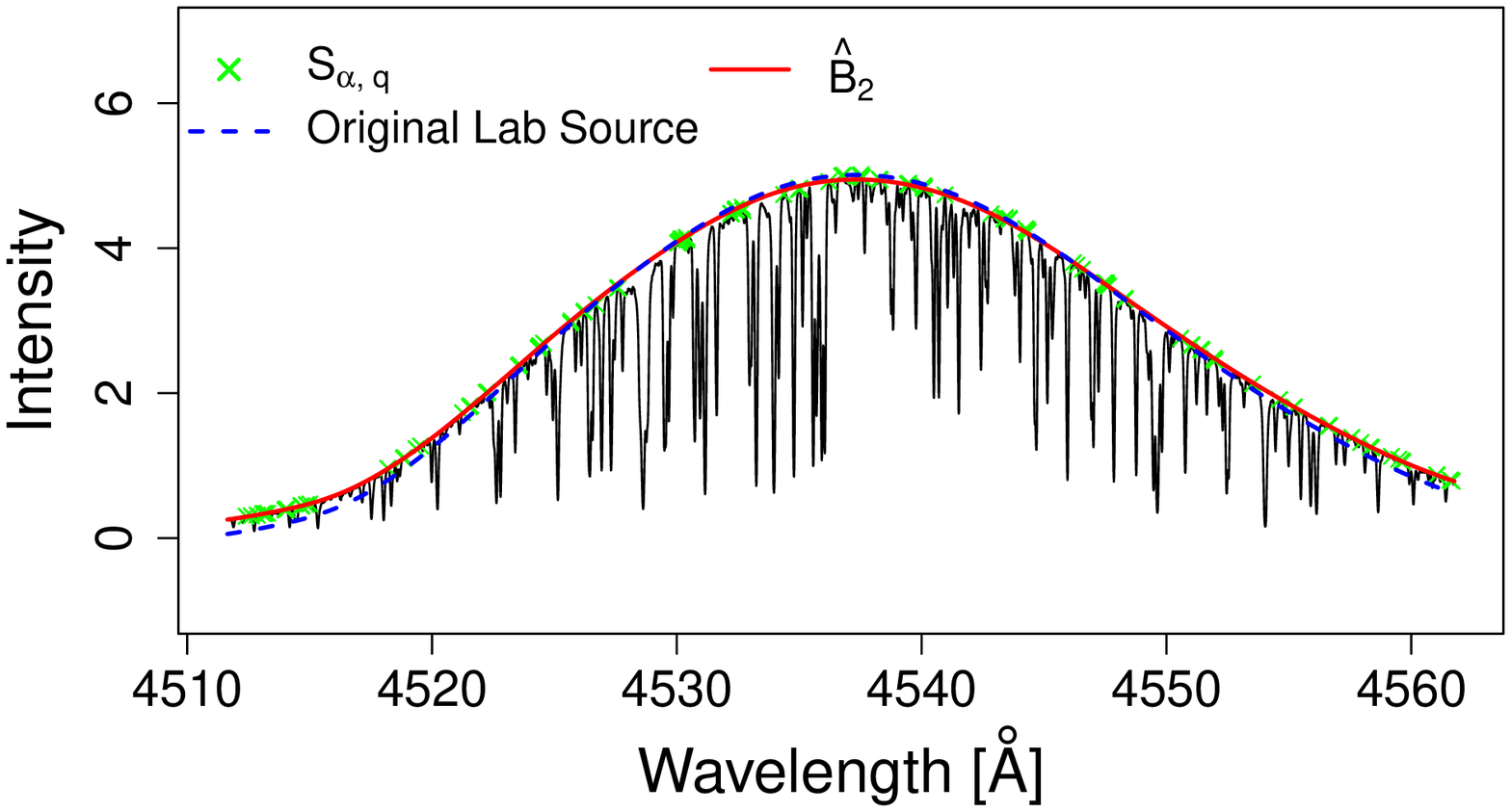}%
  \label{algo2_2}%
}
\caption{(a) Divide $y$ by $\hat{B}_1$ to get $\hat{y}^{(1)}$ (in black). Select points to the set $S_{\alpha, q}$ (green exes) both locally and globally using $\hat{y}^{(1)}$. The blue dashed line is the global quantile $Q_{2q-1}$. (b) The blue dashed line is the original lab source spectrum. Over the limited wavelength range of an order, the effective blackbody function is approximately linear, and so the lab source spectrum generally has the same shape as the true blaze function but needs linear modifications: intercept, scale, and slope parameters are used in the linear transformation to get $\hat{B}_2$. The red solid line is our final estimate for the blaze function.} \label{algo2_figure}
\end{figure}

Steps 1 and 2 are the same as the AFS algorithm. Since prior knowledge of the shape of blaze function is available, fewer points are needed for the second local polynomial fitting. Step 3 is similar to step 3 of the AFS algorithm, but we make set $S_{\alpha, q}$ more accurate by also calculating the $2q-1$ quantile of $\hat{y}^{(1)}$, denoted as $Q_{2q-1}$. $Q_{2q-1}=Q_{1-2(1-q)}$, which means the upper $2(1-q)$ quantile of $\hat{y}^{(1)}$, displayed in Figure\autoref{algo2_1}. Compared to the upper $1-q$ quantile of $\hat{y}^{(1)}$ within each window, a smaller quantile is used for the global quantile; otherwise, too many points will be excluded if the same quantile is used. 
Step 4 also departs from the AFS algorithm: for the j-th window, we select the points $\lambda_i$ where both $\hat{y}^{(1)}_i \geq$ the $q$ quantile of the window and $\hat{y}^{(1)}_i \geq Q_{2q-1}$ into set $S_{\alpha, q}$, shown in Figure\autoref{algo2_1}. 
In step 5, we use the lab source spectrum's intensity curve as a reference model, displayed in Figure\autoref{algo2_2}, to find the best linear coefficients for blaze function estimation using the points selected in the previous step. We apply a linear transformation on the lab source spectrum: $\hat{l}_i(a, b, c)=a+bl_i+c\lambda_i$, where $a$, $b$, and $c$ are intercept, scale, and slope parameters, respectively. Since our ultimate goal is to have a flat spectrum without the blaze function, we use an objective function $\sum\limits_{\substack i \in S_{\alpha, q}} \Big(\frac{l_i}{\hat{l}_i(a, b, c)}-1\Big)^2$, which measures the total distances from the removed spectrum to the constant 1. We minimize this objective function on the set $S_{\alpha, q}$ to get estimates for $a$, $b$, and $c$. Then the modified lab source spectrum $\hat{a}+\hat{b}l_i+\hat{c}\lambda_i$ is our final estimate for blaze function, displayed in Figure\autoref{algo2_2}. Step 6 is the same as the AFS algorithm, $y$ is divided by $\hat{B}_2$ to get the blaze-removed spectrum. 

\begin{algorithm}[H]
\caption{ALSFS Algorithm}\label{ALSFS_algo}
\begin{algorithmic}
\STATE Step 0: Let $\{(\lambda_i, y_i)\}_{i=1}^n$ be an observed spectrum, and $\{(\lambda_i, l_i)\}_{i=1}^n$ be the corresponding lab source.
\STATE Step 1: Let $u=\frac{max(\lambda)-min(\lambda)}{10\times max(y)}$. Multiply $y$ by $u$.
\STATE Step 2: Let $AS_{\alpha}=alpha~shape(\{(\lambda_i, y_i), i=1, \dots, n\})$ with radius value $\alpha$. and $\widetilde{AS}_{\alpha}=\{(\lambda_i, \tilde{y}(\lambda_i)): \lambda_i \in \{\lambda_i, i=1, \dots, n\}, ~\tilde{y}(\lambda_i)=\max\limits_{\substack{\forall (\lambda_i, y^*_i) \in AS_{\alpha}}}y^*_i\}$. 
\STATE Step 3: Run a local polynomial regression on $\widetilde{AS}_{\alpha}$ with $m_0$, denoted as $\hat{B}_1$. Calculate $\hat{y}^{(1)}=\frac{y}{\hat{B}_1}$. Denote $Q_{2q-1}=quantile(\hat{y}^{(1)}, 2q-1)$.
\STATE Step 4: Let $W_{\alpha}=\widetilde{AS}_{\alpha}\cap\{(\lambda_i, y_i)\}_{i=1}^n=\{(\lambda_i, y_i), i=w_1, w_2, \dots, w_{|W_{\alpha}|}\}$.
Let $S_{j, \alpha, q}=\Bigg\{w_j \leq i \leq w_{j+1}: \frac{\sum\limits_{k=w_j}^{w_{j+1}}\mathbbm{1}(\hat{y}^{(1)}_i \geq \hat{y}^{(1)}_k)}{w_{j+1}-w_j+1}\geq q\ ~and~\hat{y}^{(1)}_i\geq Q_{2q-1}\Bigg\}$. $S_{\alpha, q}=\bigcup\limits_{\substack{j=1, \dots, |W_{\alpha}|-1}} S_{j, \alpha, q}$. 
\STATE Step 5: Consider a linear transformation: $\hat{l}_i(a, b, c)=a+bl_i+c\lambda_i$, $i=1, \dots, n$. $(\hat{a}, \hat{b}, \hat{c})=\argmin\limits_{\substack a, b, c} \sum\limits_{\substack i \in S_{\alpha, q}}(\frac{l_i}{\hat{l}_i(a, b, c)}-1)^2$. $\hat{B}_2=\hat{a}+\hat{b}l_i+\hat{c}\lambda_i$, $i=1, \dots, n$. 
\STATE Step 6: Calculate $\hat{y}^{(2)}=\frac{y}{\hat{B}_2}$. Output $\{(\lambda_i, \hat{y}_i^{(2)})\}_{i=1}^n$.
\end{algorithmic}
\end{algorithm}

\subsubsection{Lab Source Smoothing by AFS Algorithm}\label{ls_smoothing}
\begin{figure} 
\centering
\subfloat[Fringing spectrum]{%
  \includegraphics[width=0.4\columnwidth]{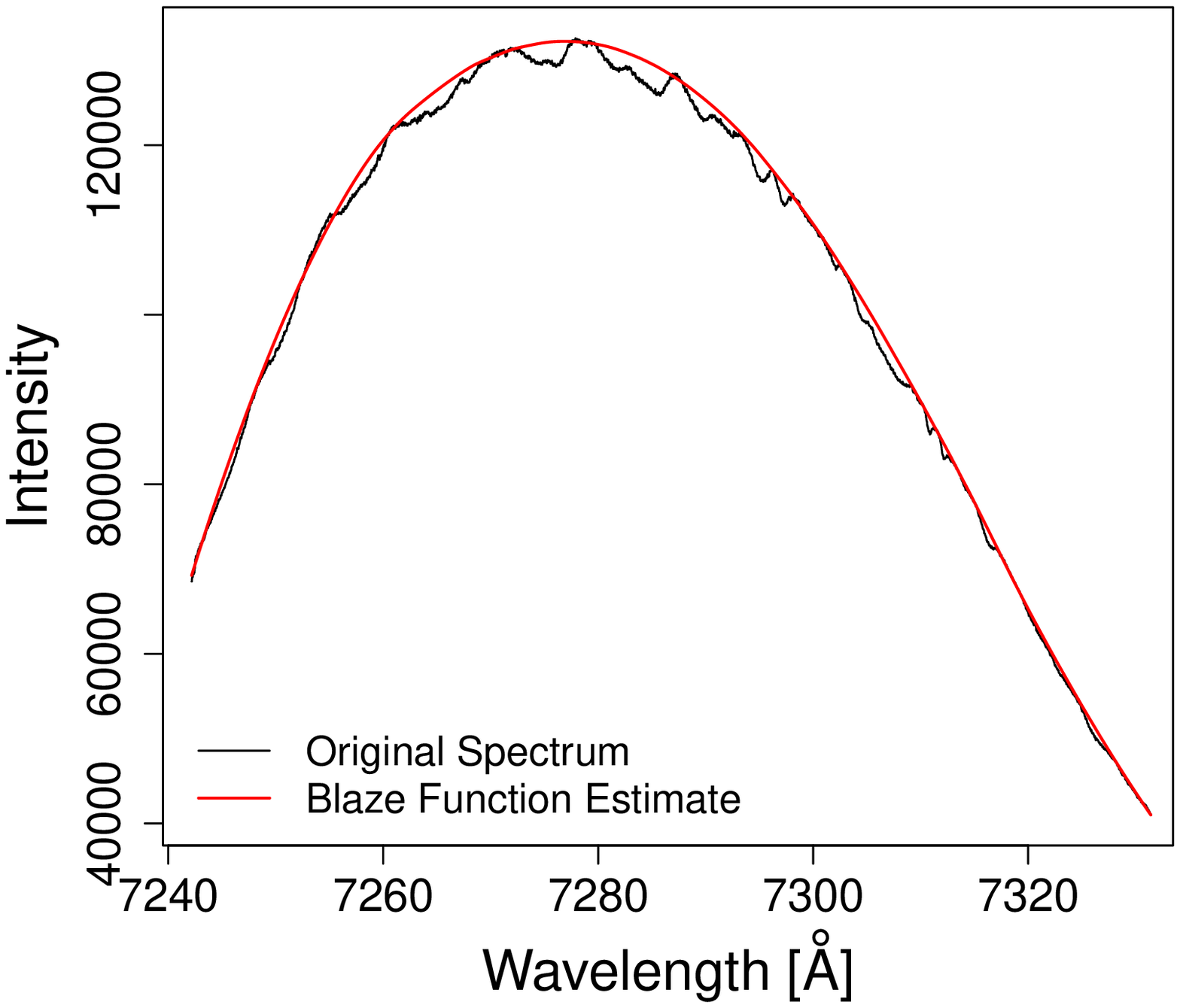}%
  \label{fringing1}%
}\qquad
\subfloat[Normalized spectrum]{%
  \includegraphics[width=0.4\columnwidth]{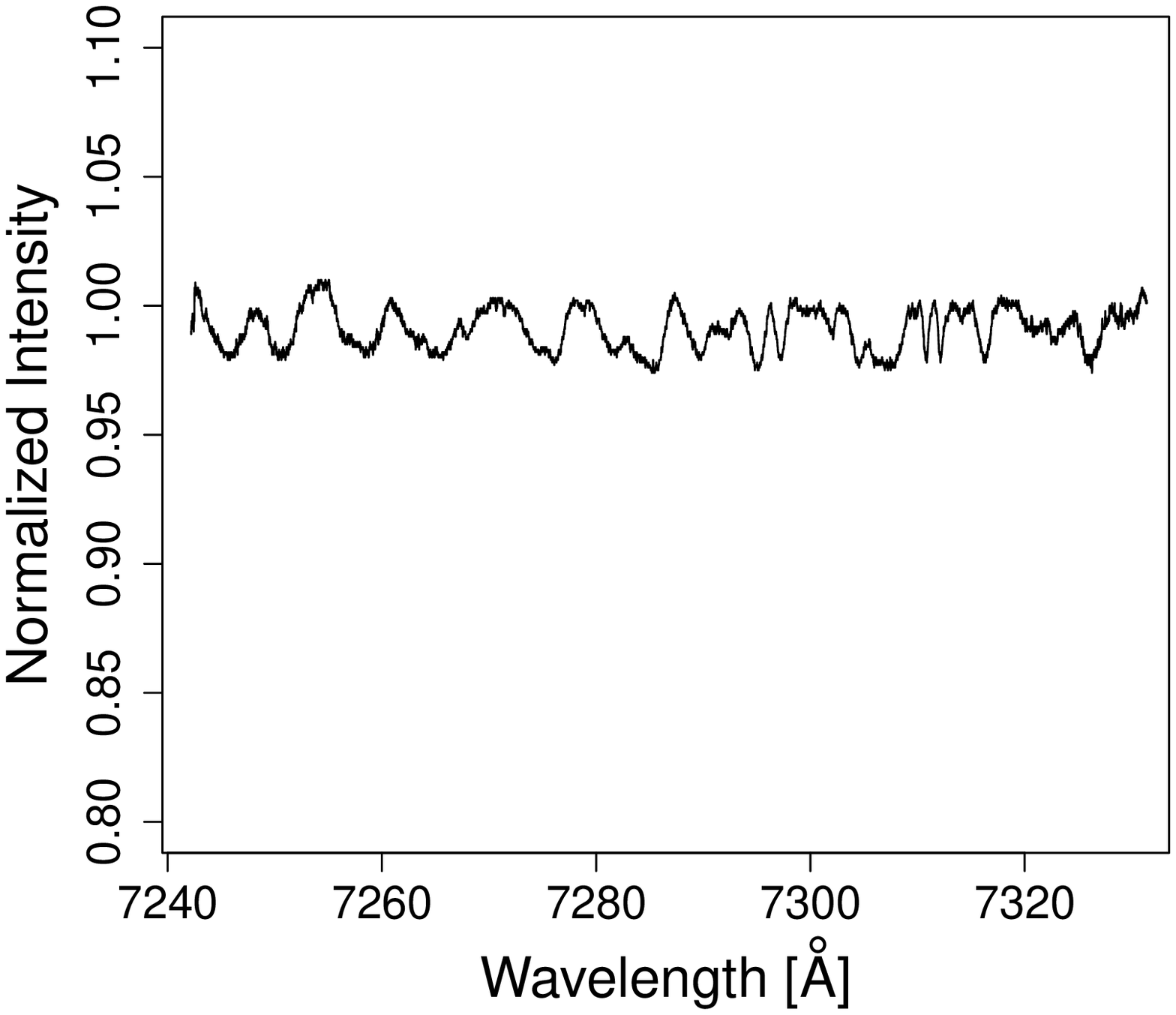}%
  \label{fringing2}%
}
\caption{(a) Echelle order of the spectrum shows fringing because the thinned silicon CCD has a thickness comparable to red wavelengths. The AFS algorithm is used to fit a smooth function across the order and division of this order of flux from the flat-field lamp produces a normalized spectrum (b). For very stable spectrographs, the stellar spectra can be divided by this normalized flat-field flux to remove fringing. }
\label{fringing}
\end{figure}

The process of flat-fielding spectra is necessary for removing pixel-to-pixel quantum efficiency (QE) variations in charge coupled devices (CCDs) that are used as detectors in astronomical spectrographs. In the case of fiber-fed echelle spectrographs, flat fielding can be carried out by extracting a featureless calibration spectrum and dividing the extracted science spectrum order-by-order. However, the flat-field source is generally not perfectly uniform in intensity over a large wavelength range, and therefore it will typically have an effective black-body temperature that does not match the stellar effective temperature. As a result, this division leaves behind residual trends. By using the AFS algorithm a model can be fitted to each order of the flat field calibration spectrum. Division of the flat field echelle orders by this fitted model will then produce a normalized spectrum that can be used to divide out the QE variation. Figure \ref{fringing} shows an example of how this process was used to create a normalized flat in red orders that exhibit fringing from interference of red wavelengths in thinned silicon detectors. This fringing can be removed by dividing stellar spectra with this normalized flat. 

\begin{figure}
\centering
\includegraphics[width=0.5\textwidth]{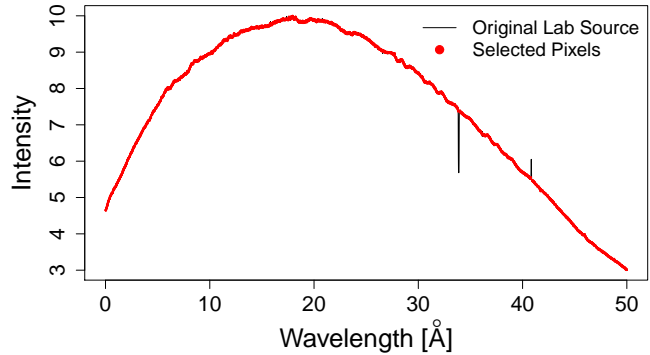}
\caption{In the smoothing process, we first use an iteration to get rid of the spikes. In the $j$-th iteration, remove pixels where $|L_i-L_{i-1}|>Q_{0.99}^{(j-1)}$ and calculate the $0.99$ quantile of the new $\Delta L$, denoted as $Q_{0.99}^{(j)}$. Iterate until $Q_{0.99}^{(j)}<Q_{q_s}$. The red points are pixels left to be used for the AFS algorithm.}
\label{algo3_1}
\end{figure}

In some orders a cosmic ray or a pixel with very low QE will create an upward or downward spike that is confined to one or a few pixels. In this case, the AFS algorithm is slightly modified to iteratively reject these pixels using outlier rejection. Let the original lab source spectrum be $\{(\lambda_i, L_i)\}_{i=1}^n$, and $\Delta L=\{|L_i-L_{i-1}|$, $i=2, \dots, n\}$. Let $Q_{q_s}$ be the $q_s$ quantile of $\Delta L$, where $q_s$ could be a number between $0.95$ and $0.99$. In the beginning, the $0.99$ quantile of $\Delta L$ is denoted as $Q_{0.99}^{(0)}$. In the $j$-th iteration, remove pixels where $|L_i-L_{i-1}|>Q_{0.99}^{(j-1)}$ and calculate the $0.99$ quantile of the new $\Delta L$, denoted as $Q_{0.99}^{(j)}$. Continue the iteration until $Q_{0.99}^{(j)}<Q_{q_s}$. The remaining pixels are used for smoothing by the AFS algorithm, displayed in \autoref{algo3_1}.

\subsection{Parameter Selection}\label{parameter}
\begin{figure}
\centering 
\subfloat[$\alpha$=$1\times$wavelength range]{%
  \includegraphics[width=0.45\columnwidth]{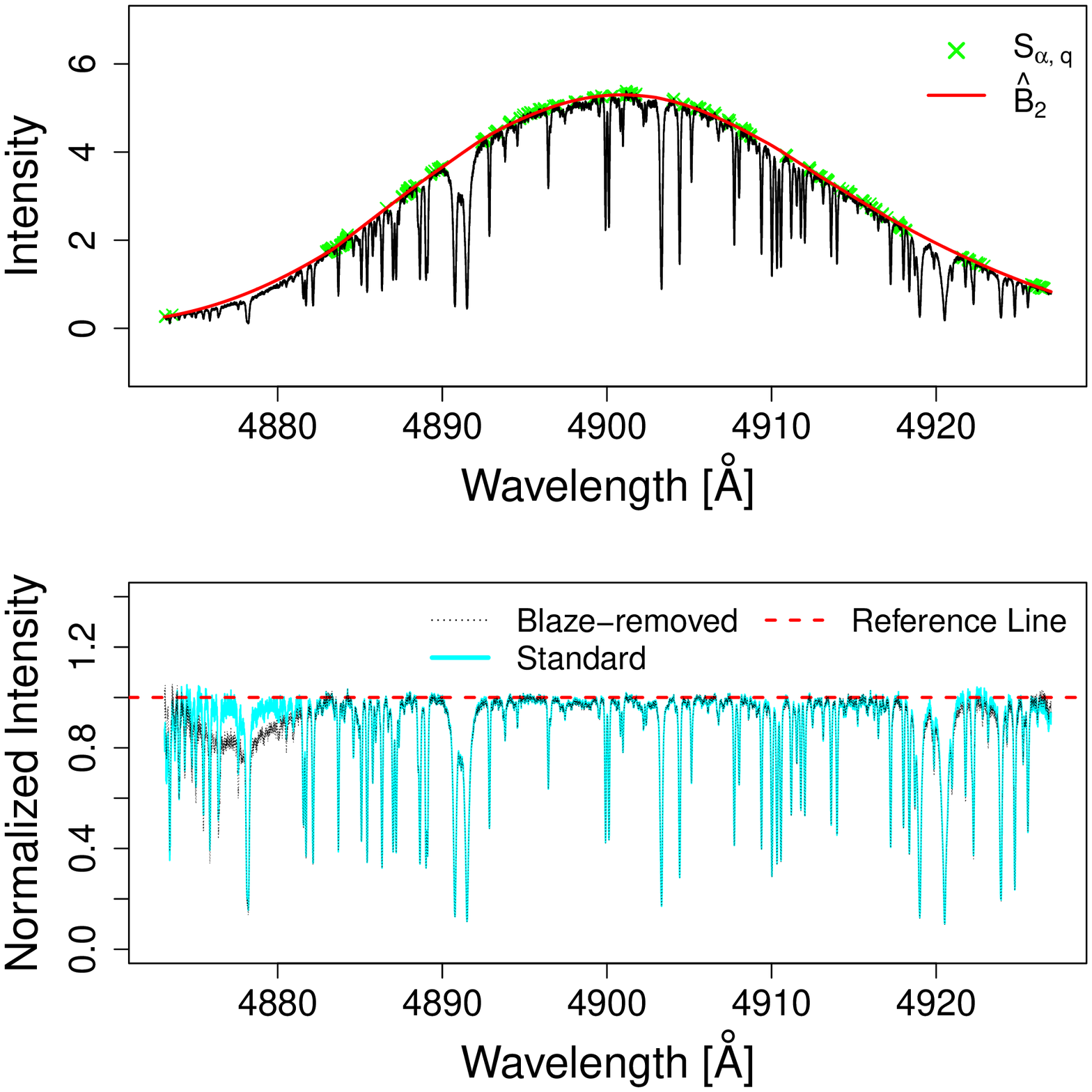}%
  \label{param_1}%
}\qquad
\subfloat[$\alpha$=$\frac{1}{50}\times$wavelength range]{%
  \includegraphics[width=0.45\columnwidth]{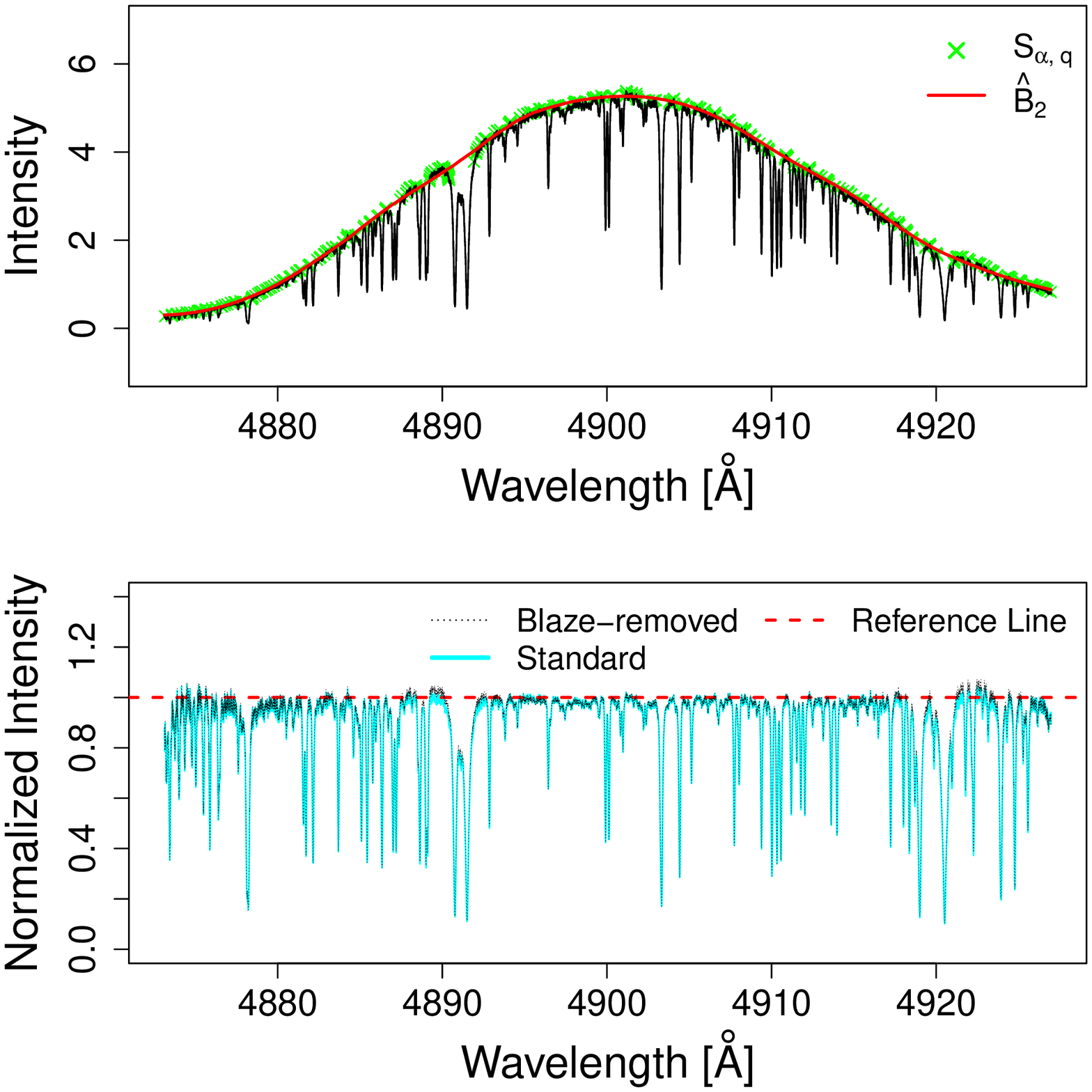}%
  \label{param_2}%
}    
\caption{Comparison of the results of the AFS algorithm using extreme values for the $\alpha$ parameter. (a) Large $\alpha$: $1\times$wavelength range. (b) Small $\alpha$: $\frac{1}{50}\times$wavelength range. The cyan spectrum shows the results from the standard value of $\alpha =\frac{1}{6}\times$wavelength range.} \label{param12}
\end{figure}

In the proposed algorithms, there are several parameters that need to be selected by users. In the AFS algorithm, there are three parameters: $\alpha$ for the alpha shape, quantile $q$ for the point selection, and $m_0$ in two local polynomial regressions. 
The $\alpha$ determines how many windows a spectrum is cut into, because the number of windows is determined by smoothness of the alpha shape, which is controlled by $\alpha$. 
Using these windows, $q$ determines how many points are selected in each window.  After selecting the points, $m_0$ determines the smoothness of local polynomial fitting on these selected points. 
Practically, the three parameters are robust within appropriate ranges. For the example spectrum in Figure\autoref{param_1}, we recommend a set of standard parameters $(\alpha =\frac{1}{6}\times$wavelength range, $q=0.95, m_0=0.25)$ as a default. In Figures 6-8, we show the blaze estimates using extreme parameter choices (top panels) and its resulting normalized spectra compared with the standard parameter choice (bottom panels). 

First, the $\alpha$ can be selected based on the shape of a blaze function. For example, in Figure\autoref{algo1_1}, we find that the blaze function increases first and then decreases. Also, it is convex first, then becomes concave, and turns to be convex in the end. We want to select an $\alpha$ that can capture the convex parts (not too large), but will not go too deep into absorption lines (not too small). The choice of $\alpha$ is mainly determined by the shape, e.g., curvature and concavity, of a blaze function. However, when there is a wide absorption line, a larger $\alpha$ may be needed than the shape would suggest. For echelle spectra orders, since each convex portion generally takes up about $\frac{1}{6}$ of an order, we recommend selecting an $\alpha$ that is $\frac{1}{6}$ of the wavelength range of the order, but have found empirically that $\alpha$ is rather robust from $\frac{1}{12}$ of the wavelength range to $\frac{1}{3}$ of the wavelength range.
Figure\autoref{param_1} shows an example of a very large $\alpha$ equal to the order's entire wavelength range. This $\alpha$ value is too large for the $\alpha$-balls to capture the shape of the spectrum near the boundaries; 
in the blaze-removed spectrum, the left part drops downward like a wide absorption feature because points in that portion of the spectrum were not selected in $S_{\alpha, q}$. 
Figure\autoref{param_2} shows an example of a small $\alpha$ of $\frac{1}{50}$ of the wavelength range. With such a small $\alpha$, the $\alpha$-balls fit into absorption lines so that points inside an absorption are selected into set $S_{\alpha, q}$. In the removed spectrum, there are regions that are above the reference line $y=1$, compared with the cyan spectrum where $\alpha =\frac{1}{6}\times$wavelength range. 

\begin{figure}
\centering 
\subfloat[$q=0.7$]{%
  \includegraphics[width=0.45\columnwidth]{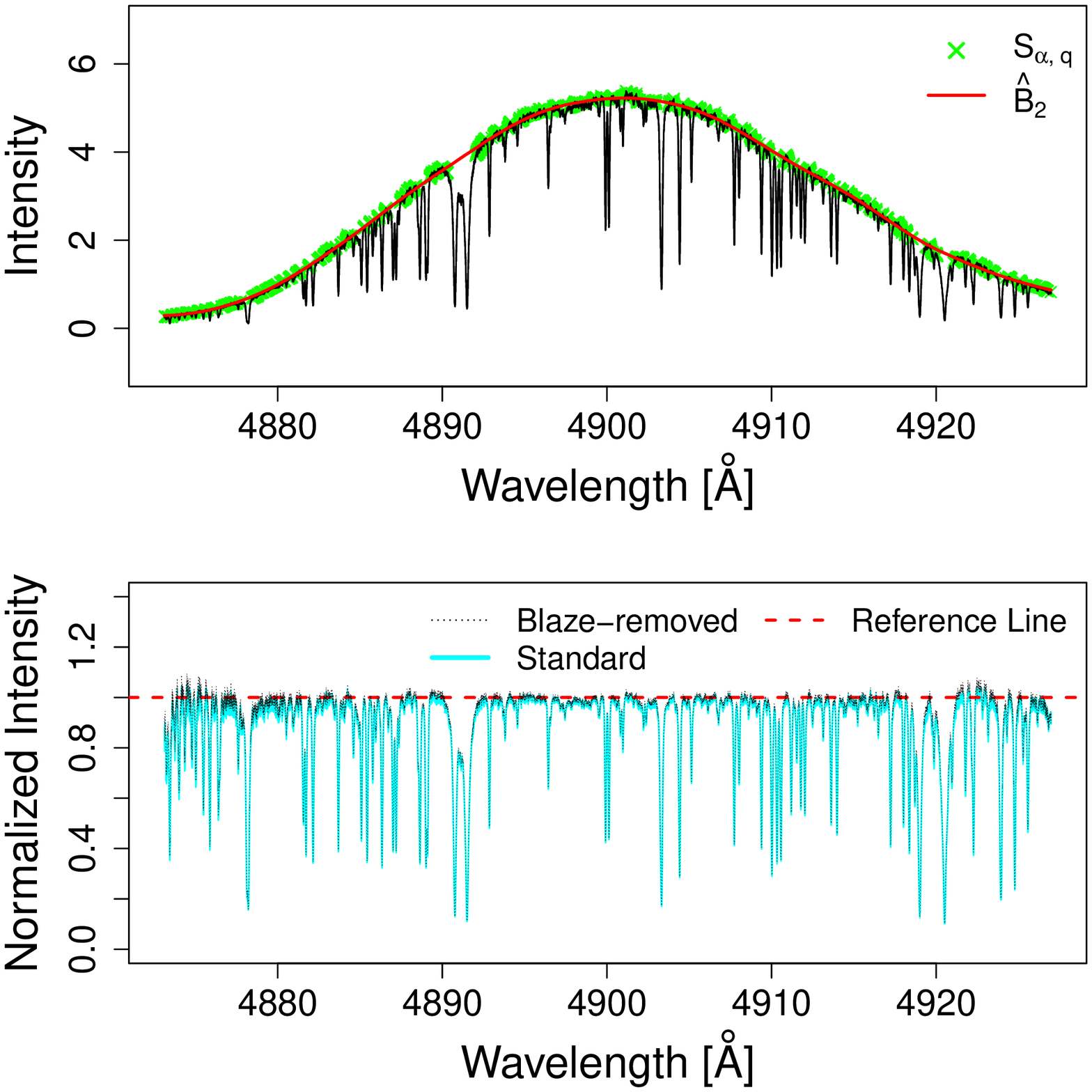}%
  \label{param_3}%
}\qquad
\subfloat[$q=0.999$]{%
  \includegraphics[width=0.45\columnwidth]{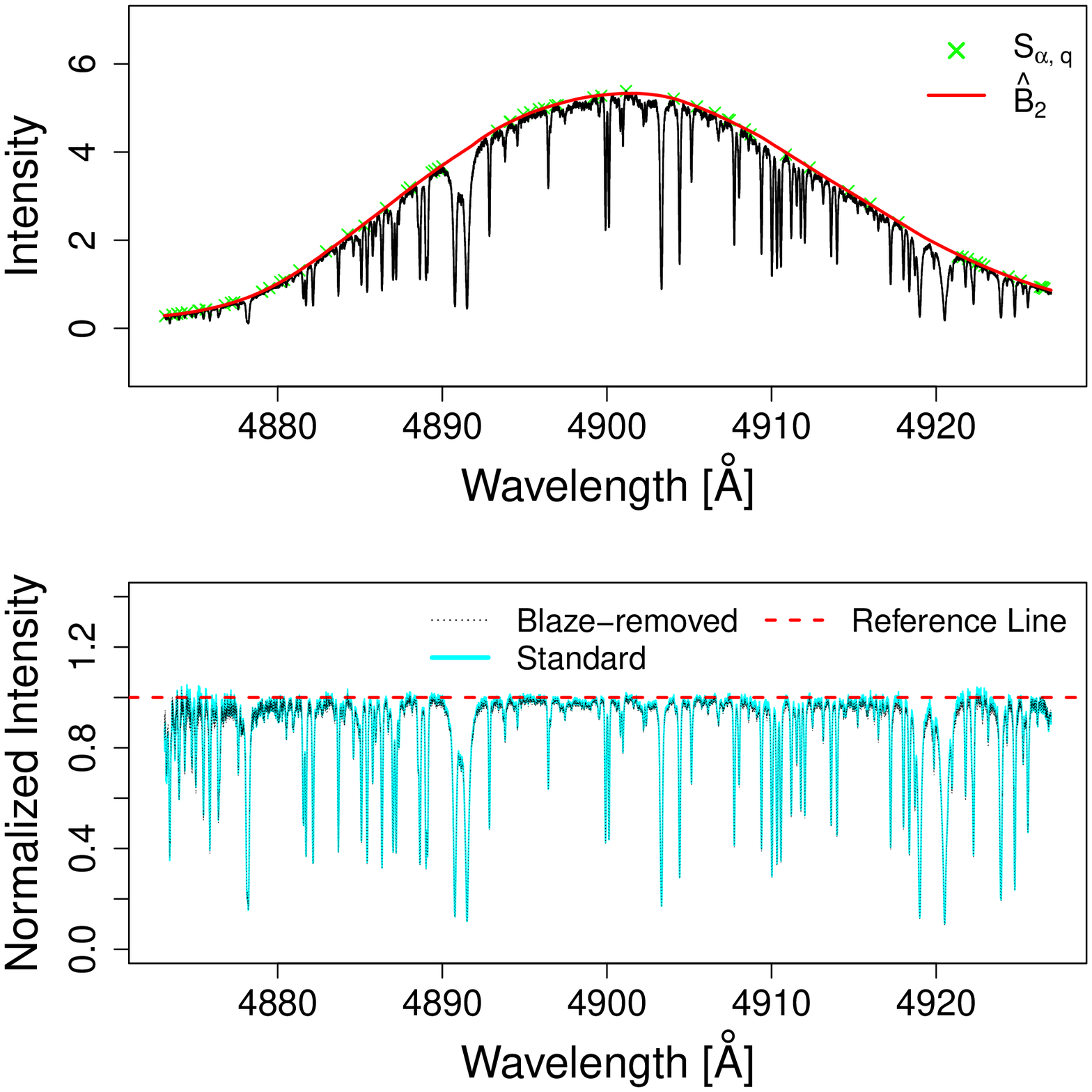}%
  \label{param_4}%
}    
\caption{Comparison of the results of the AFS algorithm using extreme values for the $q$ parameter. (a) Small $q$: $0.7$. (b) Large $q$: $0.999$. The cyan spectrum shows the results from the standard value of $q =0.95$.} 
\label{param34}
\end{figure}

The parameter $q$ depends on the S/N of a spectrum and the amount of absorption. The goal when selecting $q$ is to find points on the spectrum that do not drop into absorption lines, but instead are on the continuum. After flattening the spectrum using a preliminary estimate of its shape, we select points in the local upper $1-q$ quantile for the set $S_{\alpha, q}$ to be used in the final estimation. If we happened to know the true blaze, then after dividing the spectrum by the blaze we would expect to see points randomly scattered around $1$. 
In the proposed algorithms, these points are approximated by set $S_{\alpha, q}$. If the S/N is high or there is a large amount of absorption, a larger $q$ is needed to select points in $S_{\alpha, q}$ so that these points do not fall  in absorption lines. Conversely, if S/N is low or there is minimal absorption, a smaller $q$ is needed to get enough points into set $S_{\alpha, q}$\footnote{Alternatively, an adaptive $q$ can be used for each small window to capture the noise more accurately. An adaptive $q$ can incorporate the overall S/N and the average intensity in each window to characterize the noise variance better.}.
For an order with a similar amount of absorption as the one displayed in Figure\autoref{param_1}, a $q$ from $0.95$ to $0.99$ works for S/N 300, a $q$ from $0.85$ to $0.95$ works for S/N 150, and a $q$ from $0.5$ to $0.85$ works for S/N lower than 150. 
Figure\autoref{param_3} shows the effects of a small $q$ such as $0.7$: too many points are selected into $S_{\alpha, q}$ so that the blaze function estimate is dragged downward by the points in absorption lines, and in the removed spectrum, the left and right boundary regions are above the reference line $y=1$.  In contrast, Figure\autoref{param_4} shows a large $q$ of $0.999$. In this example $S_{\alpha, q}$ contains too few points, which makes the estimate sit almost completely above the spectrum. In the removed spectrum, almost all the pixels are under the reference line. 

\begin{figure}
\centering 
\subfloat[$m_0=0.5$]{%
  \includegraphics[width=0.45\columnwidth]{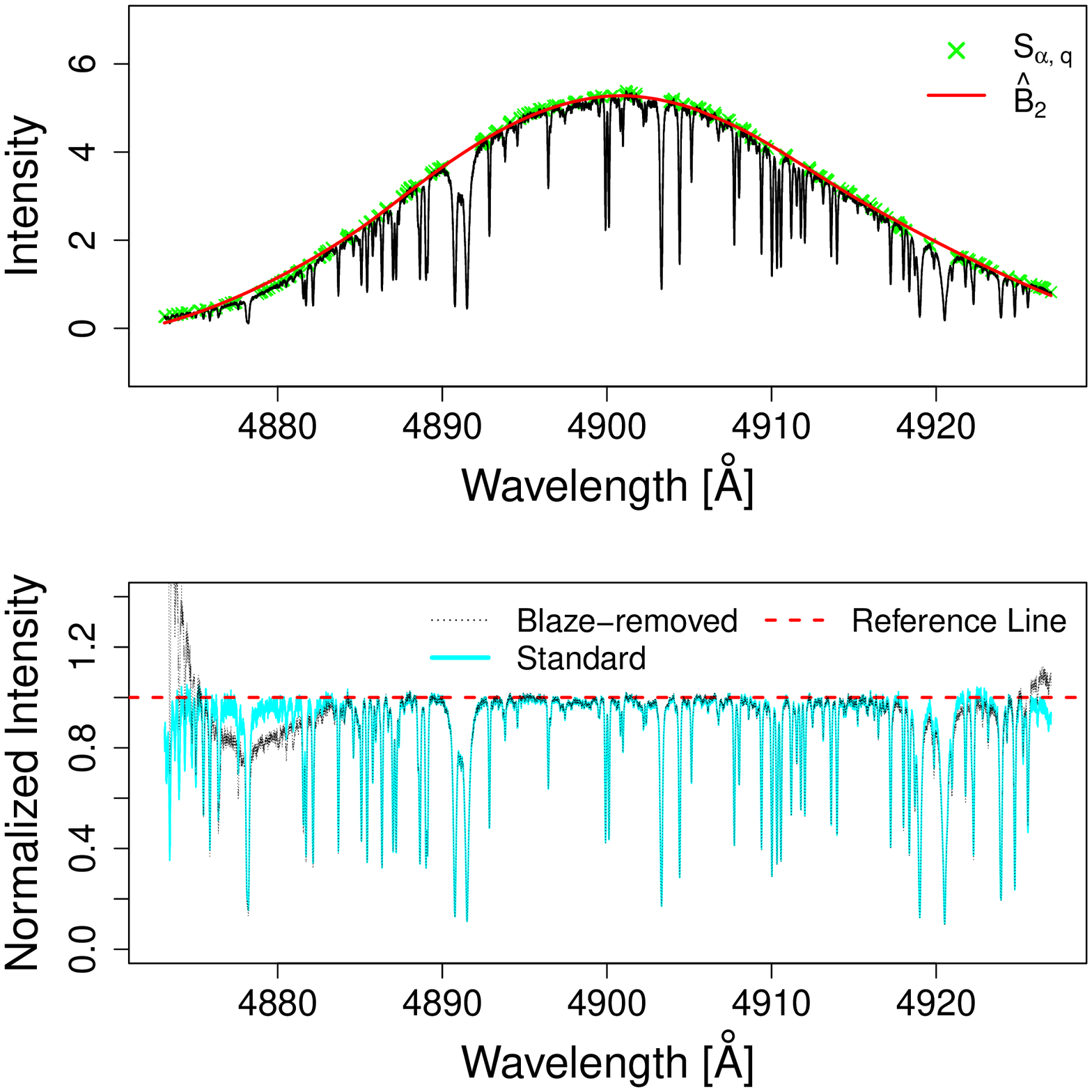}%
  \label{param_5}%
}\qquad
\subfloat[$m_0=0.1$]{%
  \includegraphics[width=0.45\columnwidth]{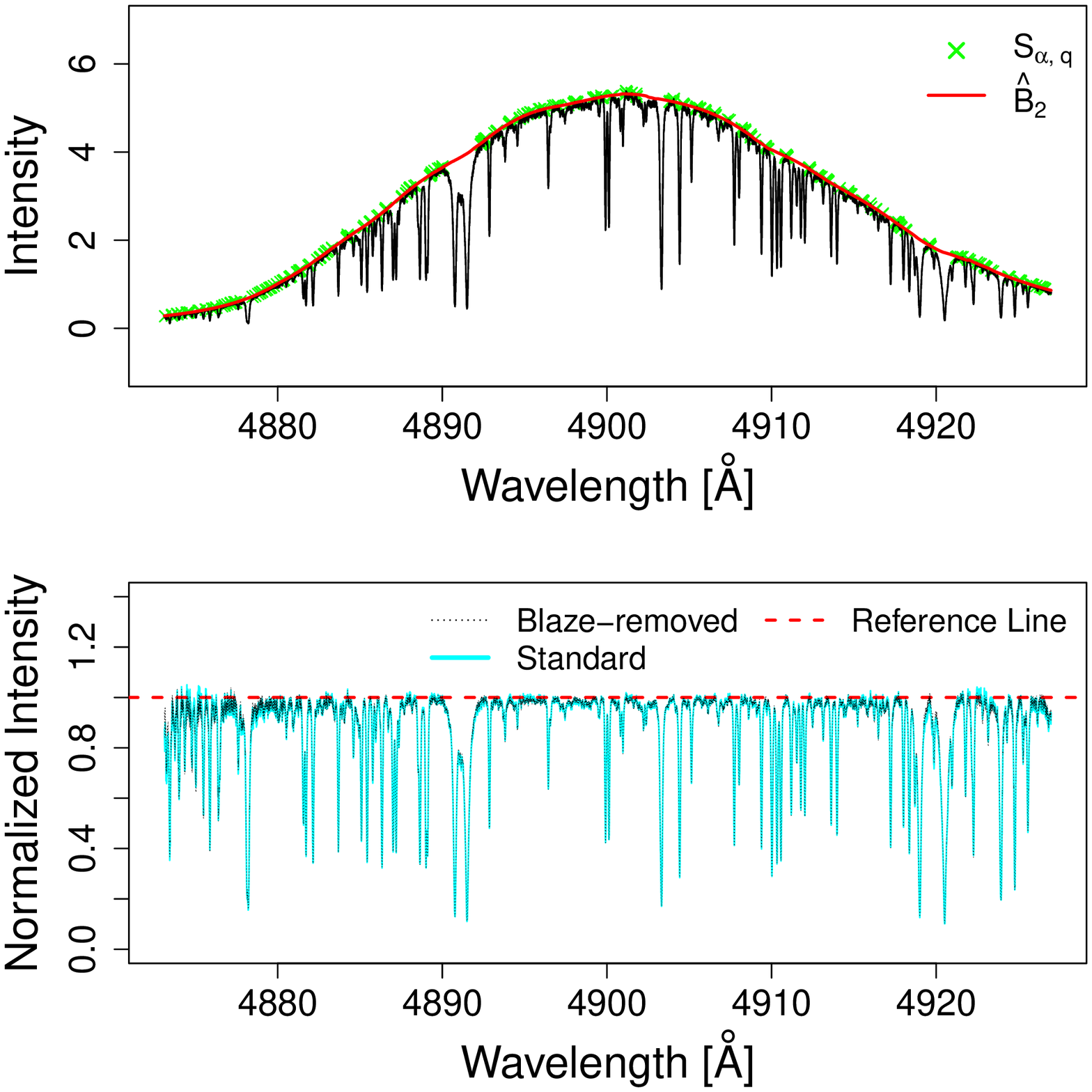}%
  \label{param_6}%
}    
\caption{Comparison of the results of the AFS algorithm using extreme values for the $m_0$ parameter. (a) Large $m_0$: $0.5$. (b) Small $m_0$: $0.1$. The cyan spectrum shows the results from the standard value of $m_0 =0.25$.} 
\label{param56}
\end{figure}

The smoothing parameter $m_0$ depends on the distribution of points in $S_{\alpha, q}$ along the spectrum, which is determined by the amount of absorption. 
If there are many absorption lines or any absorption lines that are wide, the set $S_{\alpha, q}$ has large gaps between pixels and thus a large $m_0$ is needed to get a good estimate. If there are few absorption lines or absorption lines that are narrow, a small $m_0$ is needed so that the estimation better adapts to local regions. For an order with a similar amount of absorption as the one displayed in Figure\autoref{param_1}, an $m_0$ value from $0.15$ to $0.3$ has worked well empirically. In Figure\autoref{param_5}, $m_0$ is set to be $0.5$ (too large) and the estimate is off on the left part of the spectrum, where the blaze-removed spectrum rises to above $1.5$. Figure\autoref{param_6} shows the results of a too small $m_0$ value of $0.1$: the blaze estimate has some small bumps, but the blaze-removed spectrum looks reasonable to the eye. However, since the true blaze function does not have small bumps, this blaze estimate is not as good a fit as it might appear at first glance. 

In general, the blaze function estimate is more sensitive to small changes in $q$ than to $\alpha$ and $m_0$. For an echelle spectrum order with a similar amount of absorption as the one displayed in Figure\autoref{param_1}, we can start with an $\alpha$ equal to $\frac{1}{6}$ of the wavelength range of the order, an $m_0$ equal to $0.25$, and tune the parameter $q$ within the range according to its S/N and amount of absorption. A more detailed set of recommendations for  parameters in different situations is provided in the Appendix.

The ALSFS algorithm has the same parameters: $\alpha$, $q$ and $m_0$. Because the final estimate is the reference lab source spectrum with linear modification, $m_0$ has less influence on the results than for the AFS algorithm. In the lab source spectrum smoothing process, $\alpha$, $q$, and $m_0$ operate the same as in the AFS algorithm. The quantile parameter $q_s$ in $Q_{q_s}$ depends on the particular appearances of spikes. If spikes are long, use a smaller $q_s$ such as $0.95$; if spikes are very small, use a larger $q_s$ such as $0.98$ or $0.99$.

\subsection{Complications and Corrections}\label{corrections}
We have found that the proposed algorithms work well in most cases. However, there are several special cases that can result in poorer estimates of the blaze function, for which we have developed corrections to mitigate these issues. Since the AFS algorithm relies only on the science spectrum itself, it is more susceptible to complications than the ALSFS algorithm.

\subsubsection{Boundary Correction}\label{boundary}
An order is normalized by dividing by its estimated blaze function. Since the blaze function approaches zero near the edges of the order, small errors in the blaze shape will be magnified in the divided spectrum. The ALSFS algorithm is less susceptible to this problem because of the strong constraints provided by the lab source, but other blaze estimation methods, including the AFS algorithm, can be strongly affected. 

\begin{figure}
\centering 
\subfloat[Left order]{%
  \includegraphics[width=0.3\columnwidth]{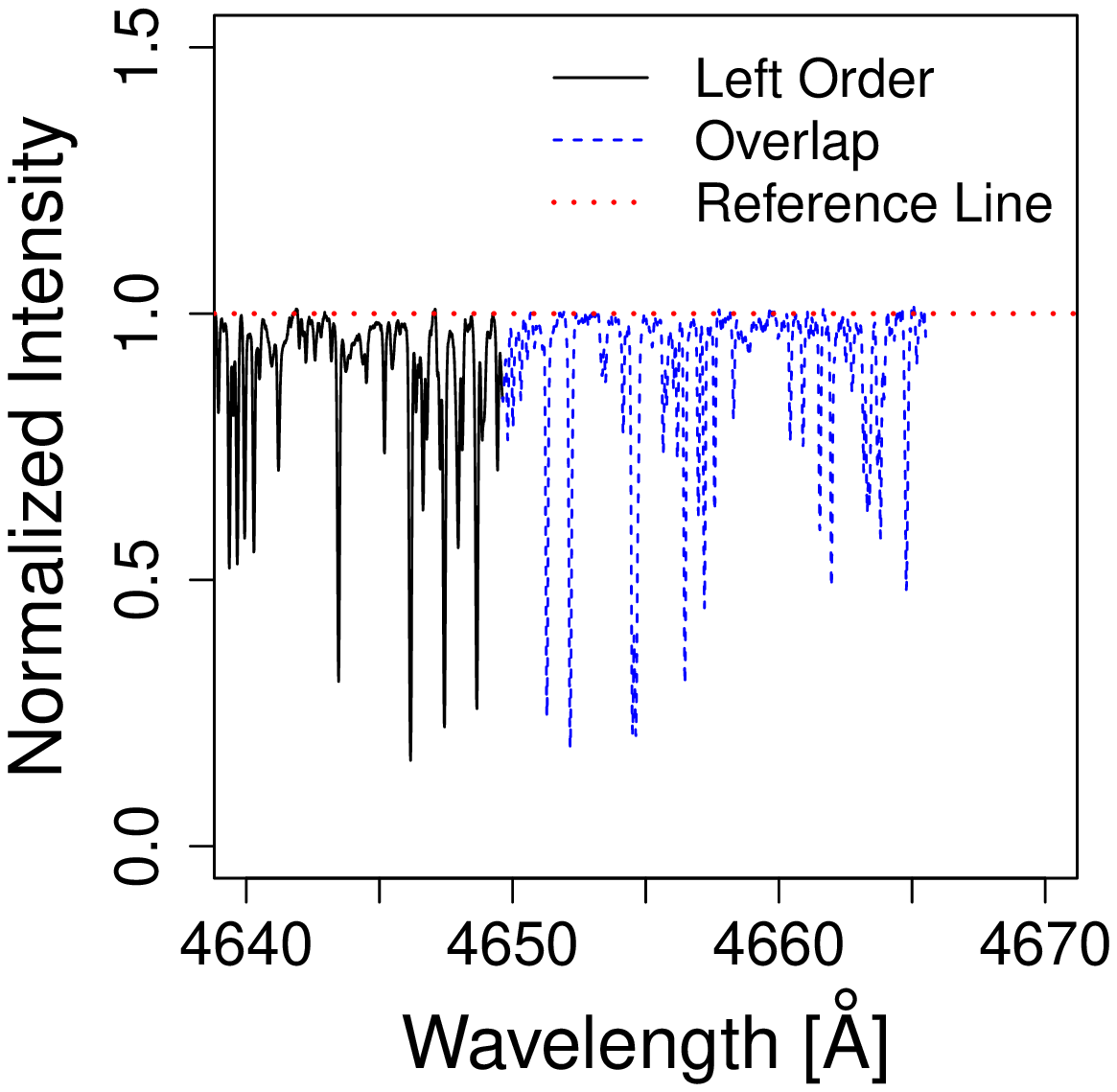}%
  \label{boundary1}%
}\qquad
\subfloat[Right order]{%
  \includegraphics[width=0.3\columnwidth]{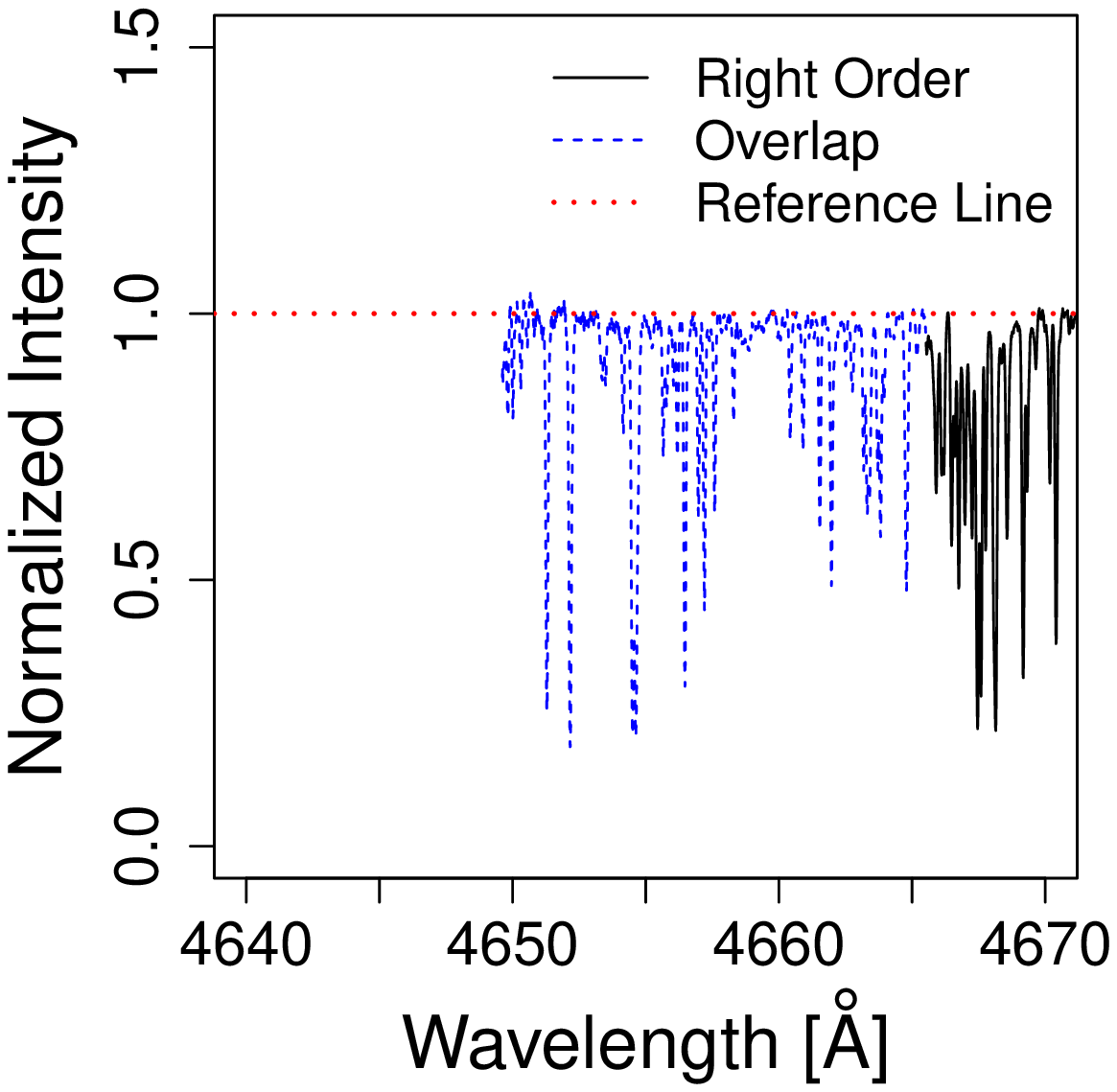}%
  \label{boundary2}%
}\qquad
\subfloat[Correction on overlap]{%
  \includegraphics[width=0.3\columnwidth]{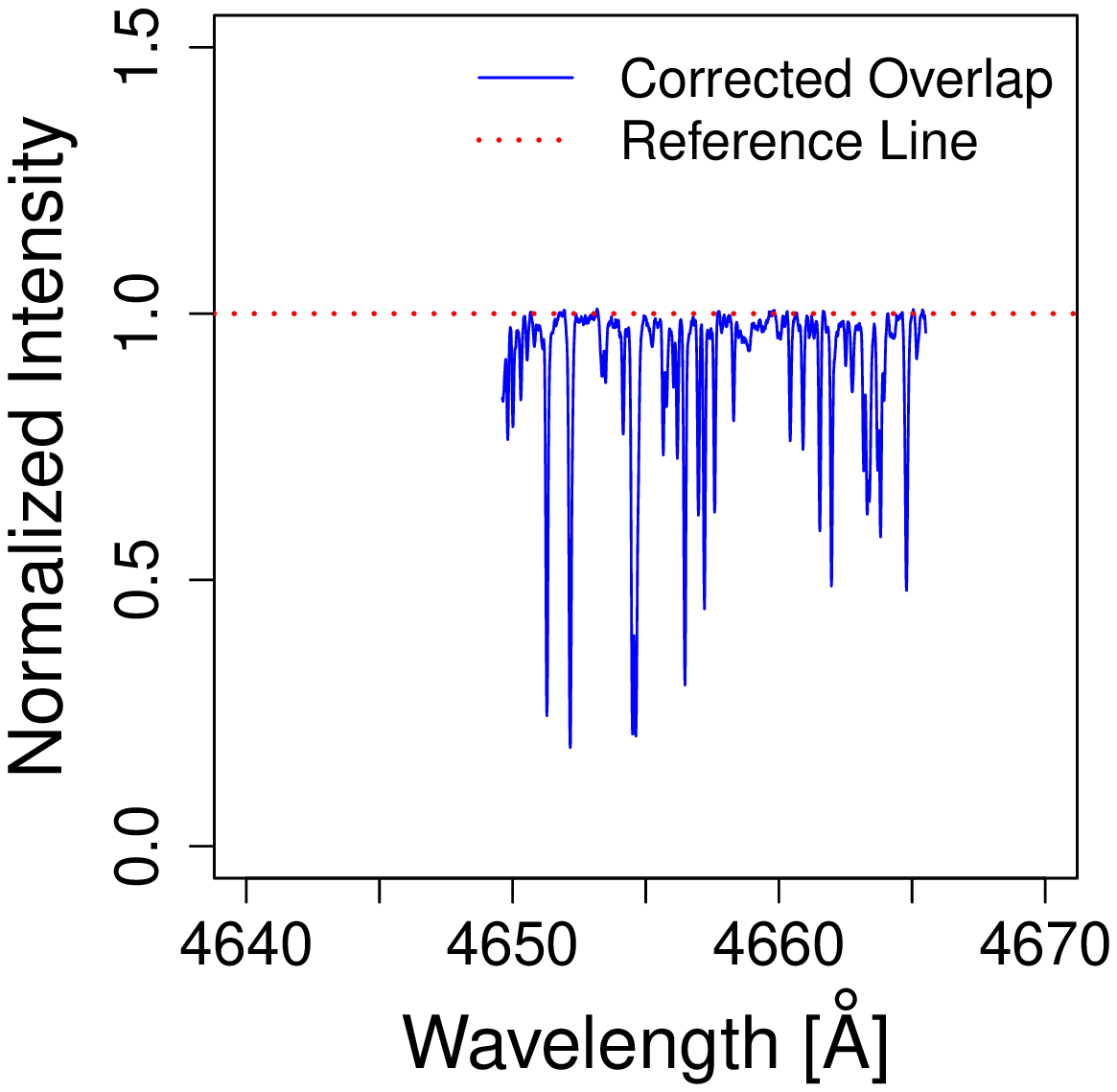}%
  \label{boundary3}%
}
\caption{Combining neighboring orders to correct boundary estimations. (a) The right part of the left order. The blue dashed line is the overlap region, which is shared by the two orders. (b) The left part of the right order, with the blue dashed line again showing the overlapping region. (c) Overlapping region after correction using a weighted average of the two orders.
}\label{boundary_figure}
\end{figure}

The AFS algorithm also has difficulty in this region in cases where the edge of an order splits an absorption line. 
Fortunately, neighboring orders often have some region of overlap that can be used to correct the boundaries. 
A weighted average of the blaze-removed spectrum of the two orders can be used as an estimate of the blaze function in the overlapping region. 
For example, \autoref{boundary_figure} shows two neighboring orders that share an overlapping region. Figure\autoref{boundary1} shows the right boundary of the left order, which looks good as a blaze-removed spectrum using AFS algorithm. Figure\autoref{boundary2} shows the left boundary of the right order, which spuriously rises above $1$. We correct the overlapping region using the following: 
\begin{equation*}
y_{corrected, l} = w_ly_{1, l} + (1-w_l)y_{2, l},  ~ ~ l=1, \dots, m,
\end{equation*}
where $y_1$ and $y_2$ are the intensities for overlapping regions from the left and right orders, respectively, $y_{corrected}$ is the array of intensities for the corrected overlapping region, $m$ is the number of pixels in the overlapping region, and $w_l=1-\frac{l}{m}$, for $l=1,\dots, m$. The result of the correction is shown in Figure\autoref{boundary3}. 
The boundary-corrected spectrum is much better than the original estimate, and we can achieve further improvement by changing the definition of $w_l$. For example, if it is known that one of the two orders has a better estimate on its boundary, we can assign more weight toward the better order. This might be the case for a pair of orders with a broad spectral feature that is cut-off on only one of the orders.

\subsubsection{Wide Absorption Lines}
\begin{figure}
\centering 
\subfloat[Blaze function and its estimate]{%
  \includegraphics[width=0.45\columnwidth]{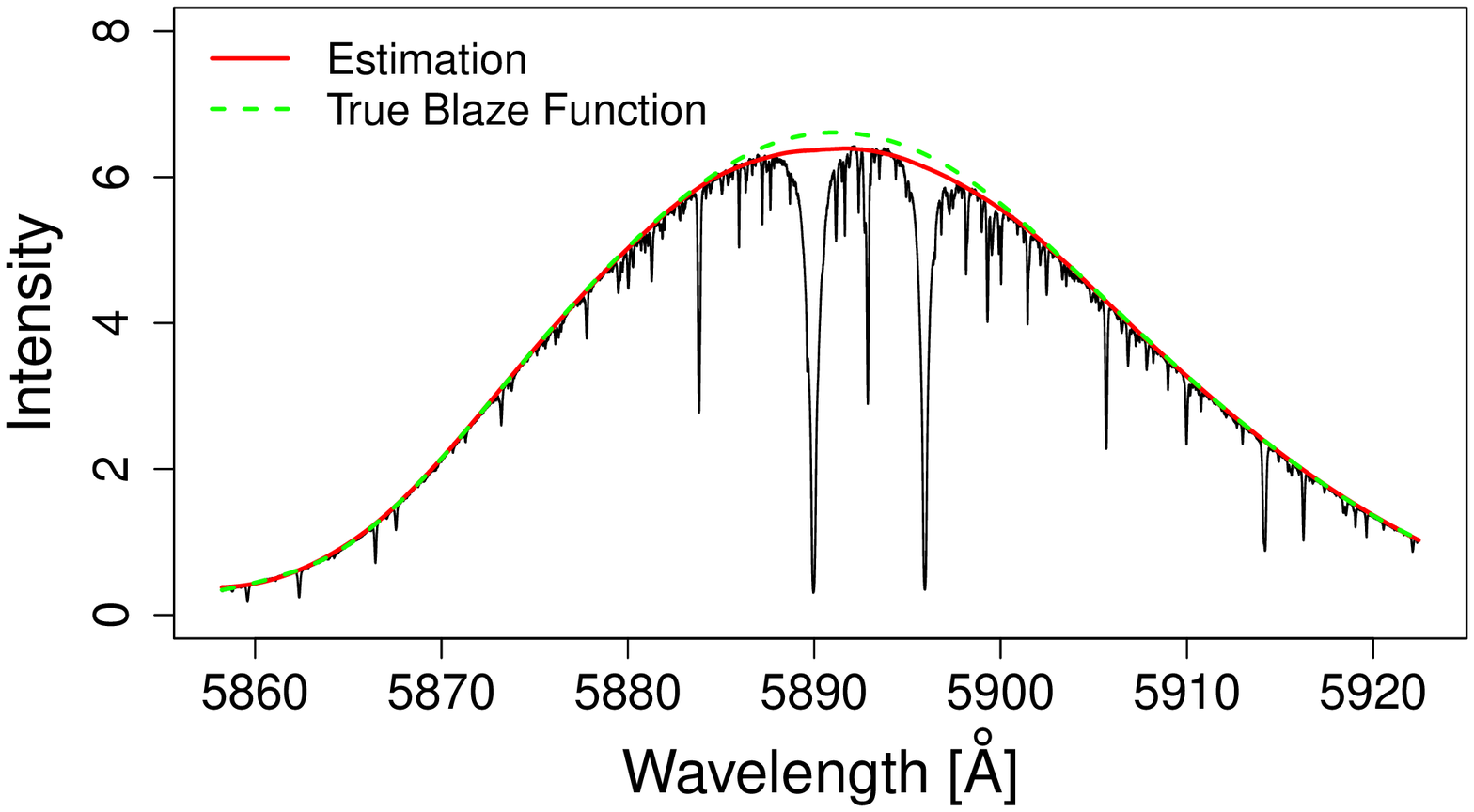}%
  \label{wide1}%
}\qquad
\subfloat[Blaze-removed spectrum]{%
  \includegraphics[width=0.45\columnwidth]{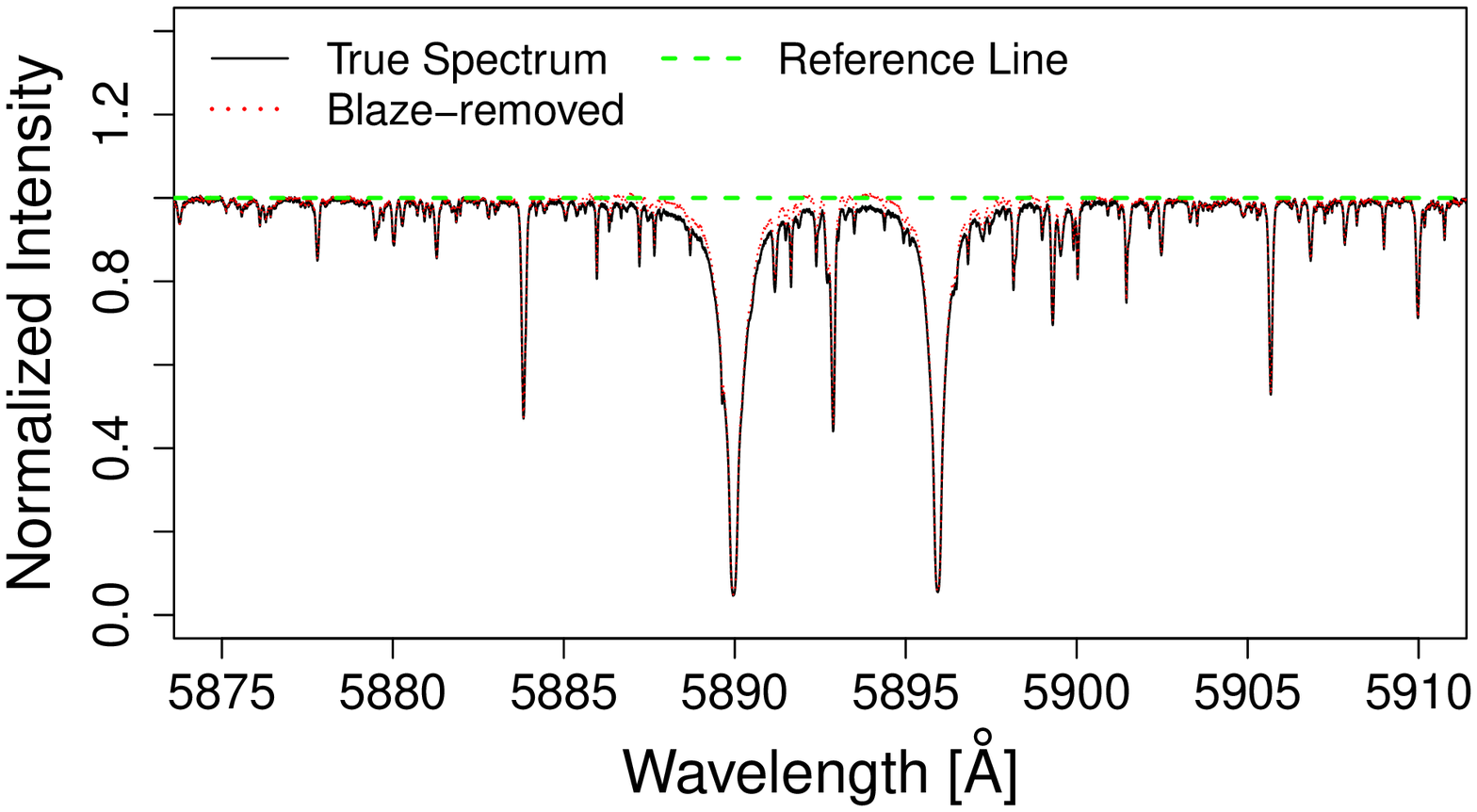}%
  \label{wide2}%
}
\caption{Wide absorption features are a problem for the AFS algorithm without prior information. (a) A spectrum with blaze function is shown in black solid line. The estimation of the AFS algorithm is shown in red solid line. The true blaze function is shown in green dashed line. (b) The true spectrum without blaze function is shown in black solid line. The spectrum after blaze function removal by the AFS algorithm is shown in red dashed line. A reference line at $y=1$ is shown as green dashed line. 
}\label{wide_figure}
\end{figure}

Sometimes an order has wide absorption lines that influence the performance of the blaze function estimation. Wide absorption lines can significantly influence the AFS algorithm performance, but only slightly influence the ALSFS algorithm. 
Figure\autoref{wide2} shows the order containing the two deep Na D lines, which are each so broad that the spectrum does not fully return to the nominal continuum level between them. When attempting to fit the blaze function, the alpha shape will dip into wide features like these and pull the final blaze function estimate downward. In Figure\autoref{wide1}, the spectrum (same as in Figure\autoref{wide2}) before the blaze function removal is displayed. 
Despite failing to return to the proper continuum level, this attempt at normalization looks reasonable to the eye. Without prior information, the proposed data-driven AFS algorithm cannot consistently determine the continuum level over regions of greatly extended absorption like this one. If we know there is a wide absorption feature before applying the algorithm, the regions can be masked in step 4 of the AFS algorithm, excluding those points from the estimate.

\subsubsection{Continuous Opacity}
If an absorption region is wider than half of the wavelength range of the target order, we refer to it as \emph{continuous opacity}. This problem impacts both the AFS algorithm and the ALSFS algorithm. Since the region is so wide, masking is not an option for the AFS algorithm.  The only way to deal with it is to use prior information about continuous opacity: location and intensity. Then the spectrum can be adjusted by accounting for the opacity to get a good estimation. 
The ALSFS algorithm can address this by connecting and adjusting neighboring orders. Since there are overlaps between neighboring orders, one can search neighboring orders until finding an order that does not contain continuous opacity as a reference. Then the orders with continuous opacity can be adjusted to the level of the reference order. An example is shown in \autoref{correction_opacity} to illustrate the continuous opacity correction.

\section{Simulations}\label{simulation}
In our simulation study, we use an integrated-disk solar flux atlas spectrum \citep{wallace2011optical} produced by the National Solar Observatory (NSO), obtained with the McMath-Pierce Solar Telescope's Fourier transform spectrometer. Since the spectra in the atlas were obtained with a Fourier transform spectrometer rather than an echelle spectrograph, they have no intrinsic blaze function. The atlas has been approximately continuum normalized. The spectral resolution of the atlas ranges from 350000 to 700000, and the spectra are essentially noiseless.

To mimic the data characteristic of EXPRES, we use the same wavelength ranges as EXPRES to divide the NSO spectrum into artificial orders. We impose a shape based on a blaze function estimated from a B-star spectrum onto each order and use our algorithms to remove the blaze function. Then simulated photon noise (Gaussian white noise, which well approximates Poisson noise for high S/N) is added corresponding to a S/N of 300. The noisy blaze-imposed spectrum is divided by the true blaze function of produce a benchmark flattened spectrum. In this simulation study, we use two orders: one from the bluer end of the spectrum and one from the redder end. The AFS algorithm and the ALSFS algorithm are tested on the two orders, respectively. Additionally, we compare our algorithms with the commonly used \textit{iterative method}. The iterative method is introduced next.

\subsection{Iterative Method}
The iterative method is commonly used to remove blaze function from a spectrum. A polynomial model is fit to a spectrum order and the fit is considered as the starting estimation for the blaze function. Next, the method iterates. In each iteration, there is a threshold curve obtained by: 
\begin{equation}
threshold(\lambda_i, t)=fit_{polynomial}^{(t)}(\lambda_i)-\frac{0.5}{t+1},
\end{equation}
where $fit_{polynomial}^{(t)}(\lambda_i)$ is the 7th-order polynomial regression estimate at $\lambda_i$ and $t$ is the iteration time. As t increases, the threshold curve increases. The method builds a subset $M_t$ containing wavelength $\lambda_i$'s with intensity larger than $threshold(\lambda_i, t)$. Then in the next iteration, it only uses spectrum points whose wavelength values are in $M_t$ to fit a polynomial model and the fit is a new estimation for continuum. To stop the iteration, a stopping time, denoted as $T$, can be set at a particular iteration.  Another option is to set a standard deviation value $sd$, such that the algorithm stops when the standard deviation of the blaze-removed spectrum is smaller than $sd$. The $sd$ cannot be too small (much smaller than the standard deviation of the true spectrum without the blaze function), otherwise the iteration will not stop. In this work, we stop the iteration if either it arrives at the $T$-th iteration or the standard deviation of the blaze-removed spectrum is smaller than $sd$. Empirically, an $sd$ value around $0.05$ works well, but the performance of the iterative method is influenced by the choice of $T$. A higher S/N requires a larger $T$ and, in general, we have found a $T$ value equal to S/N seems to well.

\subsection{Simulated Spectra}\label{simulated_spectra}
\begin{figure}
\centering 
\subfloat[Order B]{%
  \includegraphics[width=0.6\columnwidth]{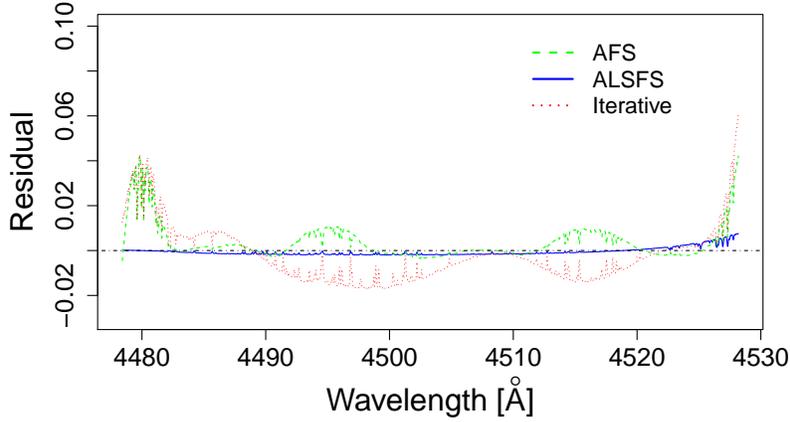}%
  \label{simu1}%
}\qquad
\subfloat[Order R]{%
  \includegraphics[width=0.6\columnwidth]{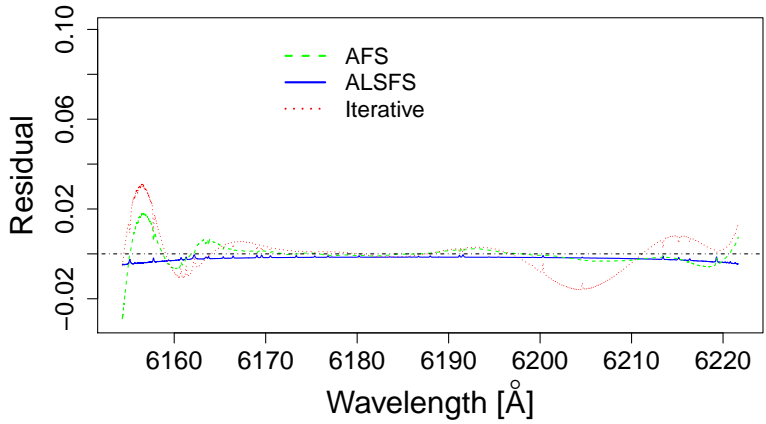}%
  \label{simu2}%
}
\caption{Two orders from NSO data, with wavelength range according to EXPRES spectrum. Order B is from wavelength range 4478 to 4528 \angstrom and the order R is from wavelength range 6154 to 6221 \angstrom. Three methods are applied to the two orders to compare residuals. (a) Residuals of order B. (b) Residuals of order R. AFS displayed here are without the boundary corrections - it could be improved further by boundary modification described in \autoref{boundary}.} \label{simu_figure}
\end{figure}

The first order (blue) has wavelengths ranging from 4478 to 4528 \angstrom, called ``order B'', and the second order (red) has wavelengths ranging from 6154 to 6221 \angstrom, called ``order R''. For this simulation, we require a realistic blaze function to add to the NSO spectrum. To do this, we estimate a blaze function of a B-star, HR 5501, observed with EXPRES \citep{jurgenson2016expres}. 
We first use ALSFS algorithm, with the corresponding LED spectrum (after using AFS algorithm on it) as the lab source, on the B-star spectrum to get the estimate of the blaze function. Then we apply this estimate as the blaze function to the two orders. The raw spectrum of the B-star spectrum is then used as the lab source reference for the ALSFS algorithm. The AFS algorithm, ALSFS algorithm, and iterative method are applied to estimate blaze functions of the two orders and residuals are calculated. The residuals of order B and R with S/N$=300$ are displayed in Figure\autoref{simu1} and \autoref{simu2}, respectively. Overall, ALSFS has the smallest residuals that are consistent across the whole order. AFS has larger residuals than ASLFS, which increase in the boundary regions, but the iterative method has larger residuals than AFS in both boundary regions and middle regions.

\subsection{Results}
\begin{figure}
\centering 
\subfloat[S/N 300, order B]{%
  \includegraphics[width=0.3\columnwidth]{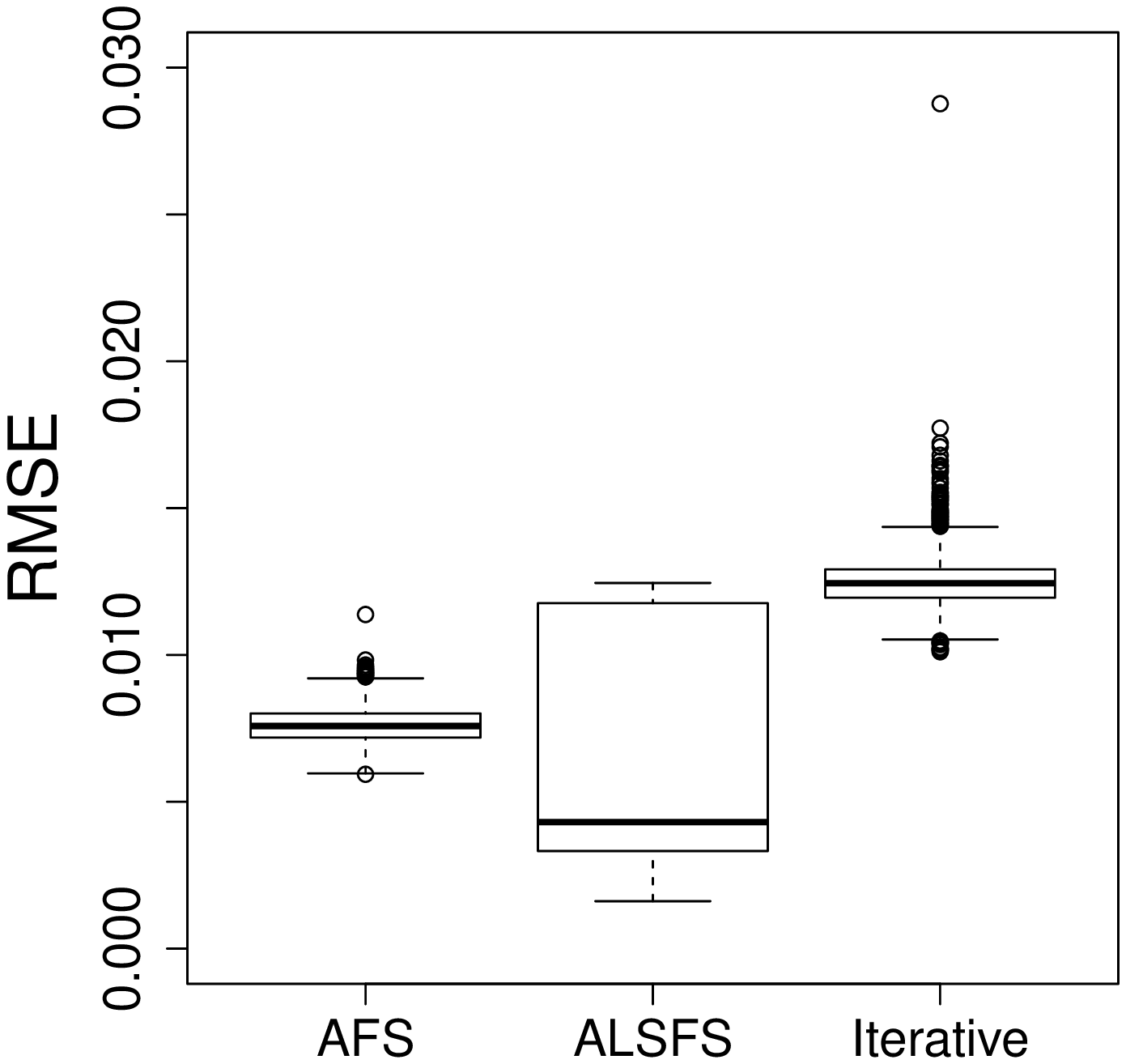}%
  \label{300_1}%
}\qquad
\subfloat[S/N 150, order B]{%
  \includegraphics[width=0.3\columnwidth]{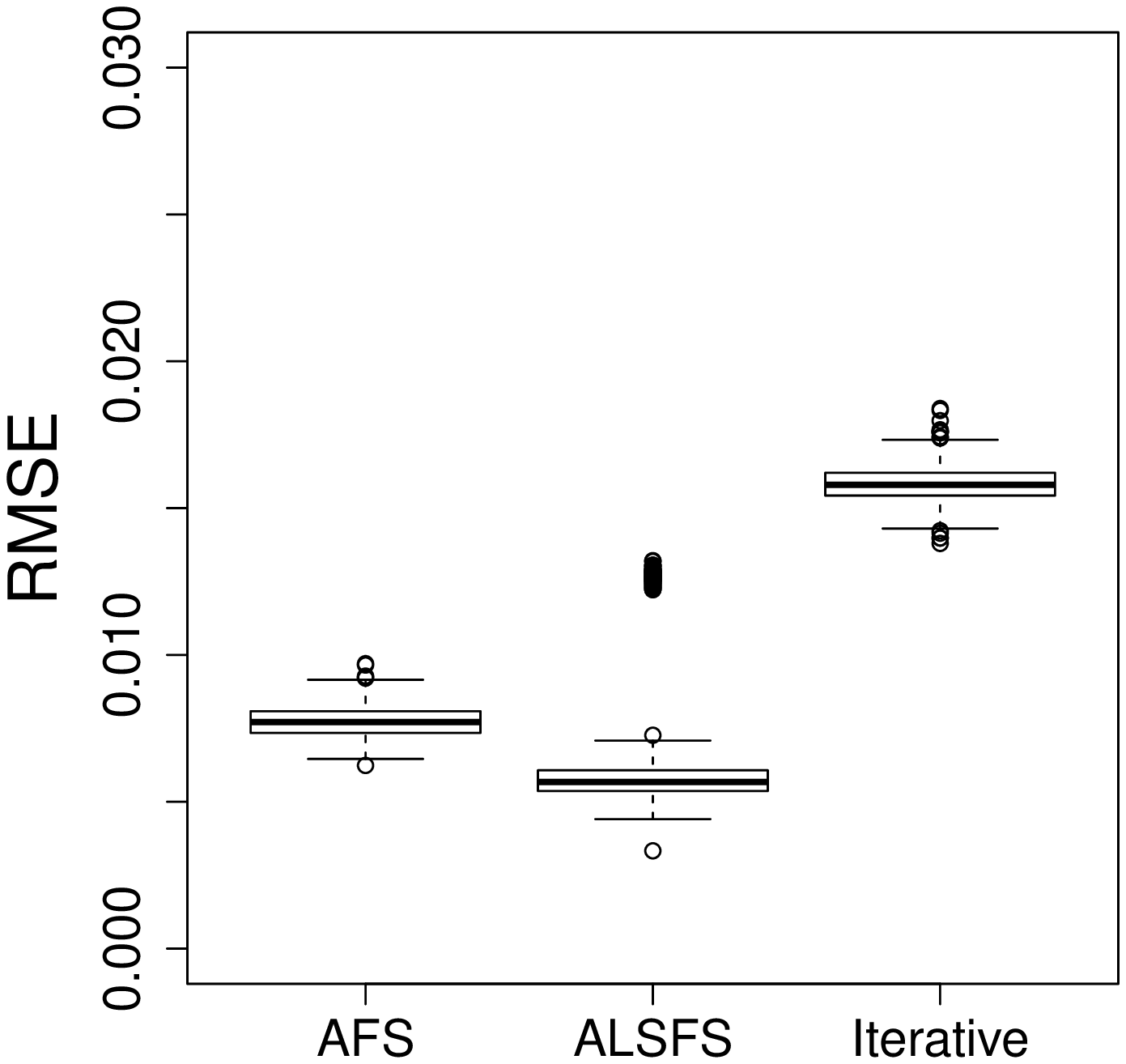}%
  \label{150_1}%
}\qquad
\subfloat[S/N 50, order B]{%
  \includegraphics[width=0.3\columnwidth]{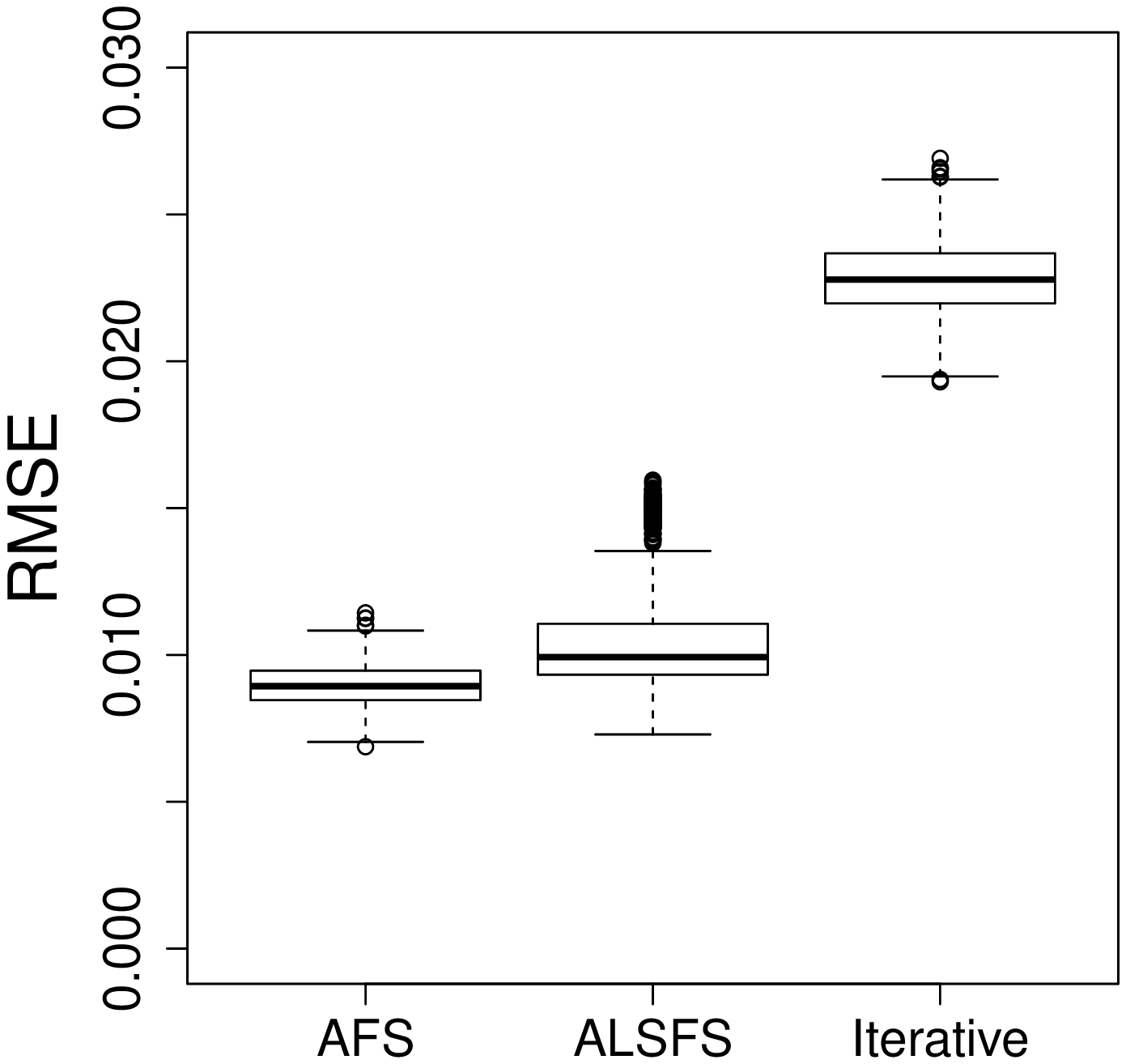}%
  \label{50_1}%
}\qquad
\subfloat[S/N 300, order R]{%
  \includegraphics[width=0.3\columnwidth]{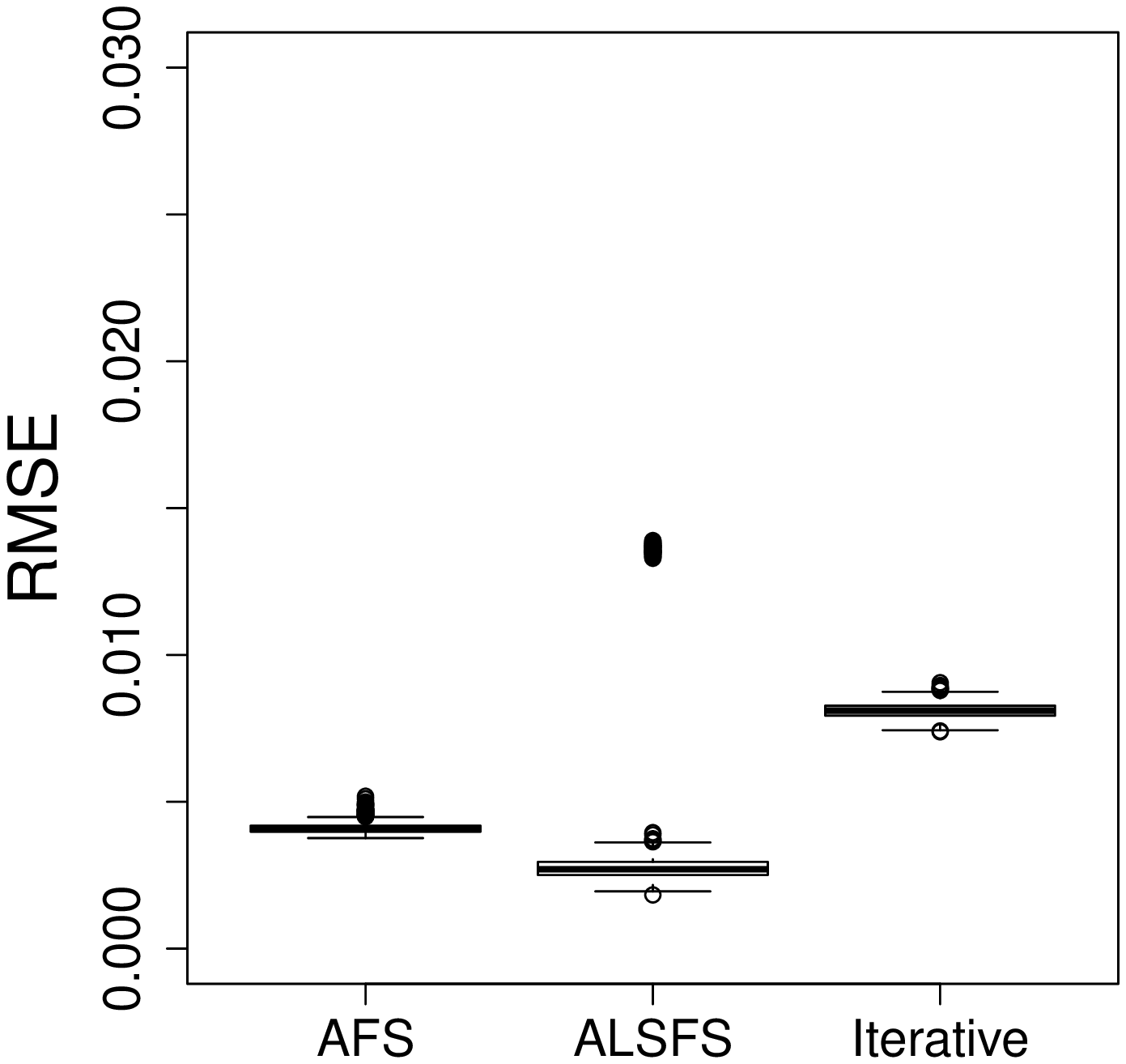}%
  \label{300_2}%
}\qquad
\subfloat[S/N 150, order R]{%
  \includegraphics[width=0.3\columnwidth]{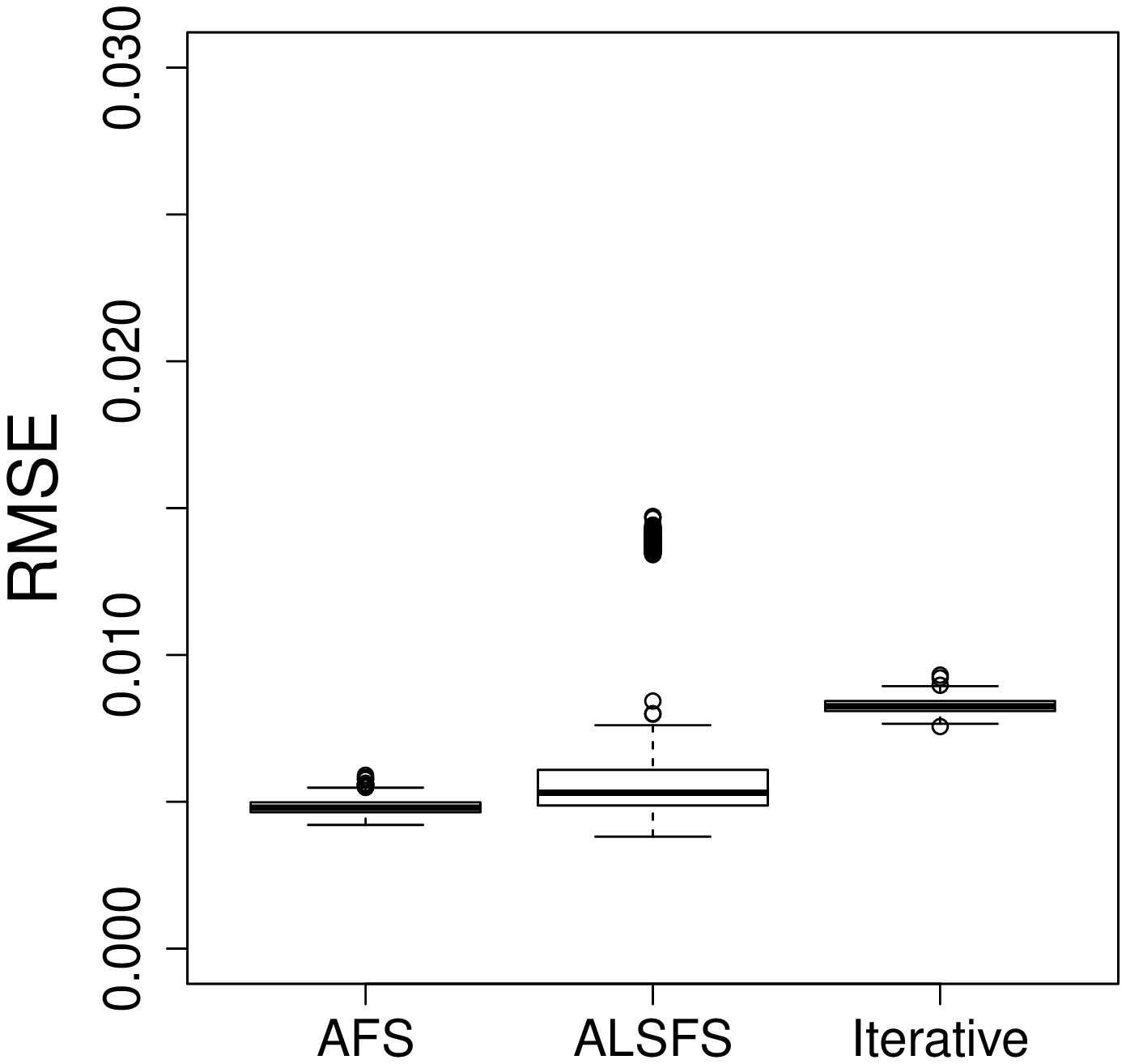}%
  \label{150_2}%
}\qquad
\subfloat[S/N 50, order R]{%
  \includegraphics[width=0.3\columnwidth]{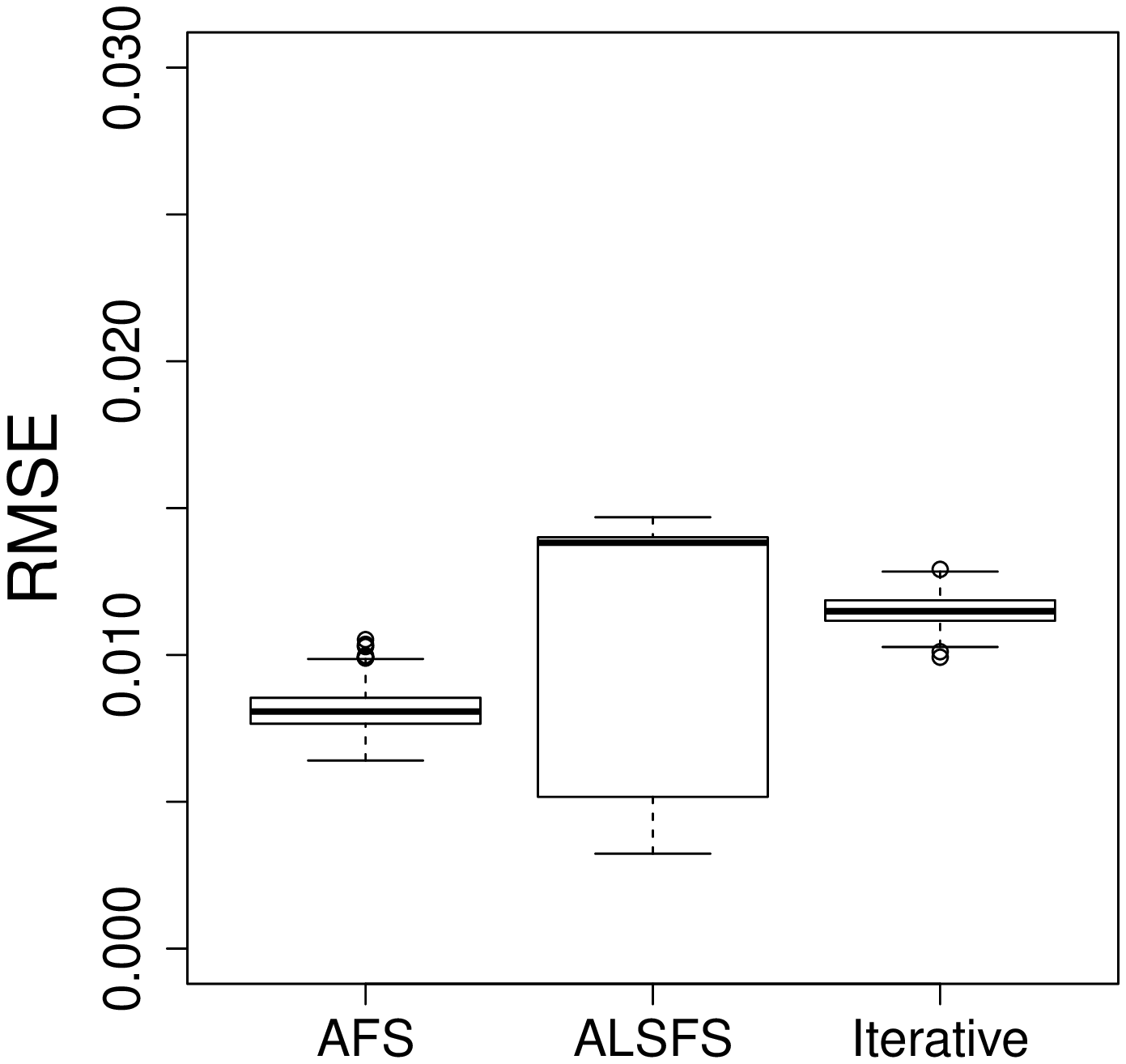}%
  \label{50_2}%
}
\caption{Distribution of RMSE over 1000 samples for S/N 300, 150, and 50. The simulation is repeated 1000 times for the three methods on the two orders, respectively. (a), (b), and (c) are results of order B. (d), (e), and (f) are results of order R. Since the true spectra are known, the RMSEs could be calculated.} \label{simu_mse}
\end{figure}

\autoref{simu_figure} displays a single realization of the noisy spectrum. We repeat the procedure by adding different realizations of noise 1000 times. For each of the 1000 realizations, the AFS algorithm, ALSFS algorithm, and the iterative method are applied to the two orders. 
This is carried out for an S/N 300, 150, and 50. The results are displayed in \autoref{simu_mse} and \autoref{table:mean}. The root mean squared error (RMSE) is calculated as $\sqrt{\frac{1}{n}\sum\limits_{i=1}^{n} r_i^2}$, where $r_i$ is the residual at pixel $i$. The ALSFS algorithm has the smallest median RMSE for three scenarios: S/N 300-order B, S/N 300-order R and S/N 150-order B. For the other three scenarios, the AFS algorithm has the smallest median RMSE. Except for S/N 50-order R, the iterative method has the largest median RMSE.

\begin{table}[ht!]
\centering
\begin{tabular}{|c|ccc|ccc|}
  \hline
   \multicolumn{1}{|c}{} & \multicolumn{3}{c|}{Order B}   &  \multicolumn{3}{c|}{Order R}   \\ 
      \hline
  \multicolumn{1}{|c|}{S/N}  & AFS & ALSFS &  Iterative  &AFS & ALSFS & Iterative  \\ 
  \hline
 300 &  7.58 & 4.31 & 12.44 & 4.06 & 2.70 & 8.10 \\ 
 150 & 7.71 & 5.67 & 15.79 & 4.80 & 5.31 & 8.25 \\ 
 50 & 8.94 & 9.93 & 22.78 & 8.07 & 13.82 & 11.49\\ 
   \hline
\end{tabular}
\caption{Unit: $1\times10^{-3}$. Medians of RMSE over 1000 samples for orders B and R with varying S/N for AFS, ALSFS, and the iterative method. Medians listed in this table are corresponding to distributions of RMSE in \autoref{simu_mse}.}\label{table:mean}
\end{table}

\subsection{Correction for Continuous Opacity}\label{correction_opacity}
\begin{figure}
\centering 
\subfloat[True NSO spectrum]{%
  \includegraphics[width=0.45\columnwidth]{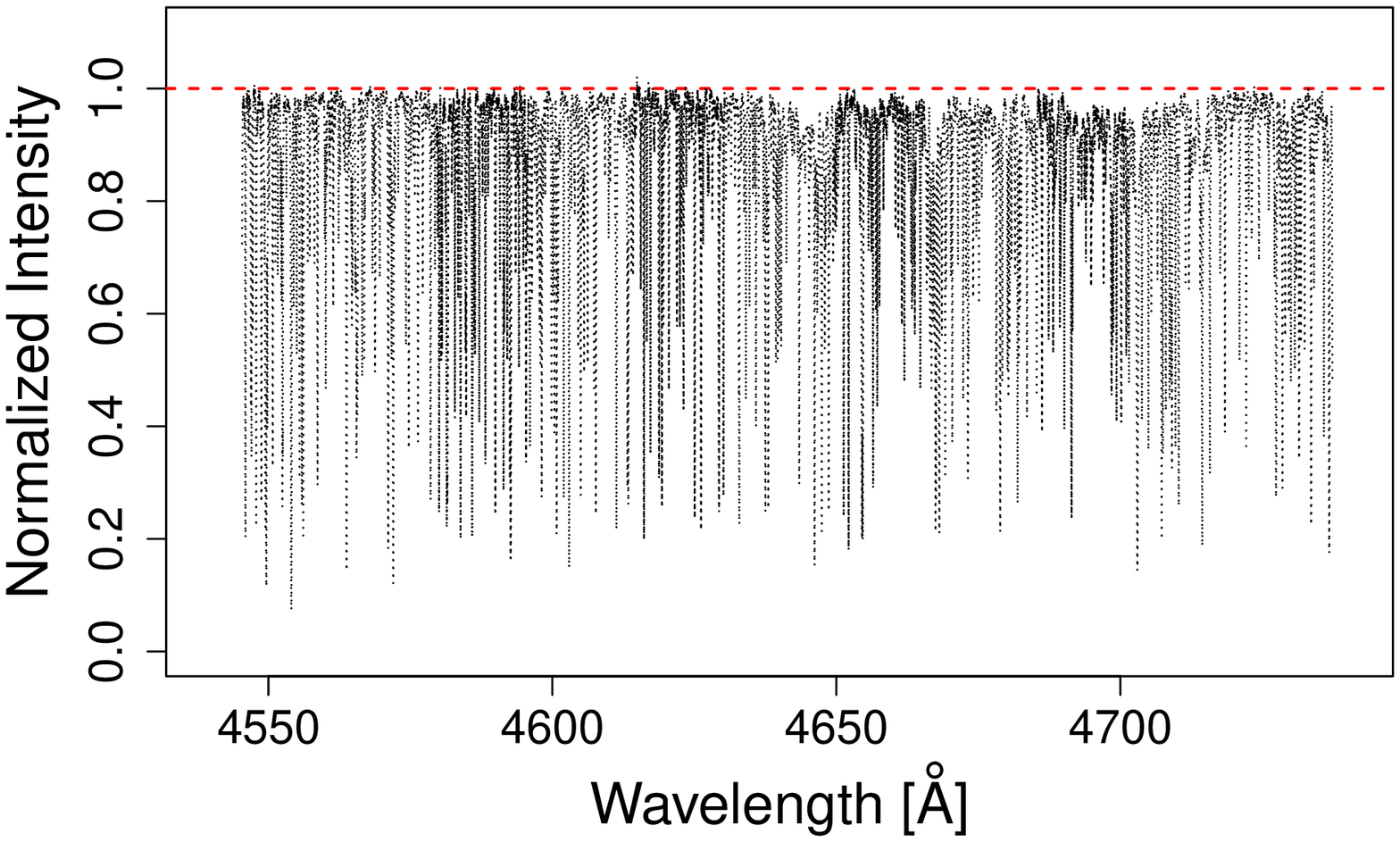}%
  \label{opacity1}%
}\qquad
\subfloat[Results of ALSFS]{%
  \includegraphics[width=0.45\columnwidth]{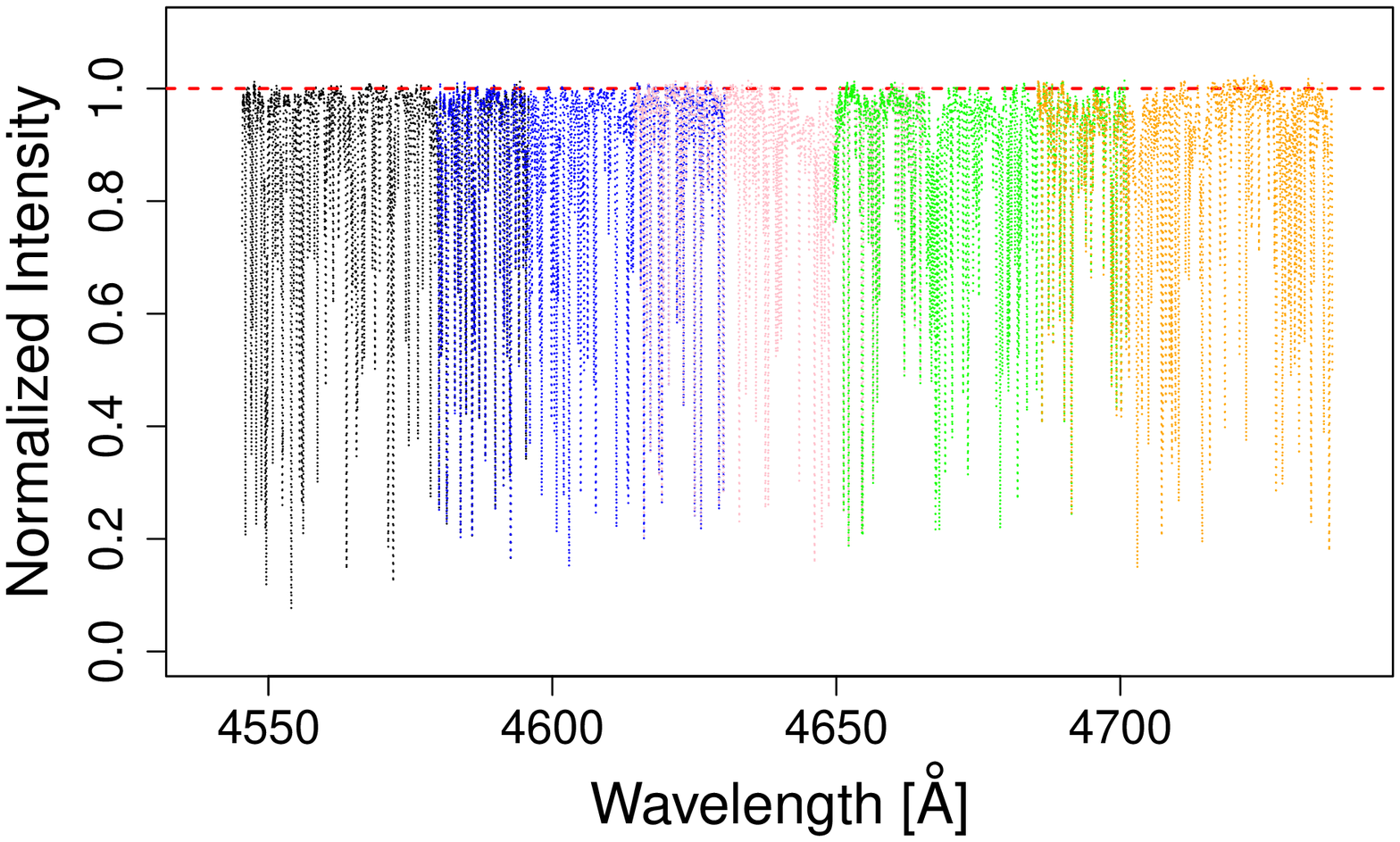}%
  \label{opacity2}%
}\qquad
\subfloat[Combined spectrum]{%
  \includegraphics[width=0.45\columnwidth]{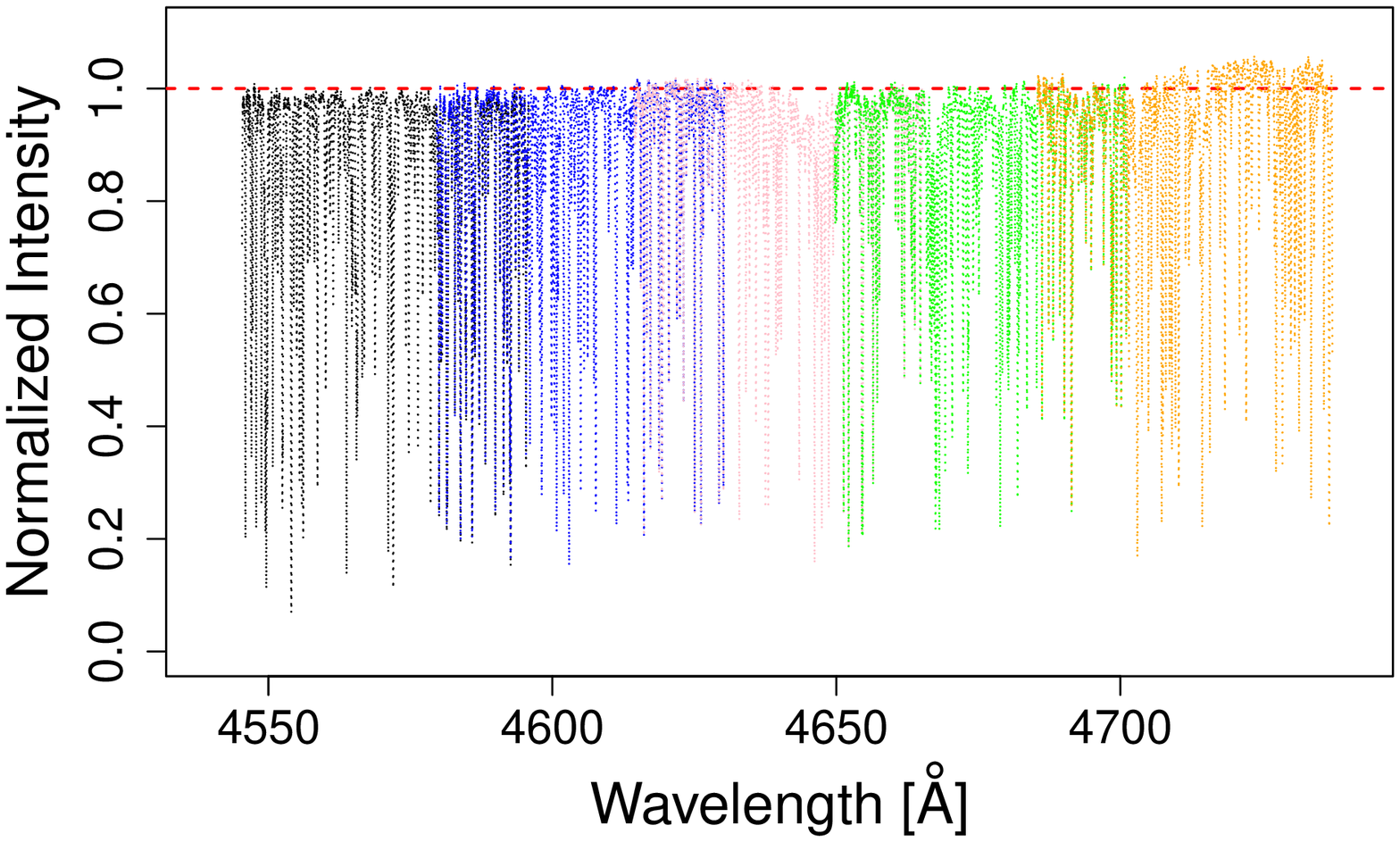}%
  \label{opacity3}%
}\qquad
\subfloat[Corrected spectrum]{%
  \includegraphics[width=0.45\columnwidth]{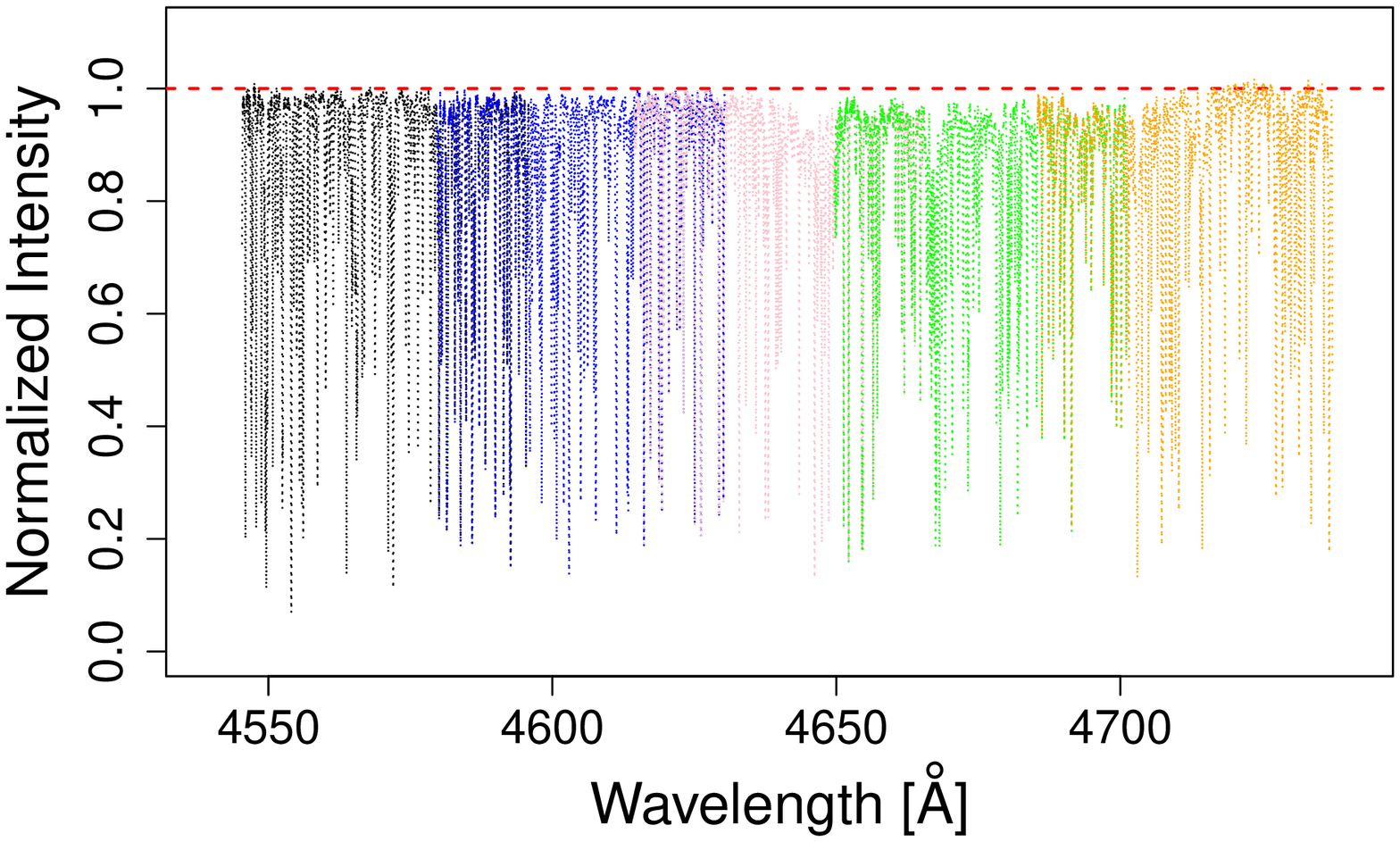}%
  \label{opacity4}%
}
\caption{(a) A segment of the full NSO spectrum, with significant continuous opacity redwards of about 4630 \angstrom. The red dashed line is a reference line at normalized intensity $=1$. (b) The segment is divided into five artificial orders with injected blaze functions. Here we show the resulting blaze-removed orders from ALSFS shown in different colors, respectively. (c) Each order is linearly adjusted with its bluewards neighboring order as a reference based on their overlapping region. The combined long spectrum is still imperfect because the slope estimation of the first order could be imperfect. (d) An ordinary linear regression is fitted to the combined spectrum to remove the extra slope. The resulting spectrum recovers the continuous opacity well.
}\label{opacity_figure}
\end{figure}

An example of continuous opacity is illustrated using the same simulation setup: we apply a blaze function obtained from a B-star spectrum to the NSO spectrum, and then we attempt to recover the underlying spectrum.
Here we examine five consecutive artificial orders. A region of continuous opacity spans about 80 \angstrom, which is captured in parts of the third, fourth and fifth orders, displayed in Figure\autoref{opacity1}. After applying the ALSFS algorithm, the resulting blaze-removed spectra are displayed in Figure\autoref{opacity2}. Although the blaze-removed spectra are flat, the ALSFS algorithm does not capture the continuous opacity correctly on its own. We know that the first two orders are not affected by the continuous opacity, and so they are used as the reference to correct the other orders: each order is linearly adjusted in intercept and slope to be align with its left neighboring order using the points within the overlapping region whose normalized intensities are in the top $\tilde{q}$ quantile. Ideally the combined segment should recover the continuous opacity in the normalized intensities after the blaze function removal. However, error in the slope estimation of the first order is amplified: the combined spectrum in Figure\autoref{opacity3} is not perfectly horizontal and it goes upward from left to right. We fit another linear regression using the points in the combined spectrum whose normalized intensities are in the top $\tilde{q}$ quantile to remove the extra slope. The resulting spectrum recovers the continuous opacity well, as displayed in Figure\autoref{opacity4}. 
\subsection{Cosmic Rays}
\begin{figure}
\centering 
\subfloat[Blaze function and its AFS estimate]{%
  \includegraphics[width=0.45\columnwidth]{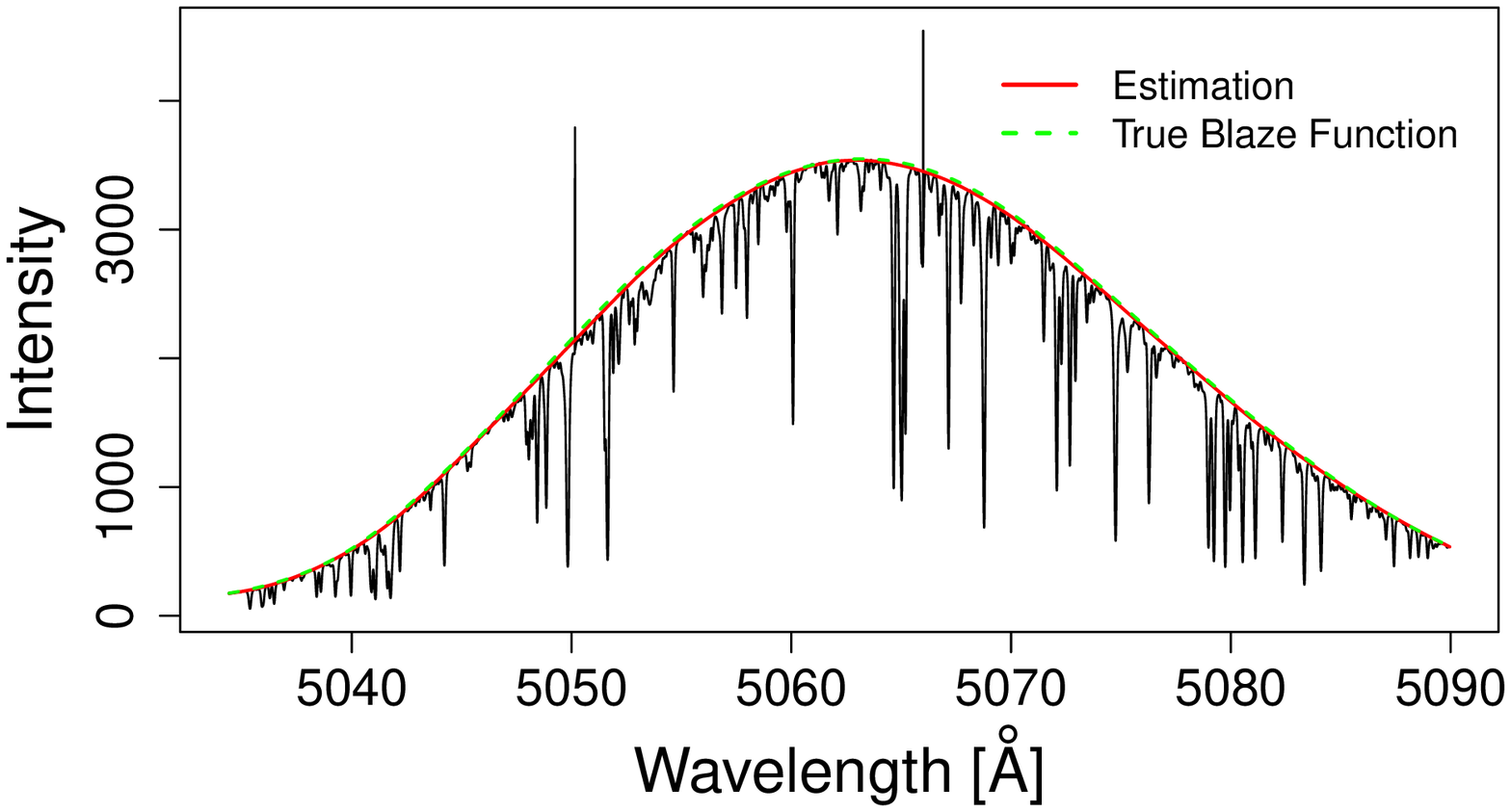}%
  \label{cosmic1}%
}\qquad
\subfloat[Blaze-removed spectrum]{%
  \includegraphics[width=0.45\columnwidth]{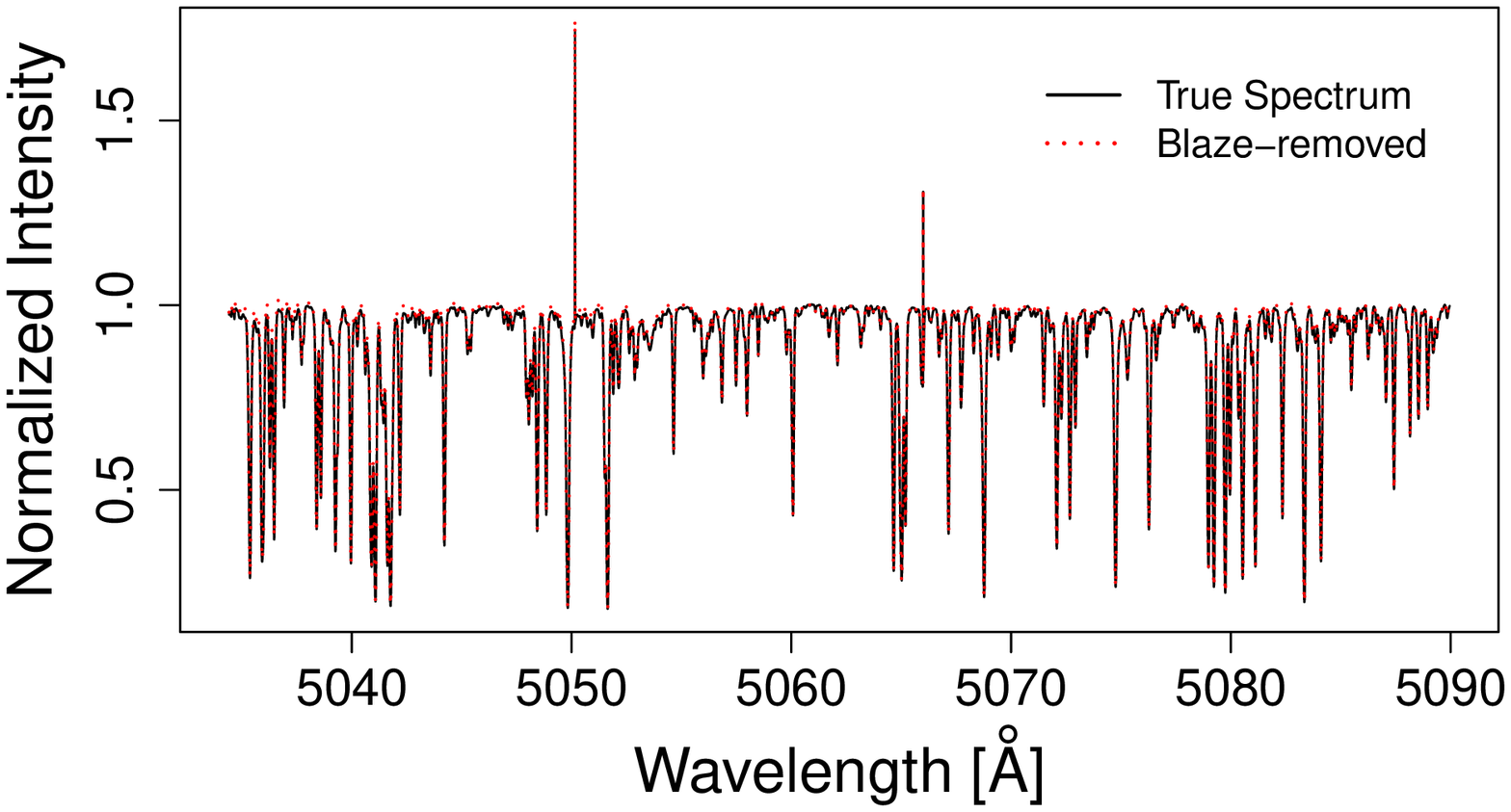}%
  \label{cosmic2}%
}
\caption{(a) An order with simulated cosmic rays. The black solid line is the raw spectrum, the red solid line is the blaze function estimate from the AFS algorithm, and the green dashed line is the true blaze function. (b) The blaze-removed spectrum using the AFS algorithm. The black solid line is the true spectrum without the blaze function and red dashed line is the blaze-removed spectrum.} \label{cosmic_rays}
\end{figure}

The proposed methods can also be used for spectra with cosmic rays. In particular, the modified AFS algorithm for lab source smoothing, described in \autoref{ls_smoothing}, can be used directly to deal with the presence of cosmic rays. To demonstrate this, an order with simulated cosmic rays is displayed in Figure\autoref{cosmic1}, which contains two upward spikes. The blaze function estimate from the AFS algorithm is shown in Figure\autoref{cosmic1}, and the blaze-removed spectrum is displayed in Figure\autoref{cosmic1} in comparison to the true spectrum. The ALSFS algorithm can be modified similarly to deal with cosmic rays.

\section{Applications and Discussions}

The AFS algorithm can be useful for studying telluric and micro-telluric absorption lines. Telluric lines, originating in the Earth's atmosphere, create time-varying and humidity-dependent perturbations to the shapes of stellar lines (Leet et al. 2019, in prep). It is a particularly acute problem in the field of high-precision exoplanet radial velocity detection where uncorrected telluric lines contribute a radial-velocity error of $\sim$ 0.2 to 1 meters per second in optical wavelengths \citep{cunha2014impact} and as much as a few meters per second in near infrared wavelengths \citep{bean2010crires}. 

Telluric lines can be measured using spectroscopic observations of B-stars, which are bright, rapidly rotating young stars whose spectra are devoid of all but the strongest absorption lines because of extreme rotational broadening. A B-star acts as a background against which narrow telluric lines can be observed as a calibration tool for radial velocity measurements of other stars. Fitting and removing the blaze function and continuum of B-stars allows the depth of the telluric lines to be measured, which folds into current and proposed methods to mitigate their effects (e.g., Leet et al. 2019, in prep).

\begin{figure}
\centering 
\subfloat[Order B, AFS]{%
  \includegraphics[width=0.45\columnwidth]{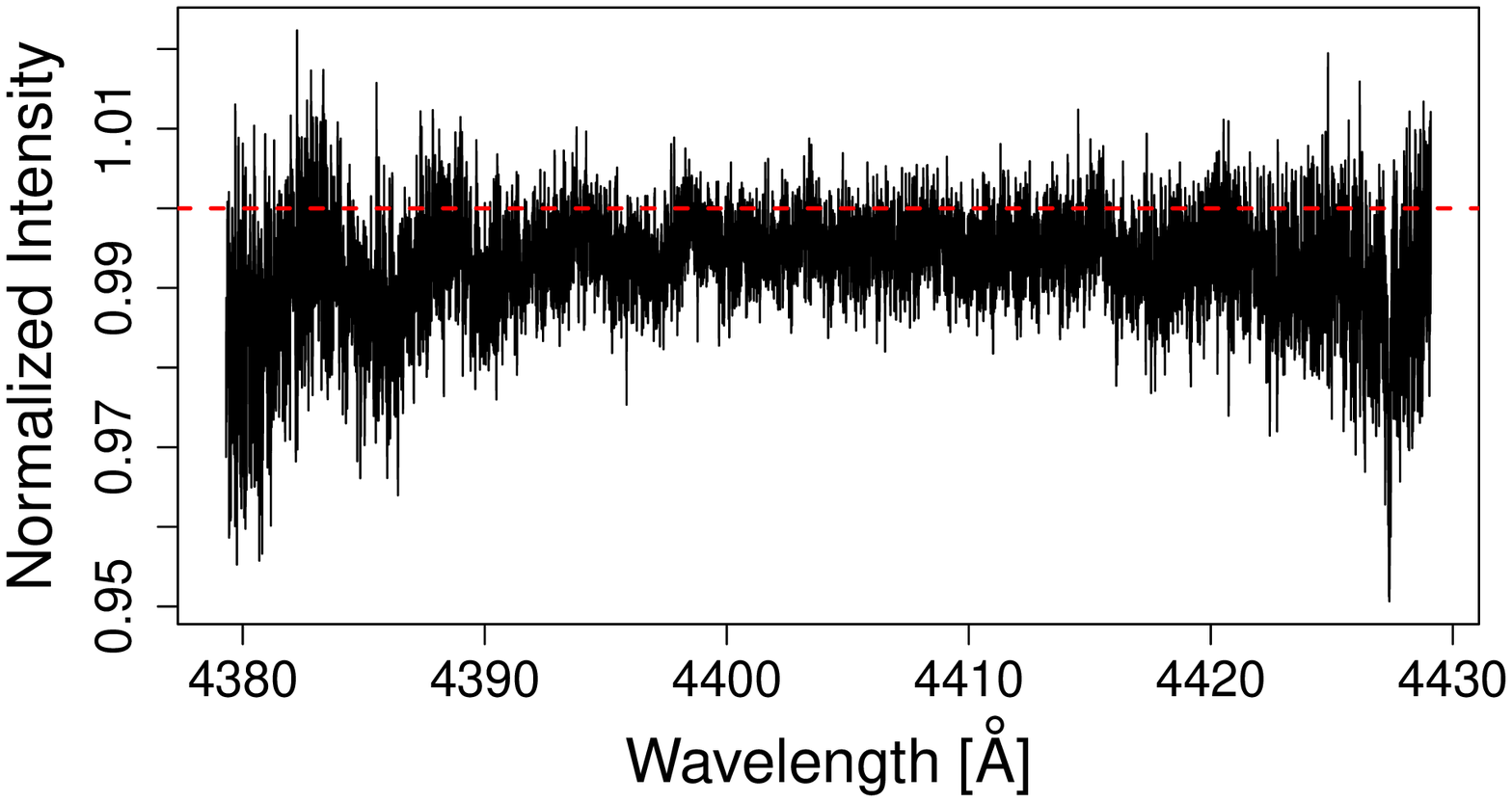}%
  \label{discuss1}%
}\qquad
\subfloat[Order B, Iterative]{%
  \includegraphics[width=0.45\columnwidth]{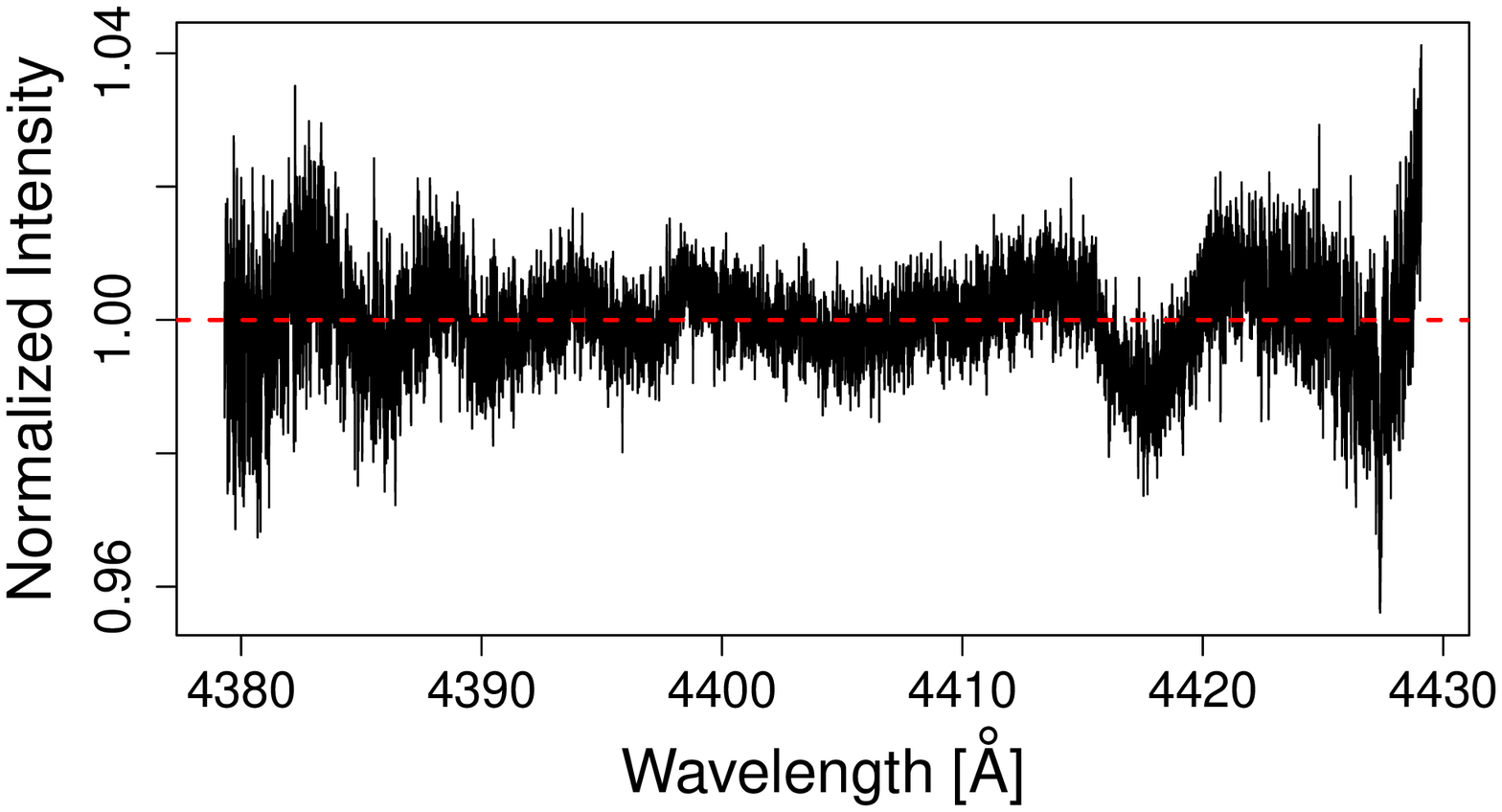}%
  \label{discuss2}%
}\qquad
\subfloat[Order R, AFS]{%
  \includegraphics[width=0.45\columnwidth]{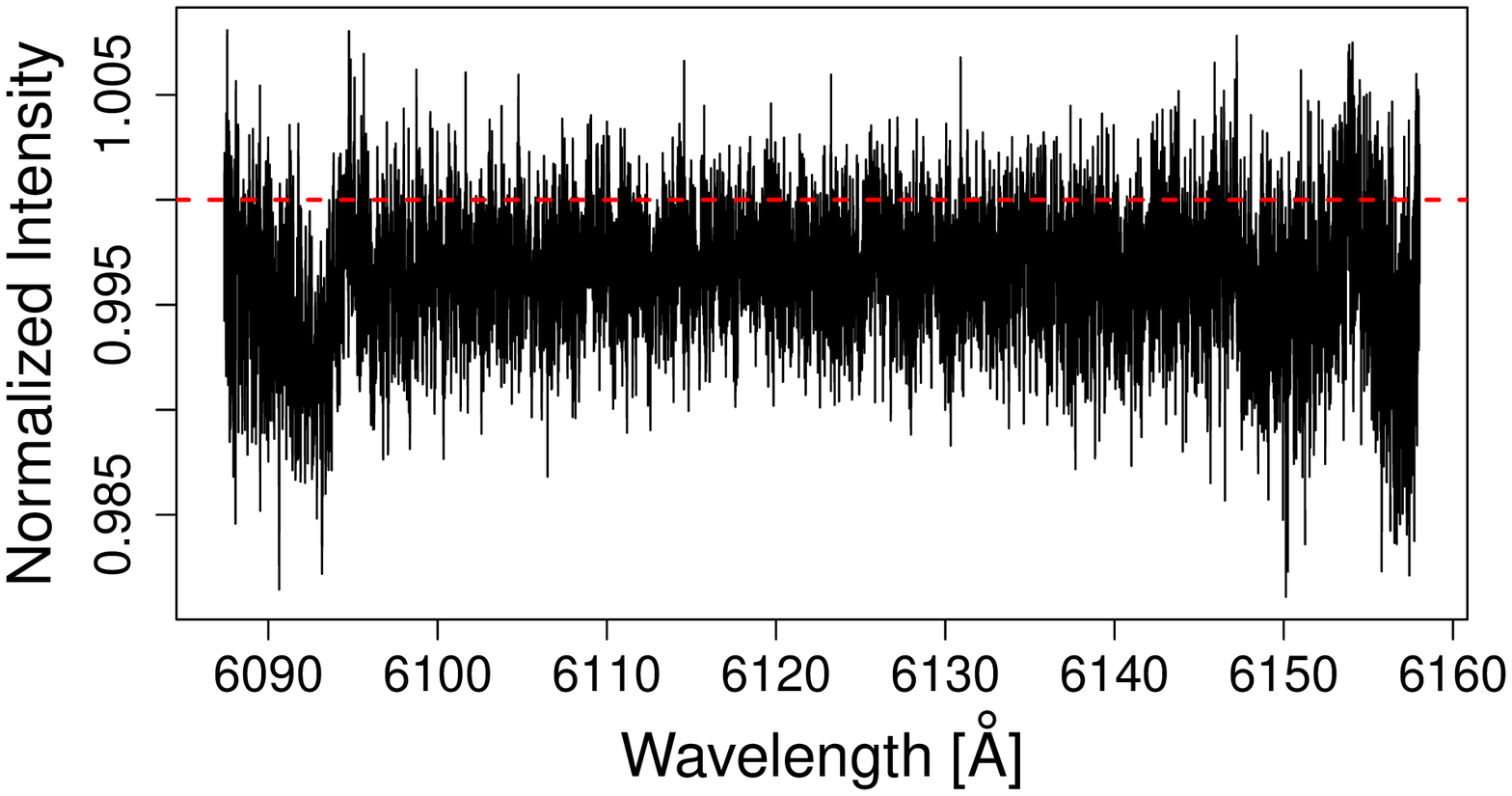}%
  \label{discuss3}%
}\qquad
\subfloat[Order R, Iterative]{%
  \includegraphics[width=0.45\columnwidth]{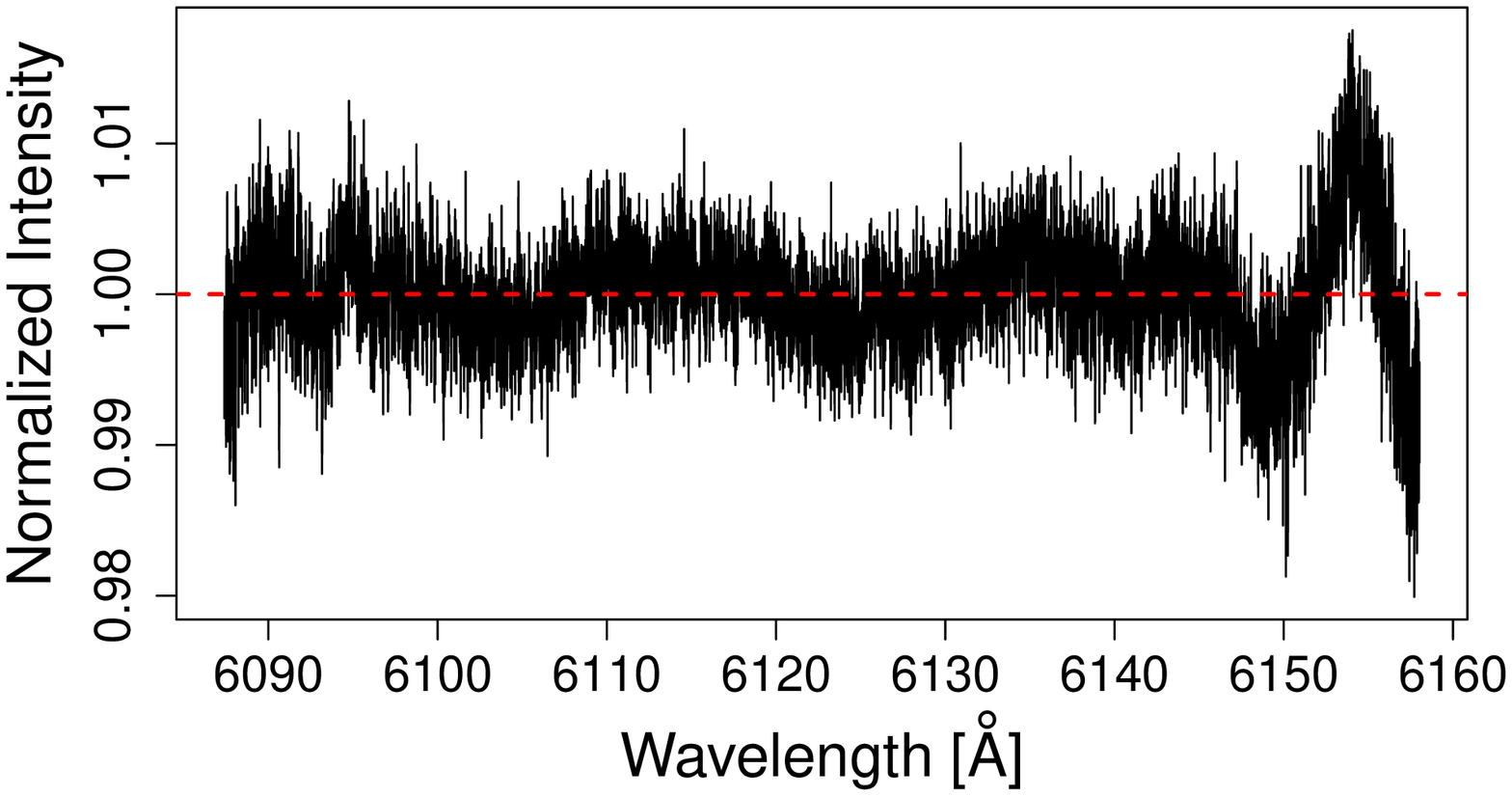}%
  \label{discuss4}%
}
\caption{A B-star spectrum, HR 8634, observed with EXPRES on July 7, 2018. The AFS and iterative method are applied to a blue order and a red order respectively. (a) and (c) are the blue and red order flattened by AFS. (b) and (d) are the blue and red order flattened by the iterative polynomial fitting method, which leaves a zigzag shape to the continuum.} \label{discussion_Bstar}
\end{figure}

 To illustrate the applicability of the proposed algorithms, the AFS and, for comparison, the iterative method are applied to a B-star spectrum,  HR 8634, which was observed with EXPRES \citep{jurgenson2016expres} on July 7, 2018. A blue order (order B: 4376 to 4431 \angstrom) and a red order (order R: 6085 to 6160 \angstrom) of this spectrum were selected to show the effect of the blaze function removal. The flattened spectra are displayed in \autoref{discussion_Bstar}, with a reference line at normalized intensity $=1$. Though the ground truth is unknown, it appears that the proposed AFS works well on flattening the spectra, while the spectra flattened by the iterative method have a zigzag pattern.

The ALSFS method can be useful for estimating the blaze function of the more complex spectra of late-type stars by incorporating information from a lab source spectrum to obtain an initial guess. Late-type stars are the primary targets of Extreme Precision Radial Velocity (EPRV) planet searches, which aim to suppress radial velocity measurement errors below $\sim$ 1 meter per second. Among the greatest challenges hindering EPRV is the problem of stellar activity: magnetically-driven motions within the stellar atmosphere lead to time-varying features, such as spots and faculae, that create line-profile distortions that skew the measured centroids of the lines leading to imprecise RV measurements \citep[summarized in][Section 4.2]{fischer2016state}.

Work to address the problem of stellar activity is ongoing, but one encouraging approach used by several teams is to investigate the sensitivity of individual spectral lines to activity, in order to obtain activity-free RVs \citep{davis2017insights, wise2018new, dumusque2018measuring}. These methods all utilize some sort of continuum-fitting method, because the depths of the lines must be known with precision in order to search for correlations with activity over time. By providing flatter, more uniform blaze function estimates, the ALSFS algorithm will permit more precise measurements of the individual line depths and line profile shapes that are correlated with stellar activity.

\begin{figure}
\centering 
\subfloat[Order B, ALSFS]{%
  \includegraphics[width=0.45\columnwidth]{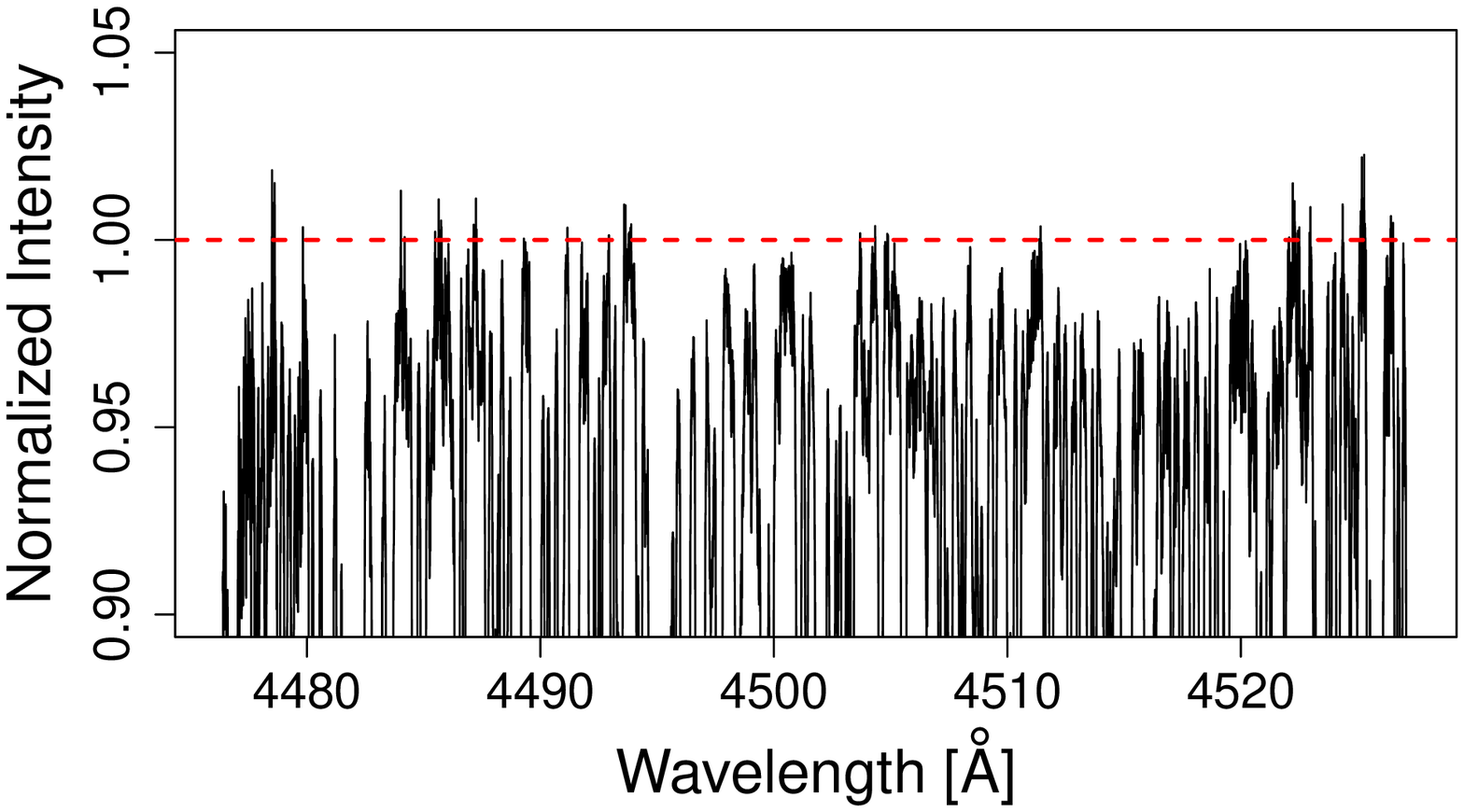}%
  \label{discuss5}%
}\qquad
\subfloat[Order B, Iterative]{%
  \includegraphics[width=0.45\columnwidth]{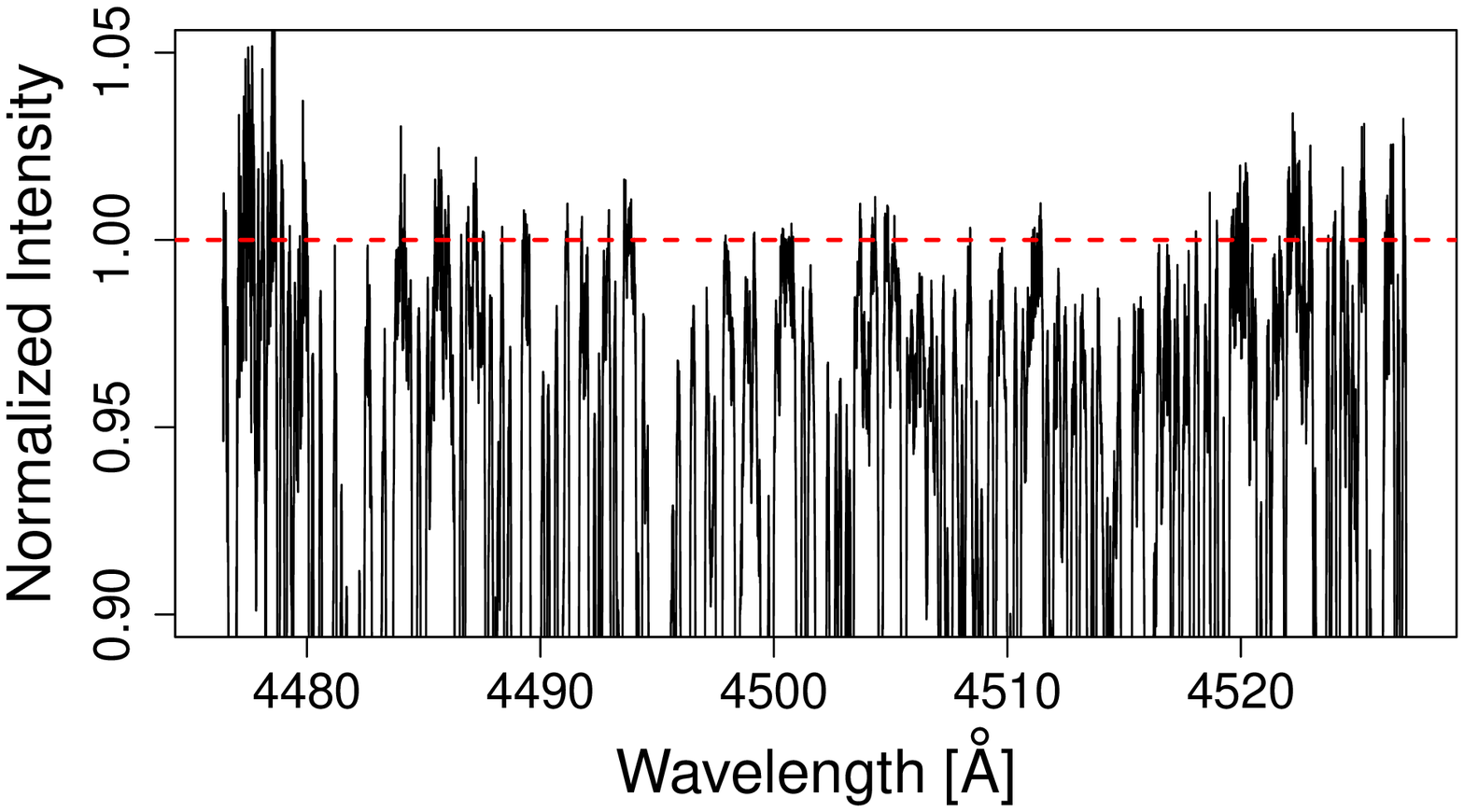}%
  \label{discuss6}%
}\qquad
\subfloat[Order R, ALSFS]{%
  \includegraphics[width=0.45\columnwidth]{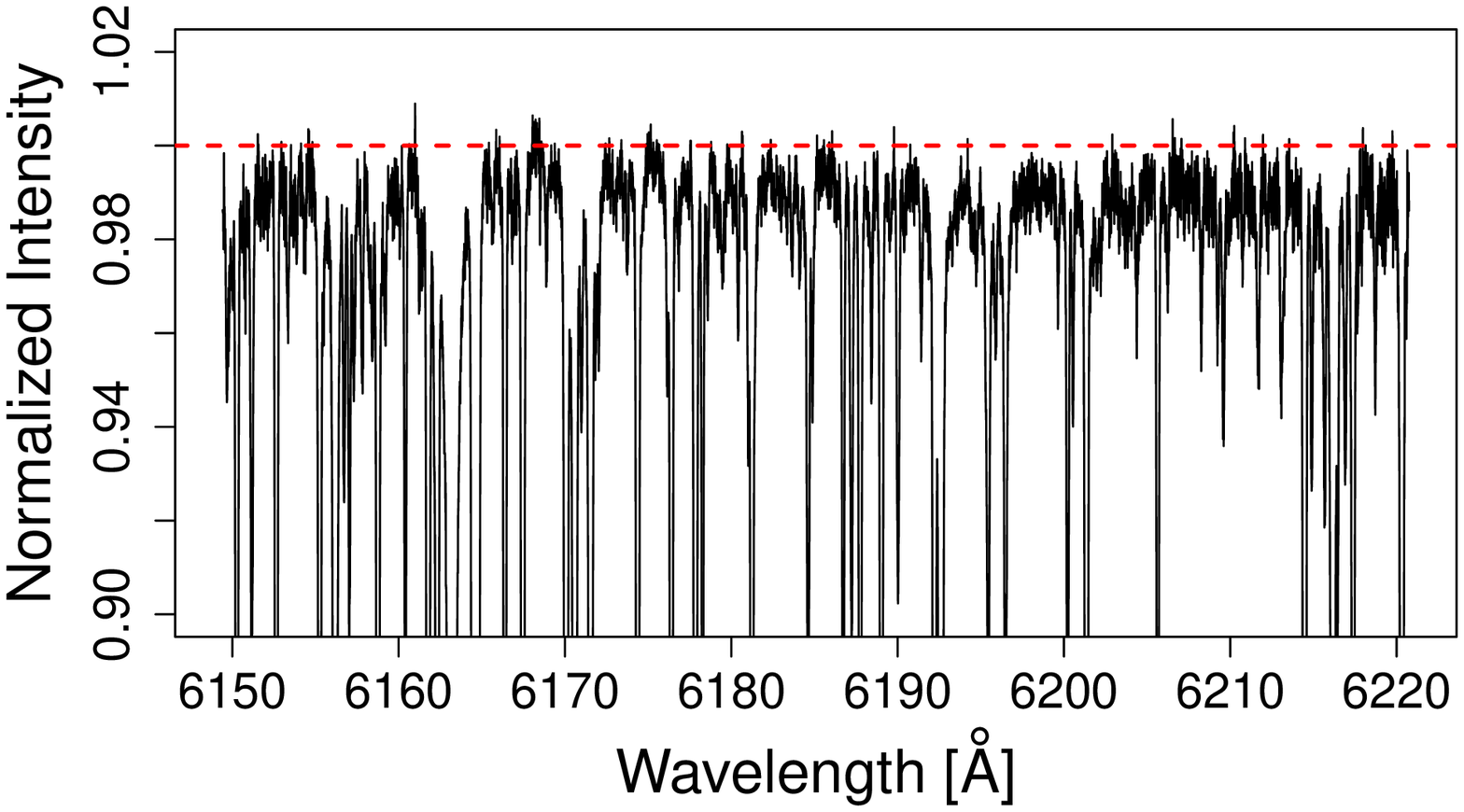}%
  \label{discuss7}%
}\qquad
\subfloat[Order R, Iterative]{%
  \includegraphics[width=0.45\columnwidth]{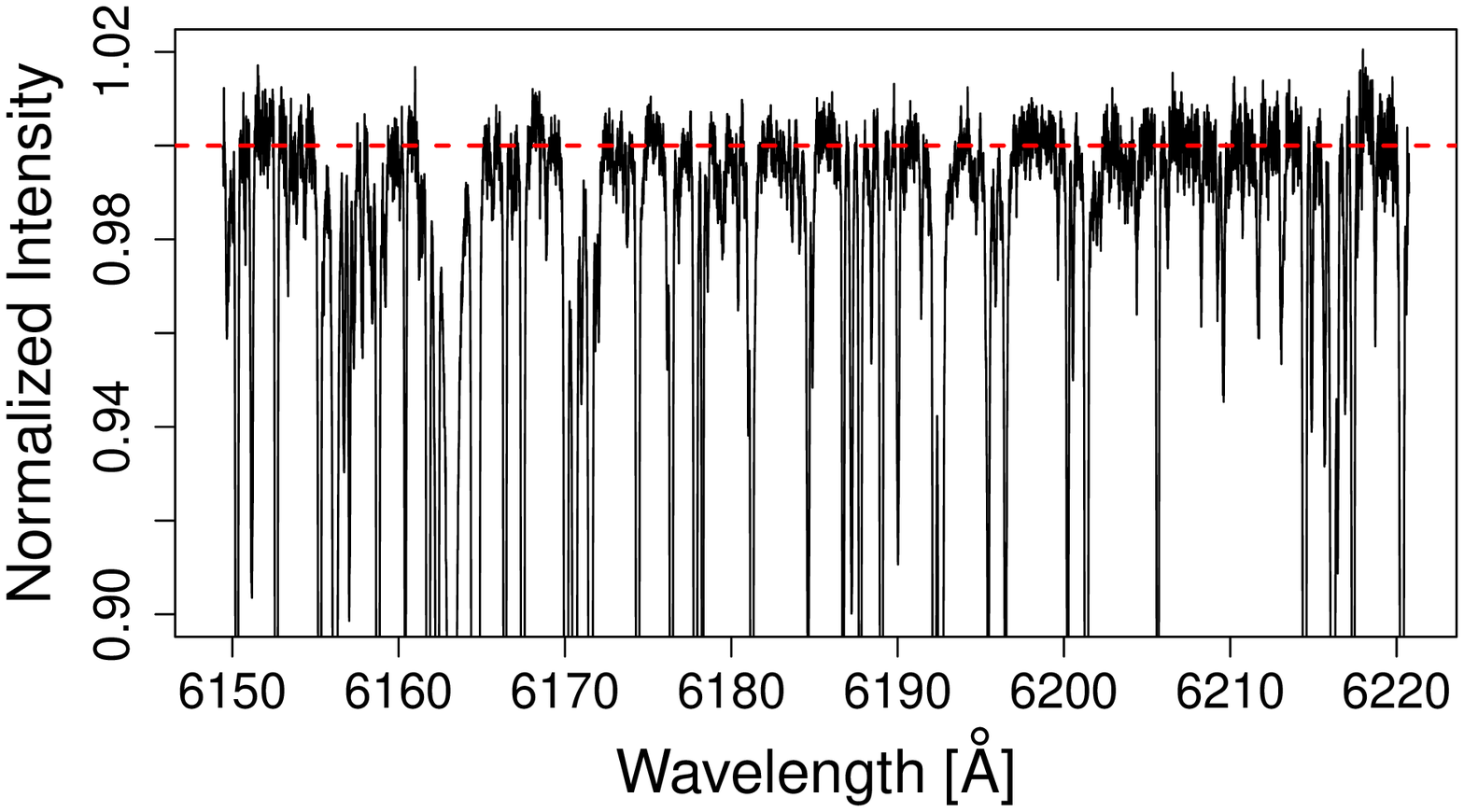}%
  \label{discuss8}%
}
\caption{ALSFS and the iterative method are applied to a blue order and a red order of the star 51 Pegasi, observed with EXPRES on July 8, 2018. (a) and (b) are the blue order flattened by ALSFS and the iterative method, respectively. (a) is relatively flat and (b) has severe boundary issues. (c) and (d) are the red order flattened by ALSFS and the iterative method. Both (c) and (d) are pretty flat, while (d) has a scale issue that the normalized intensities are higher than expected.} \label{discussion_51Peg}
\end{figure}

The ALSFS and iterative method are applied to a blue order (order B: 4473 to 4529 \angstrom) and a red order (order R: 6147 to 6223 \angstrom) of the star 51 Pegasi, observed with EXPRES on July 8, 2018. Amplified figures of the flattened orders are displayed in \autoref{discussion_51Peg}. For order B, the spectrum from ALSFS is mostly flat, while the spectrum from the iterative method has much higher intensities on boundary regions. For order R, both methods works well in flattening, but the iterative method is inaccurate in scale so that the peaks are above the reference line at Normalized Intensity $=1$.

We did not test the methods on cooler stars, where the continuum is poorly defined, such as M dwarfs. Since the sun is a relatively metal rich star, as is 51 Peg (metallicity of +0.2; \citealt{frasca2009rem}), our methods are expected to perform at least as well on stars with lower metallicities and higher temperatures.

\section{Conclusions}
In this work, we presented two data-driven algorithms, AFS and ALSFS, for removing the blaze function from spectra obtained from echelle spectrographs. 
The key aspects of the algorithms are the use of alpha shapes to provide an initial guess of the blaze function's shape, and the use of local polynomial regression to refine this guess. The two algorithms are designed for two scenarios: the AFS algorithm operates without a reference spectrum and may be applied directly to stellar spectra containing even a high number of absorption lines, while the ALSFS algorithm also incorporates additional information from a reference continuum spectrum to inform its initial guess. As an application of the AFS algorithm, a continuum lab source reference spectrum - such as an LED or quartz lamp spectrum - could be corrected and smoothed to be used in the ALSFS algorithm. 

A simulation study was presented to illustrate the performance of the proposed algorithms compared to the commonly used iterative method for spectral normalization. In general, our algorithms have smaller RMSE than the iterative method. 
Overall, the ALSFS algorithm has the smallest median RMSE when S/N is high. Moreover, our algorithms are able to capture the edge effects better than the iterative approach. ALSFS is relatively robust to edge effects, and we have also developed a method of boundary correction for the AFS algorithm. 
Furthermore, detailed discussion regarding the applications of the algorithms was presented with examples of B-star and star 51 Pegasi spectra, which are observed with EXPRES.

This work proposes methodology to correct the continuum of an echelle spectrum by modeling the blaze function of individual orders. 
A flattened echelle spectrum obtained from the proposed methods works better than its original form in studying physical and astronomical properties of a star, e.g., blaze-removed B-star spectra for understanding telluric lines, more precise absorption line depths for studying stellar activity.

\begin{figure}
\centering 
\subfloat[A blue order of 55 Cancri, S/N 130, temperature 5165 K, metallicity +0.27 dex: $q=0.99$, $m_0=0.35$]{%
  \includegraphics[width=0.45\columnwidth]{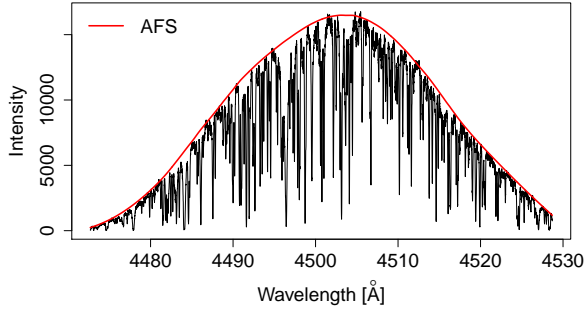}%
  \label{appendix_1}%
}\qquad
\subfloat[A red order of 55 Cancri, S/N 290, temperature 5165 K, metallicity +0.27 dex: $q=0.97$, $m_0=0.25$]{%
  \includegraphics[width=0.45\columnwidth]{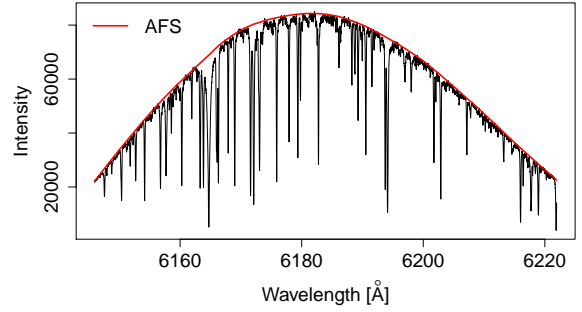}%
  \label{appendix_2}%
}\qquad
\caption{Examples for parameter selection.} \label{appendix1}
\end{figure}

\begin{figure}
\centering 
\subfloat[A blue order of NSO simulated spectrum, S/N 300, temperature 5778 K, metallicity 0 dex: $q=0.99$, $m_0=0.3$]{%
  \includegraphics[width=0.45\columnwidth]{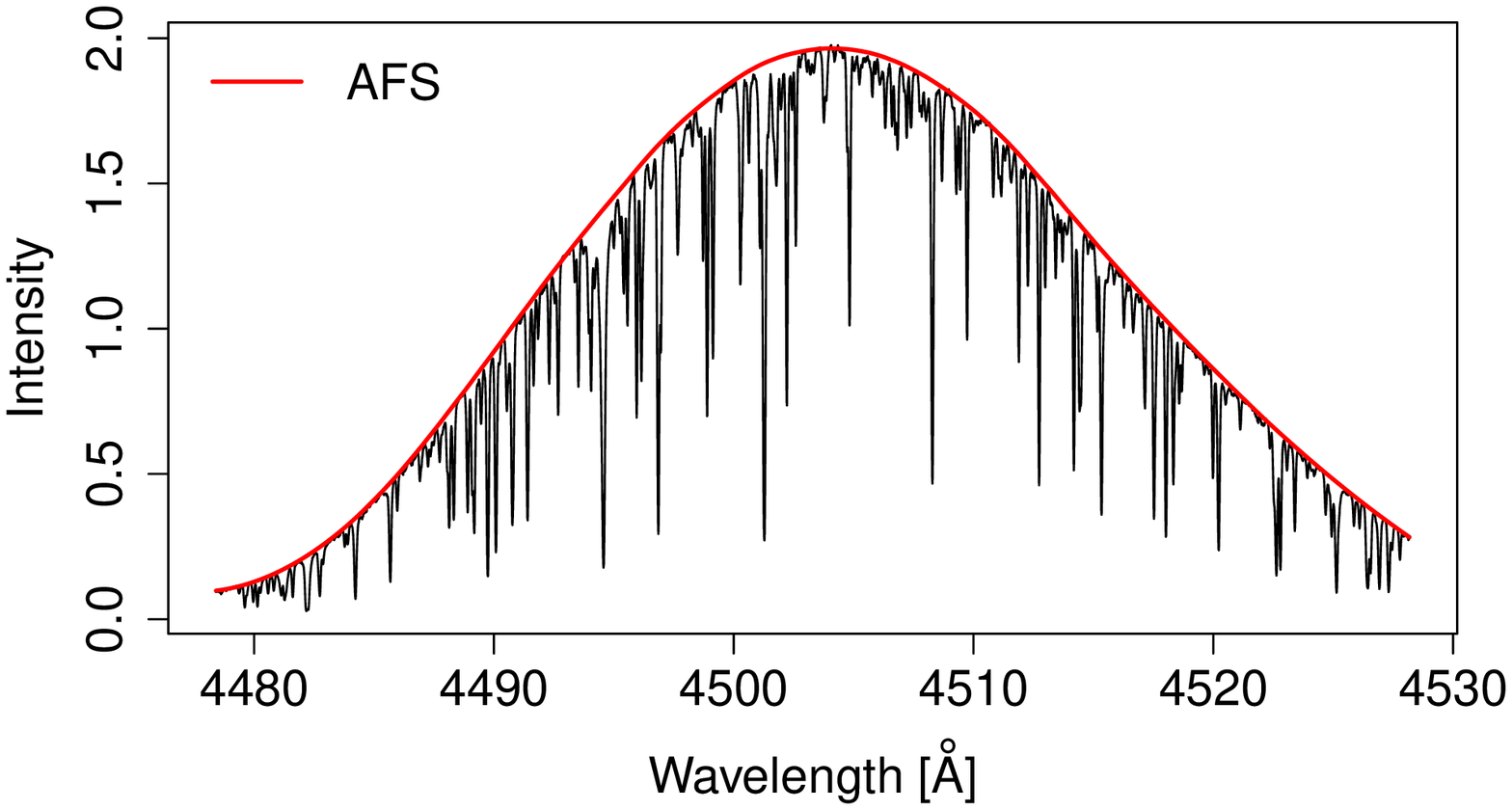}%
  \label{appendix_3}%
}\qquad
\subfloat[A red order of NSO simulated spectrum, S/N 300, temperature 5778 K, metallicity 0 dex: $q=0.95$, $m_0=0.25$]{%
  \includegraphics[width=0.45\columnwidth]{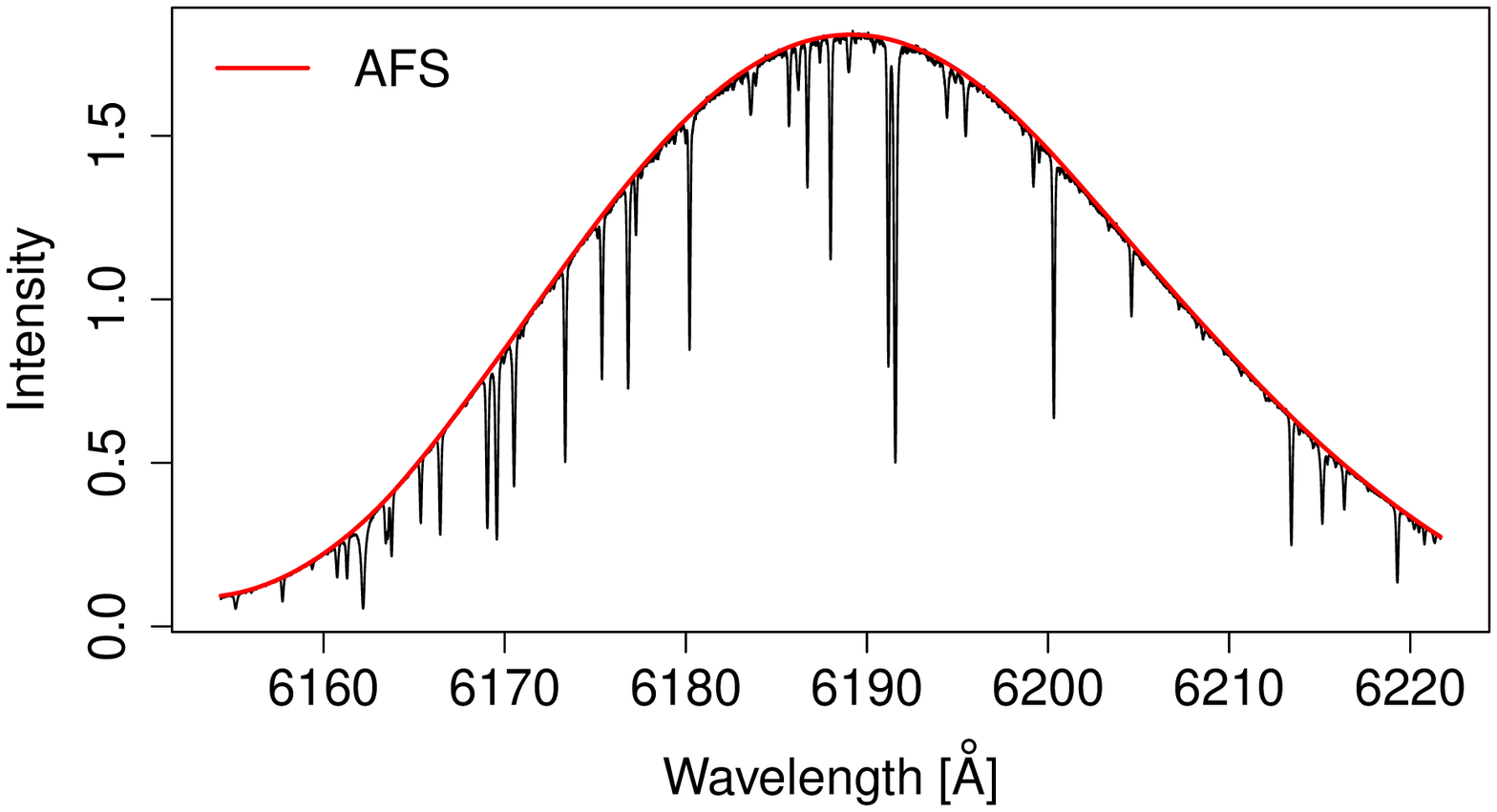}%
  \label{appendix_4}%
}\qquad
\subfloat[A blue order of NSO simulated spectrum, S/N 150, temperature 5778 K, metallicity 0 dex: $q=0.95$, $m_0=0.25$]{%
  \includegraphics[width=0.45\columnwidth]{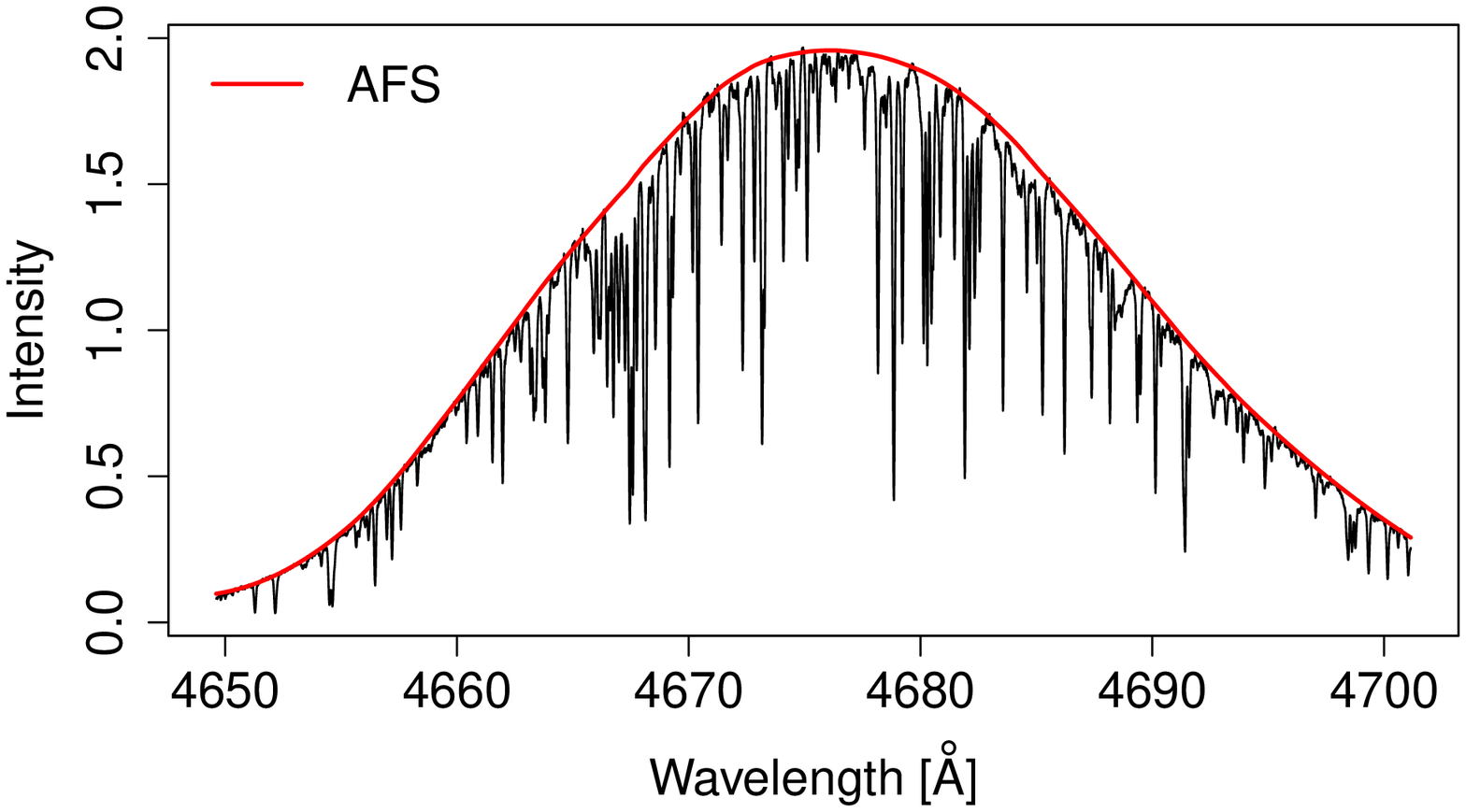}%
  \label{appendix_5}%
}\qquad
\subfloat[A red order of NSO simulated spectrum, S/N 150, temperature 5778 K, metallicity 0 dex: $q=0.8$, $m_0=0.25$]{%
  \includegraphics[width=0.45\columnwidth]{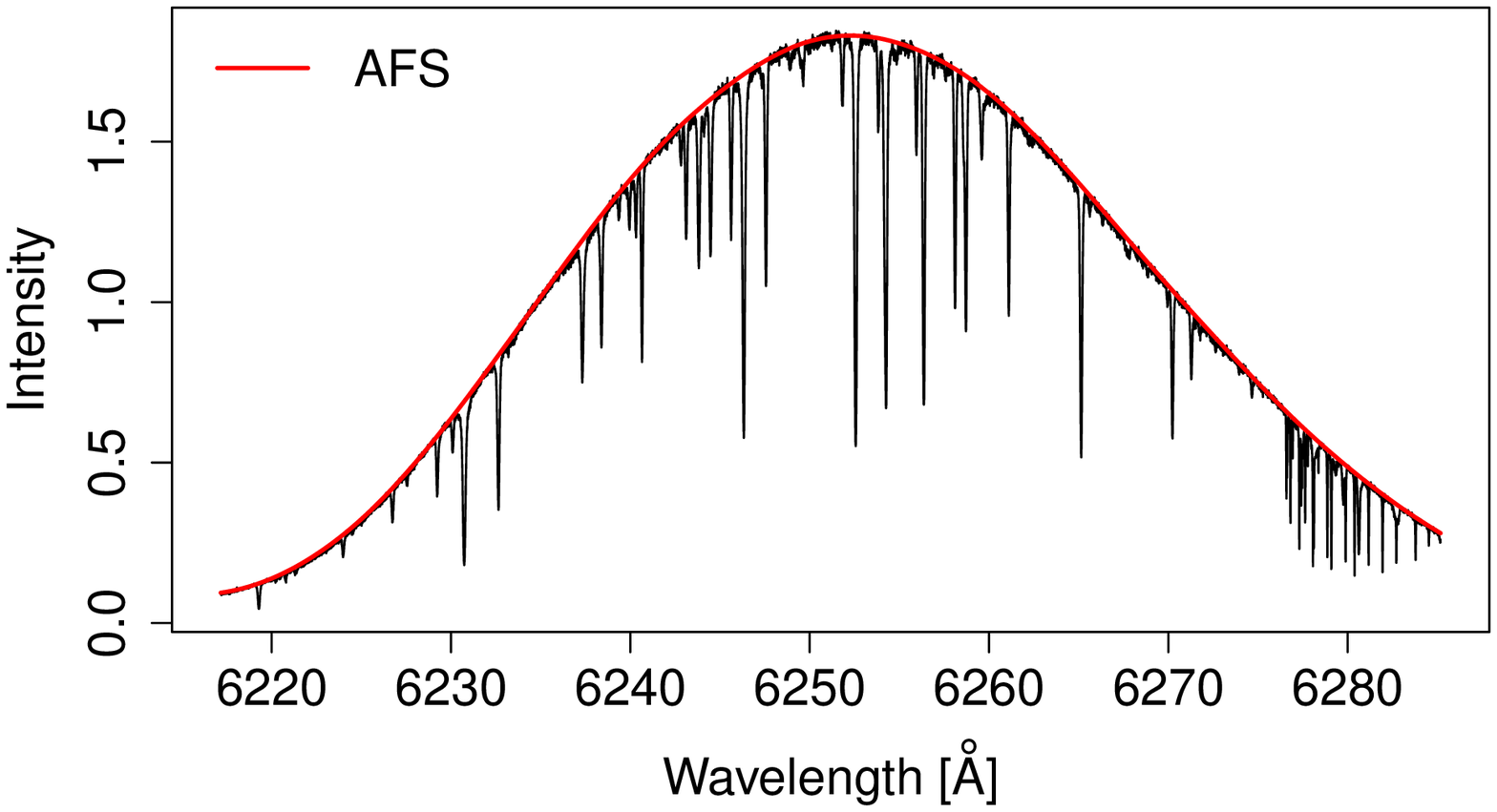}%
  \label{appendix_6}%
}\qquad
\subfloat[A blue order of NSO simulated spectrum, S/N 50, temperature 5778 K, metallicity 0 dex: $q=0.8$, $m_0=0.25$]{%
  \includegraphics[width=0.45\columnwidth]{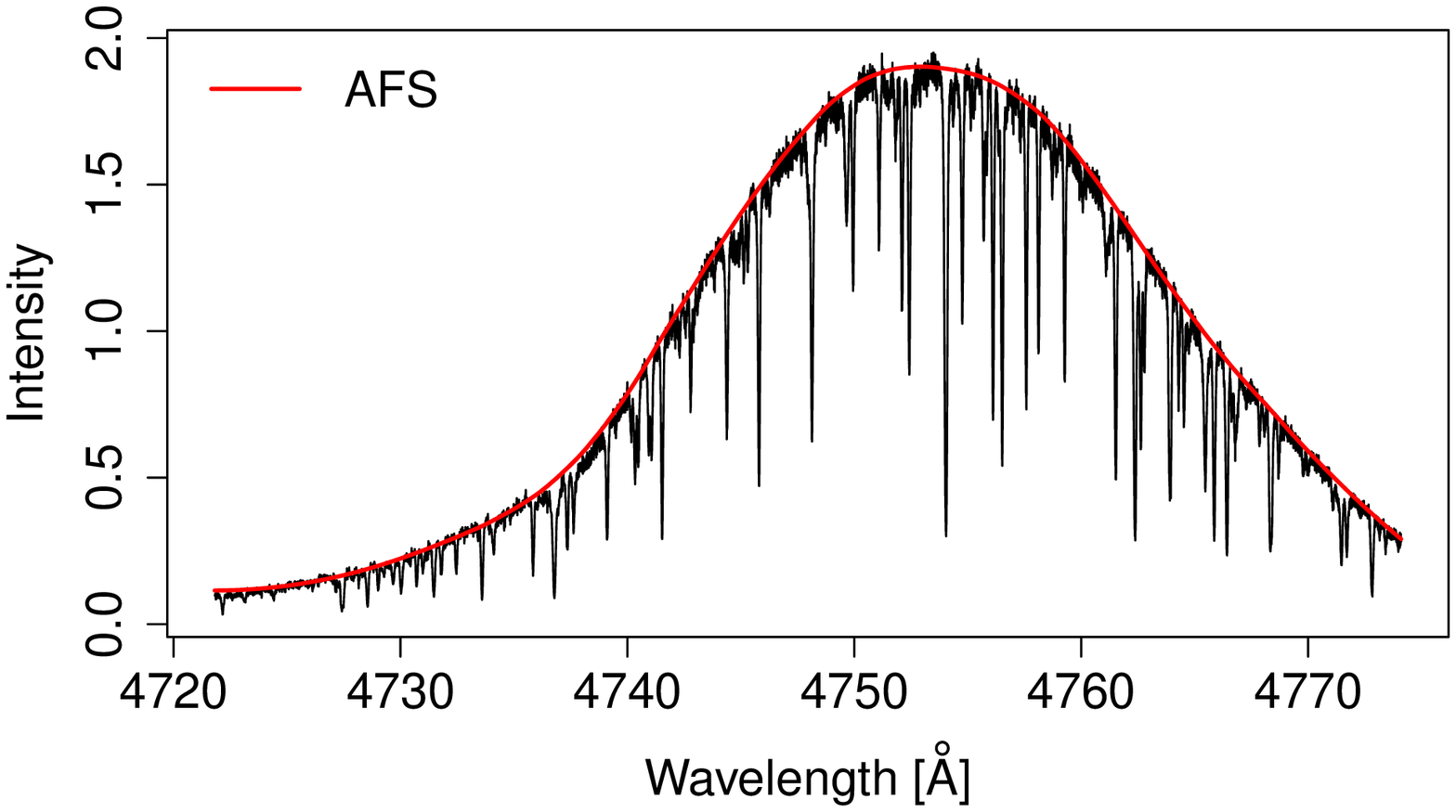}%
  \label{appendix_7}%
}\qquad
\subfloat[A red order of NSO simulated spectrum, S/N 50, temperature 5778 K, metallicity 0 dex: $q=0.5$, $m_0=0.25$]{%
  \includegraphics[width=0.45\columnwidth]{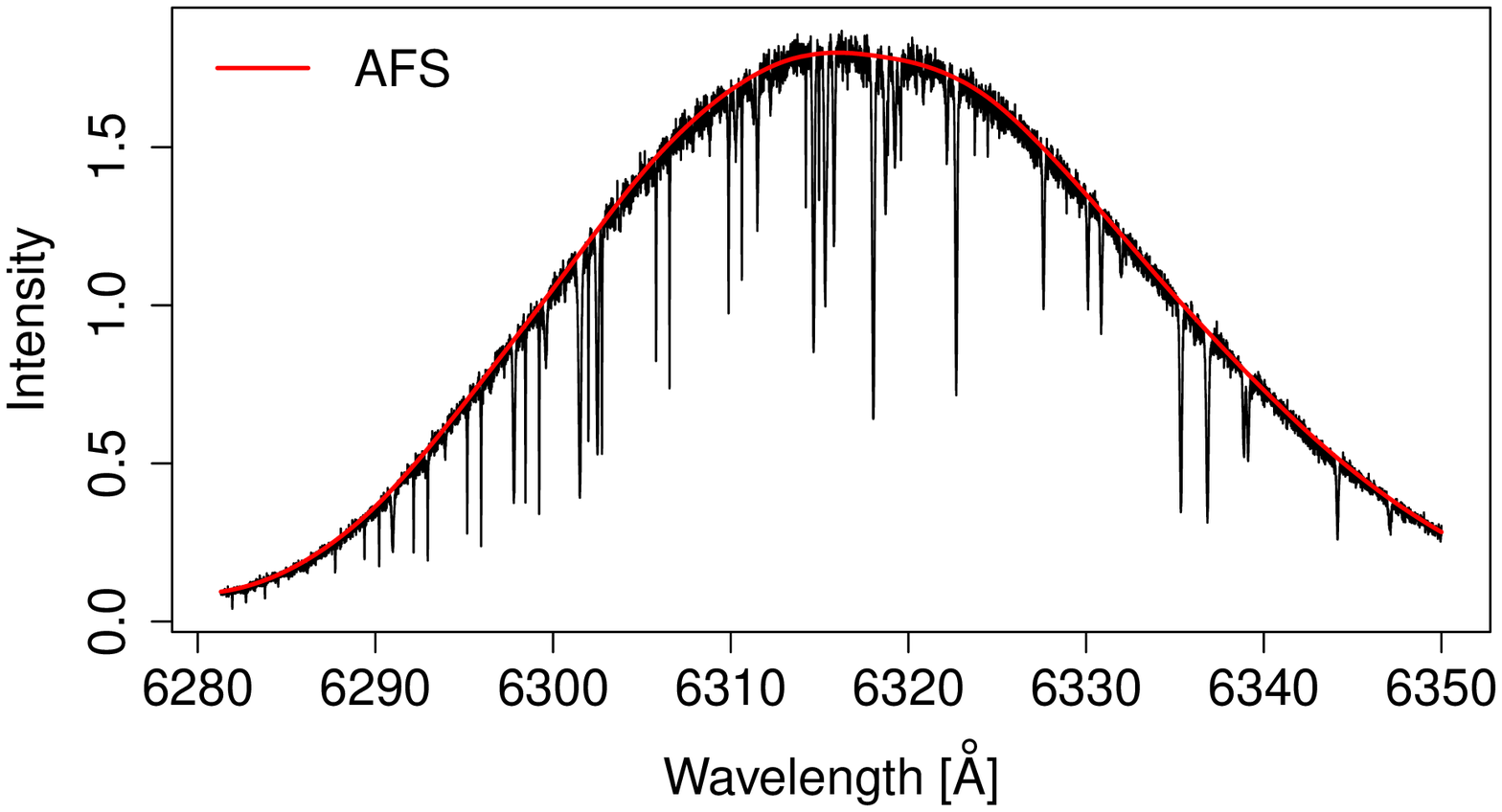}%
  \label{appendix_8}%
}\qquad
\caption{Examples for parameter selection.} \label{appendix2}
\end{figure}

\appendix\label{appendix}
The selection of the parameter $\alpha$ is discussed in \autoref{parameter} so in this section we focus on the selection of $q$ and $m_0$. The $m_0$ depends on the amount of absorption of an order, and $q$ depends on both the S/N and the amount of absorption. 
While S/N can be estimated, the amount of absorption is influenced by multiple factors such as wavelength, temperature, surface gravity, and stellar metallicity. Instead of recommending parameter values based on these factors individually, we provide several example orders to give an idea of the rough ranges of parameters to use. The examples below are all echelle spectra, and an $\alpha =\frac{1}{6}\times$wavelength range is used for all of them. The selected $q$ and $m_0$ values are listed under each figure. 
\autoref{appendix1} displays two orders of EXPRES spectra for the G8 V star 55 Cancri, which has a temperature of about 5165 K and a metallicity of +0.27 dex \citep{marcy2002planet}. 
\autoref{appendix2} displays six simulated orders from the NSO solar spectrum, as described in \autoref{simulation}. The noise is added to each order with S/N set to be $300$, $150$, or $50$.
We suggest these examples be used as a guide for parameter selection.


\newpage
\bibliography{paper}

\end{document}